\documentclass[journal]{IEEEtran}
\usepackage[caption=false]{subfig}
\usepackage{url}
\usepackage{graphicx}
\usepackage{balance}
\usepackage{makecell}
\usepackage{stfloats}
\usepackage{color,soul}
\usepackage{amsmath}
\usepackage{multirow}
\usepackage{hhline}
\usepackage{cite}
\usepackage{amssymb}
\usepackage{amsthm}
\usepackage{mathtools}
\usepackage{multicol}
\usepackage{comment}
\usepackage{adjustbox}
\usepackage{blindtext}
\usepackage{hyperref}
\definecolor{Gray}{gray}{0.9}
\usepackage{tikz}
\newcolumntype{M}[1]{>{\centering\arraybackslash}m{#1}}

\usepackage{booktabs, makecell, tabularx}
\newcolumntype{C}{>{\centering\arraybackslash}X} 

\usepackage{stfloats}
\usepackage{siunitx}
\usepackage{xcolor,colortbl}
\usepackage{enumitem}
\usepackage{blindtext}
\definecolor{Gray}{gray}{0.9}
\usepackage{tikz}
\newcolumntype{M}[1]{>{\centering\arraybackslash}m{#1}}

\DeclareMathOperator*{\argmin}{arg\,min}

\begin{document}

\title{\textcolor{black}{Applications of Distributed Machine Learning for the Internet-of-Things: A Comprehensive Survey}}

\author{Mai~Le, Thien~Huynh-The, Tan~Do-Duy, Thai-Hoc~Vu, Won-Joo~Hwang, and Quoc-Viet~Pham \vspace{-0.25cm}
\thanks{Mai Le is with the Department of Information Convergence Engineering, Pusan National University, Busan 46241, Korea, and also with the School of Computer Science and Statistics, Trinity College Dublin, The University of Dublin, D02 PN40, Ireland (e-mail: maile2108@gmail.com).}
\thanks{Thien Huynh-The and Tan Do-Duy are with the Department of Computer and Communication Engineering, Ho Chi Minh City University of Technology and Education, Vietnam (e-mail: \{thienht, tandd\}@hcmute.edu.vn).}
\thanks{Thai-Hoc Vu is with the Department of Electrical, Electronic and Computer Engineering, University of Ulsan, Ulsan 44610, Korea (e-mail: vuthaihoc1995@gmail.com).}
\thanks{Won-Joo Hwang is with the Department of Biomedical Convergence Engineering, Pusan National University, Yangsan 50612, Korea (e-mail: wjhwang@pusan.ac.kr).}
\thanks{Quoc-Viet Pham is with the School of Computer Science and Statistics, Trinity College Dublin, The University of Dublin, Dublin, D02 PN40, Ireland (e-mail: viet.pham@tcd.ie).}
}

\maketitle

\begin{abstract}
The emergence of new services and applications in \textcolor{black}{emerging wireless networks (e.g., beyond 5G and 6G)} has shown a growing demand for the usage of artificial intelligence (AI) in the Internet of Things (IoT). However, the proliferation of massive IoT connections and the availability of computing resources distributed across future IoT systems have strongly demanded the development of distributed AI for better IoT services and applications. Therefore, existing AI-enabled IoT systems can be enhanced by implementing distributed machine learning (aka distributed learning) approaches. 

This work aims to provide a comprehensive survey on distributed learning for IoT services and applications in \textcolor{black}{emerging networks}. In particular, we first provide a background of machine learning and present a preliminary to typical distributed learning approaches, such as federated learning, multi-agent reinforcement learning, and distributed inference. Then, we provide an extensive review of distributed learning for critical IoT services (e.g., data sharing and computation offloading, localization, mobile crowdsensing, and security and privacy) and IoT applications (e.g., smart healthcare, smart grid, autonomous vehicle, aerial IoT networks, and smart industry). From the reviewed literature, we also present critical challenges of distributed learning for \textcolor{black}{IoT} and propose several promising solutions and research directions in this emerging area.
\end{abstract}


\IEEEpeerreviewmaketitle
\section{Introduction}
\label{sec:Introduction}
\begin{figure*}[t]  
	\centering
	\subfloat[Number of IoT connections worldwide from 2019 to 2030. \label{Fig:SR_vs_NoGUs}]{\includegraphics[width=0.53\linewidth]{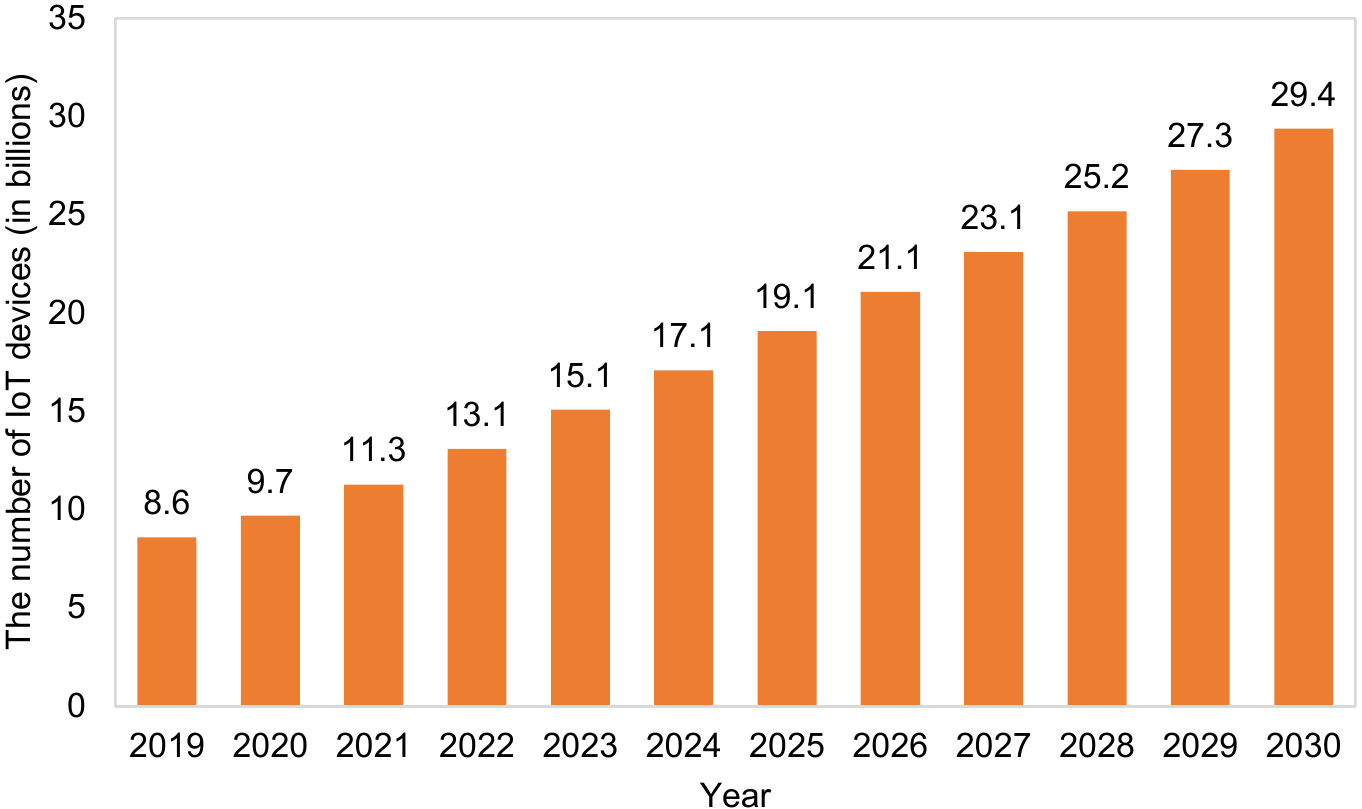}}\quad\;
	\subfloat[Number of related publications of distributed learning for IoT.\label{Fig:SR_vs_Pm_MBS}]{\includegraphics[width=0.435\linewidth]{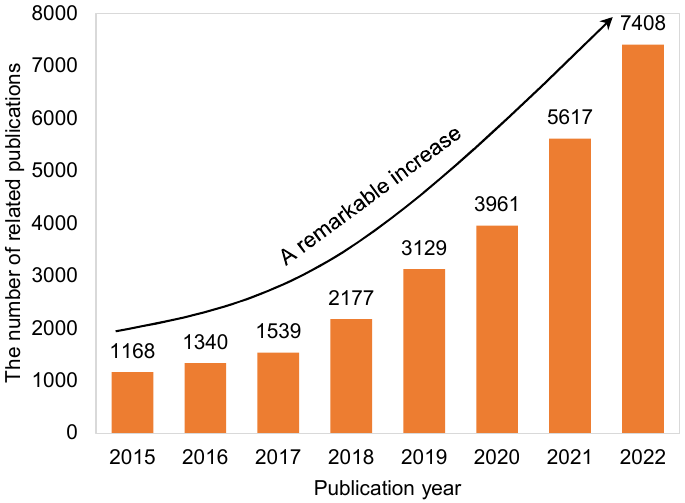}}
	\caption{Distributed learning in the future IoT (data as of March 08, 2023). }
	\label{Fig:IoT_DtL_trend}
\end{figure*}

The emergence of new services and applications, such as Metaverse, autonomous vehicles, and satellite networks,  has shown an explosive demand for the usage of artificial intelligence (AI) techniques in the Internet-of-Things (IoT) and the upcoming sixth-generation (6G) wireless networks \cite{letaief2019roadmap, liu2022distributed, xu2022full}. Besides, such demands have also been strongly driven by the significant proliferation of numerous IoT connections, recent innovations in IoT computing hardware, and the remarkable success of AI in various engineering disciplines, from military and computer vision to smart healthcare and mobile and communication networking. For example, Fig.~\ref{Fig:IoT_DtL_trend}(a) shows that the number of IoT connections in 2030 will reach $29.4$ billion, virtually $3.5$ times over $8.6$ billion connections in 2019\footnote{www.statista.com/statistics/1183457/iot-connected-devices-worldwide/}, and will offer a considerable amount of IoT data from countless applications. On the one hand, IoT devices are equipped with increasingly powerful computing, storage, sensing, and communication capabilities to process more complex learning tasks. 
On the other hand, massive IoT devices create massive data in many varied formats, which come from not only the physical world (e.g., healthcare, vehicular, and industry) but also the virtual environment (e.g., digital twin and Metaverse) \cite{xu2022full}. Therefore, appropriate AI techniques must be developed to exploit IoT devices' increasing capabilities and efficiently learn from the data at the same time so as to satisfy the requirements of 6G IoT \cite{de2021survey}. 

Nevertheless, sharing the raw data and information between mobile users and IoT devices via a cloud for central storage and learning purposes has a lot of potential shortcomings and risks. Specifically, transmitting a large amount of data consumes \textcolor{black}{excessive wireless resources, such as power budget and bandwidth, which are usually limited in practice}. Taking content caching at the network edge as an example \cite{wang2022learning}, network parameters are transmitted from distributed IoT devices to a central entity, which is then responsible for converting the received data to gray-scale images and deploying a convolutional neural network (CNN) architecture to accurately predict caching locations. Another example is \cite{doan2020learning}, where the central entity (e.g., radio base station) centrally collects radio characteristics from users, converts them to gray-scale constellation images, and deploys a CNN model, namely FiF-Net, for constellation-based modulation classification. Moreover, privacy issues become critical in centralized AI when sharing the user's sensitive information, such as gender, blood type, bank accounts, habits, and social activities. Aside from that, processing large amounts of data for learning tasks at \textcolor{black}{centralized servers} might cause some limitations on the performance of centralized AI, e.g., data bottlenecks, data information controls, user scheduling, or delays in service responses. \textcolor{black}{Furthermore, the rapid growth of massive IoT connections and the availability of more enhanced computing capabilities of IoT devices have also rendered the need to use outsourced supplies with proper storage and computing resources and communication-efficient designs to realize some learning tasks for distributed IoT devices in centralized AI.} 

As such, there has been a remarkable increase in the use of distributed learning for IoT in future 6G networks. Unlike centralized AI, distributed learning can exploit the availability of distributed computing resources across the network, from massive IoT devices to edge and cloud computing \cite{duan2022distributed}. For example, the performance of Gboard (also known as Google Android) on Android can be improved by distributed federated learning (FL) without asking for any historical data in the cloud \cite{gbroad}. Herein, mobile devices with Gbroad directly exploit the local histories to train their respective local model, while the cloud is only responsible for a weighted aggregation operation. Due to massive mobile devices (in order of billions as of 2019), Gbroad using distributed FL has shown a competitive performance (e.g., Top-1 recall and click-through rate) over the centralized learning approach \cite{hard2018federated}. That notable success has promoted the widespread use of distributed FL for various engineering fields, such as smart healthcare \cite{nguyen2022federated}, smart vehicles \cite{du2020federated}, and wireless networking \cite{qin2021federated}. \textcolor{black}{Moreover, different from centralized learning, which primarily focuses on the computing aspect (e.g., data preprocessing and model training), distributed learning also takes into account the communication between the cloud and distributed devices. In many scenarios in distributed learning, the communication overhead (e.g., transmission energy consumption or completion latency) is even higher than the computing overhead as multiple communication rounds are performed during the learning process before the learning model becomes converged. Therefore, it is important to design efficient and effective resource allocation schemes for distributed learning in IoT systems.}

Enormous computing, storage, and caching resources distributed across massive IoT devices and network nodes (e.g., edge servers and centralized clouds) help expedite the implementation of distributed learning solutions. In the meanwhile, distributed learning is a primary tool to enable the deployment of numerous new/emerging IoT services and IoT applications in 6G networks. On the one hand, \textcolor{black}IoT, with a massive number of distributed IoT devices, helps to distribute the data storage, model training, and learning inference of centralized AI, which mostly relies on the central clouds. On the other hand, distributed AI is expected to enable new IoT services and applications, which are difficult to support with conventional AI solutions. As an example in Fig.~\ref{Fig:IoT_DtL_trend}(b), there is a remarkable increase in the annual number of related studies on distributed learning for IoT. The result is based on an advanced search in IEEE Xplore\footnote{https://ieeexplore.ieee.org/search/advanced} with the two sets of keywords \{distributed learning, FL, split learning (SL), distributed inference, multi-agent reinforcement learning (MARL)\} and \{Internet\} and the OR combination function, over the publication period from 2015 to 2022. Note that the year 2022 also includes publications in early access. This analysis result shows that distributed learning will remain a powerful AI approach in the future of IoT.

\subsection{State-of-the-Arts and Contributions}
\label{sec:Introduction_Contributions}
Due to the importance of distributed learning in IoT services and applications in future networks, there have been several surveys on this topic. Aiming at providing a vision on 6G IoT, the work in \cite{nguyen20226g} first presents the fundamental technologies, such as edge AI, intelligent surfaces, and blockchain, and then discusses the representative applications, such as smart healthcare, industrial IoT (IIoT), the Internet of Vehicles (IoV), and aerial access networks (e.g., unmanned aerial vehicle (UAV) and satellite communications). The work in \cite{al2020survey} emphasizes the use of ML and DL to overcome a variety of IoT security threats and attack surfaces. A survey on deep reinforcement learning (DRL) for 4C (Communication, Control, Caching, and Computing) problems of IoT vertical applications is presented in \cite{chen2021deep}. The potential and applications of FL for IoT are reviewed in \cite{nguyen2021federated} and \cite{khan2021federated}. Meanwhile, the use of MARL for wireless networks is presented in \cite{feriani2021single, li2022applications}. An early review paper on distributed learning is \cite{predd2006distributed}, where its application for wireless sensor networks with a fusion center and with in-network processing is presented. There have been several surveys focusing on distributed learning, for example, distributed ML in \cite{verbraeken2020survey} and distributed learning for wireless networks in \cite{chen2021distributed} and \cite{qian2022distributed}. An overview of distributed ML in wireless communication is carried out in \cite{hu2021distributed}, where the use of FL for 5G applications, such as network topology control, power management, resource allocation, quality-of-service (QoS) provisioning, and multi-access edge computing (MEC), is reviewed. The work in \cite{ma2022trusted} discusses security and privacy issues in distributed learning from the communications perspective. More recently, the confluence of pervasive computing and AI (i.e., pervasive AI) with a focus on computation and communication challenges is reviewed in \cite{baccour2022pervasive}.
The existing surveys mentioned above and our contributions are summarized in Table~\ref{Table:Summary_ExistingSurveys}.

\begin{table*}[ht!]
    \renewcommand{\arraystretch}{1.10}
	\caption{Summary of related surveys on distributed learning and future networks. Acronyms: DtL = Distributed Learning.}
	\label{Table:Summary_ExistingSurveys}
	\centering
	\begin{tabular}{|p{0.95cm}|c|c|c|p{7.45cm}|p{5.80cm}|}
		\hline 
		\multirow{1}{*}{\textbf{Refs.}}  & \multicolumn{3}{c|}{\textbf{Review Topics}}  & \multirow{1}{*}{\textbf{Key Contributions}} & \multirow{1}{*}{\textbf{Limitations}}  \\ 
		
		\cline{2-4}
		{} & \textbf{DtL} & \textbf{IoT} & \textbf{6G} & {} & \\
		\hline
		\hline

        \multirow{2}{*}{\cite{duan2022distributed}} 
		& \checkmark &  & 
		& A survey on optimization techniques, security and privacy issues, and applications of end-edge-cloud computing for distributed AI.
		& The discussions of distributed AI for IoT services and applications in 6G are limited.
		\\ \hline
		
		\multirow{2}{*}{\cite{nguyen20226g}} 
		&  & \checkmark & \checkmark
		& A comprehensive survey enabling technologies and representative applications of IoT in future 6G networks.
		& This paper does not focus on the use of distributed learning for 6G IoT.
		\\ \hline
		
		\multirow{1}{*}{\cite{al2020survey}} 
		&  & \checkmark & 
		& A review of IoT security issues and attack surfaces and the use of ML/DL to design enhanced IoT security mechanisms. 
		& This paper is limited to conventional ML and DL approaches for security issues in IoT.
		\\ \hline
		
		\multirow{1}{*}{\cite{chen2021deep}} 
		&  & \checkmark & 
		& A review of DRL solutions for IoT vertical applications, including smart grids, blockchain, intelligent transportation, MCS, and IIoT.
		& This work only presents DRL for 4C problems and does not highlight distributed learning and 6G IoT. 
		\\ \hline
		
		\multirow{1}{*}{\cite{nguyen2021federated}, \cite{khan2021federated}} 
		& \checkmark & \checkmark & 
		& Surveys on FL approaches for various issues, services, and applications in IoT systems.
		& The papers only focus on FL for IoT, while recent advancements in distributed learning are ignored. 
		\\ \hline

        \multirow{1}{*}{\cite{feriani2021single}, \cite{li2022applications}} 
		& \checkmark & \checkmark & 
		& Extensive reviews on the use of single-agent RL and MARL for wireless networks.
		& The papers only focus on a specific learning approach (i.e., MARL), while their use in future IoT services and applications is not discussed. 
		\\ \hline
		
		\multirow{1}{*}{\cite{predd2006distributed}} 
		& \checkmark &  & 
		& A review of non-parametric distributed learning for wireless sensor networks with a fusion center and in-network processing. 
		& This work has been published for a long time and does not focus on 6G IoT. 
		\\ \hline
		
		\multirow{1}{*}{\cite{verbraeken2020survey}} 
		& \checkmark &  & 
		& An extensive review of challenges and opportunities of distributed ML over conventional ML. 
		& This work does not focus on communications and IoT aspects. 
		\\ \hline
		
		\multirow{1}{*}{\cite{chen2021distributed}} 
		& \checkmark &  & \checkmark
		& A survey on several distributed learning frameworks and an extensive discussion on how they can be leveraged in wireless networks. 
		& The applications of distributed learning for 6G IoT are not presented. 
		\\ \hline

		\multirow{1}{*}{\cite{qian2022distributed}} 
		& \checkmark &  & \checkmark
		& A review of the application of distributed learning for wireless networks in the physical, media access control, and network layers.
		& The applications of distributed learning for 6G IoT networks are not presented. 
		\\ \hline
		
		\multirow{1}{*}{\cite{hu2021distributed}} 
		& \checkmark &  & 
		& A review of distributed ML for 5G applications, such as power management, resource allocation, MEC, and QoS provisioning.
		& The discussions of distributed learning for 6G IoT networks and IoT services are not provided. 
		\\ \hline
		
		\multirow{1}{*}{\cite{ma2022trusted}} 
		& \checkmark &  & 
		& A definition of the four-level distributed learning framework and a survey on its security and privacy issues. 
		& This work only focuses on security and privacy aspects in distributed learning. 
		\\ \hline

        \multirow{1}{*}{\cite{baccour2022pervasive}} 
		& \checkmark & \checkmark & 
		& A survey on the computation and communication challenges of AI/DL for pervasive computing and IoT systems. 
		& This survey does limitedly discuss the use of pervasive AI for future IoT services and applications.
		\\ \hline
		
		\multirow{2}{0.9cm}{Our paper} 
		& \checkmark & \checkmark & \checkmark
		& A comprehensive survey of distributed learning for IoT services and applications. In particular,
		\begin{itemize}
		    \item We present the fundamentals of ML and recent advancements in distributed learning. 
		    
		    \item We provide a comprehensive review of distributed learning for IoT services and IoT applications in future 6G networks.
		    
		    \item We provide key challenges arising from the reviewed literature and propose several potential directions for future research on distributed learning for 6G IoT. \vspace{-0.25cm}
		\end{itemize}
		 & - 
		\\ \hline
	\end{tabular}
\end{table*}

From the summary and explanation above, it is apparent that there is a complete lack of a comprehensive survey on distributed learning for \textcolor{black}IoT. In particular, recent advancements in distributed learning (e.g., FL or Multi-agent DRL) for IoT services and applications in \textcolor{black}{emerging networks} have yet to be explored \cite{duan2022distributed, feriani2021single, li2022applications, predd2006distributed, verbraeken2020survey, chen2021distributed, qian2022distributed, hu2021distributed, ma2022trusted, baccour2022pervasive}. Moreover, several works (e.g., \cite{nguyen2021federated, khan2021federated, feriani2021single, li2022applications}) focus only on a specific kind of learning methods, while there is a distinct lack of applications to solve key issues in \textcolor{black}{IoT}, such as IoT localization, mobile crowdsensing (MCS), autonomous vehicles, and aerial networks.  
As a result, \textcolor{black}{we aim to complete} a comprehensive survey on the usage of distributed learning approaches for IoT services and applications in the context of future networks. In particular, we first present a preliminary overview of different distributed learning approaches, including MARL, FL, SL, and distributed inference. A key contribution of this work lies in an extensive discussion on the use of distributed learning for IoT services (e.g., computation offloading, IoT localization, MCS, security and privacy) and IoT applications (e.g., smart healthcare, smart grid, autonomous driving, aerial IoT network, and smart industry). \textcolor{black}{Specifically, we investigate the adoption of distributed AI approaches for the realization of IoT services and applications by covering important issues in distributed AI, such as overcoming IoT system dynamics and heterogeneity, and optimization of limited communication resources and finite IoT computing capabilities.} Further, we summarize our review, based on which we present several critical challenges in the literature and highlight important research directions. In sum, the features and contributions presented by this work are as follows. 
\begin{itemize}
    \item \textbf{Background of distributed learning}: To signify our contributions, this work first provides the background of conventional ML and an introduction to recent advancements in distributed learning, including FL, SL, MARL, and distributed inference. 
    
    \item \textbf{Distributed learning for IoT services}: The opportunities offered by distributed learning for IoT services (e.g., computation offloading, IoT localization, MCS, and IoT security and privacy) in future 6G networks are presented. 
    
    \item \textbf{Distributed learning for IoT applications}: We perform an extensive review of the use of distributed learning for 6G IoT applications, such as IoT-enabled healthcare, smart grid, autonomous vehicles, aerial IoT networks, and IoT-enabled smart industry.
    
    \item \textbf{Discussion of challenges and potential research}: From the extensive review of distributed learning for IoT services and applications, we identify critical challenges in the literature on this topic and present a number of potential directions for future research.
\end{itemize}

\begin{table}[ht!]
    \renewcommand{\arraystretch}{1.025}
    \caption{List of frequently used acronyms.}
    \centering
    \begin{tabular}{|p{1.10cm}|p{5.75cm}|}
        \hline
        \textbf{Acronym} & \textbf{Description}\\
        \hline
        {3D}    & {Three-dimensional} \\ \hline
        {5G}    & {Fifth-generation} \\ \hline
        {6G}    & {Sixth-generation} \\ \hline
        {AAN}   & {Aerial access network} \\ \hline 
        {AI}    & {Artificial intelligence} \\ \hline
        {AIoT}  & {Aerial Internet of Things} \\ \hline
        {CAV}   & {Connected autonomous vehicle} \\ \hline
        {CNN}   & {Convolutional neural network} \\ \hline 
        {DDPG}    & {Deep deterministic policy gradient} \\ \hline
        {DL}    & {Deep learning} \\ \hline
        {DRL}   & {Deep reinforcement learning } \\ \hline 
        {DQN}    & {Deep Q-network} \\ \hline
        {FedAvg}  & {Federated averaging} \\ \hline
        {IID}  & {Independent and identically distributed} \\ \hline
        {IIoT}  & {Industrial Internet-of-Things} \\ \hline
        {IoMT}   & {Internet-of-Medical-Things} \\ \hline
        {IoV}   & {Internet of Vehicles} \\ \hline
        {IoT}   & {Internet-of-Things} \\ \hline
        {MARL}  & {Multi-agent reinforcement learning} \\ \hline
        {MADRL}  & {Multi-agent deep reinforcement learning} \\ \hline
        {MEC}   & {Multi-access edge computing} \\ \hline
        {ML}    & {Machine learning} \\ \hline
        {RL}    & {Reinforcement learning} \\ \hline 
        {SL}    & {Split learning} \\ \hline
        {UAV}   & {Unmanned aerial vehicle} \\ \hline 
    \end{tabular}
    \label{Tab:acronym}
\end{table}

\subsection{Paper Organization}
\begin{figure}[t]
	\centering
	\includegraphics[width=0.995\linewidth]{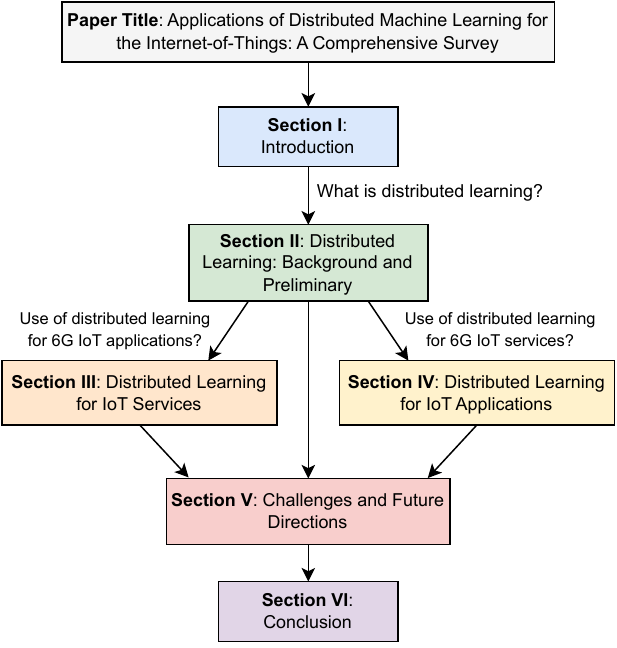}\vspace{-0.25cm}
	\caption{The outline of this paper with the main topic: distributed learning for IoT in emerging networks.}
	\label{Fig:Outline}
\end{figure}

This paper is organized as follows. In Section~\ref{Sec:Preliminary}, we present the background of ML and the preliminary to typical distributed learning approaches. Then, our main focus on the use of distributed learning for IoT services in future 6G networks is presented in Section~\ref{Sec:IoTservices}. Next, Section~\ref{Sec:IoTapplications} provides an extensive review of distributed learning for IoT applications, including smart healthcare, smart grid, autonomous vehicles, aerial IoT networks, and smart industry. In Section~\ref{Sec:challenges}, we summarize the two main sections (\ref{Sec:IoTservices} and \ref{Sec:IoTapplications}) and provide our discussions on the existing challenges and future research directions. Finally, we conclude this survey paper on distributed learning for 6G IoT in Section~\ref{sec:Conclusion}. 
The list of frequently used acronyms is shown in Table~\ref{Tab:acronym}. For ease of following the discussion, the paper outline is shown in Fig.~\ref{Fig:Outline}. 

\section{Distributed Learning: Preliminary}
\label{Sec:Preliminary}
In this section, we present the background of conventional ML and a preliminary to typical distributed learning approaches: FL, SL, MARL, and distributed inference. This section serves as a useful guide to general readers to understand the following sections of 6G IoT services and applications.

\subsection{Background of Machine Learning}
\begin{figure}[t]
	\centering
	\includegraphics[width=0.975\linewidth]{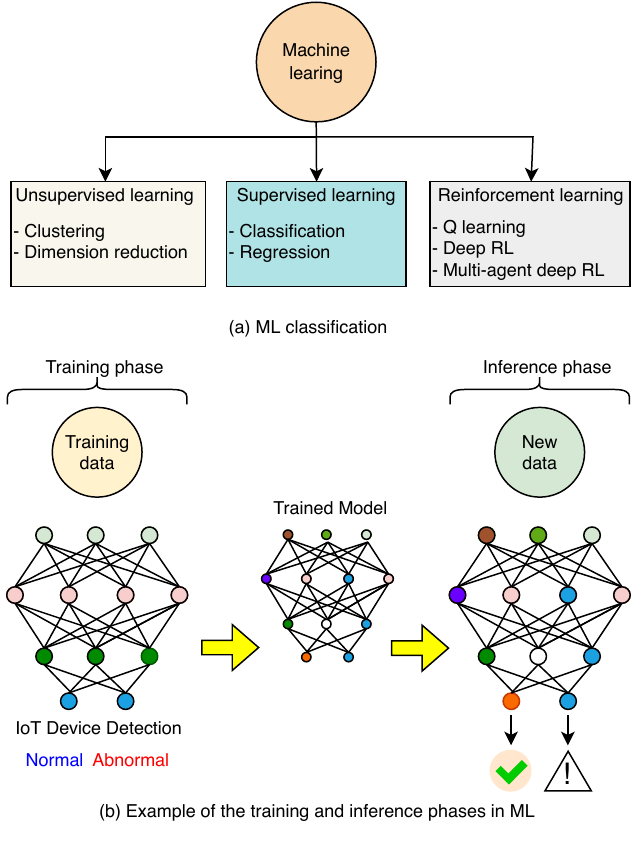}
	\caption{Machine learning: a) a general classification of ML and b) a general ML process including the training phase and the inference phase.}
	\label{Fig:ML}
\end{figure}

ML is a branch of AI that teaches a computer system to make accurate predictions with data. In general, ML can be categorized into three approaches, including supervised learning, unsupervised learning, and RL, as shown in Fig.~\ref{Fig:ML}(a). The training data in supervised learning includes both inputs and labeled outputs, and its goal is to estimate the unknown model that maps known inputs to known labels. The two main kinds of supervised learning are classification and regression. The former learns to predict a discrete class label (e.g., object classification), whereas the latter learns to predict a continuous quantity (e.g., price prediction). Unlike supervised learning, the training data does not include labels in unsupervised learning, and its goal is to learn a more efficient representation of a set of unknown inputs. A well-known example of unsupervised learning is k-means clustering, which can be applied to numerous clustering problems in IoT systems, such as user clustering and placement optimization. In RL, the agent is not told what actions to take but continuously interacts with the learning environment and tries to find the good policy/action that generates the best cumulative reward.

As shown in Fig.~\ref{Fig:ML}(b), an ML problem can be divided into two phases: the training phase and the inference phase. In the training phase, the ML model (i.e., prediction or classification) is built based on the input data, which in turn can be either labeled or unlabeled. In conventional AI, the training data is collected from users for central training in a centralized cloud with powerful storage and computing resources. The second phase of the ML problem is inference when the live data is put into the trained model to produce the output actions. As an example, the trained ML model is used to identify abnormal IoT devices and detect new and unknown attacks in IoT systems \cite{vu2020learning}. In this example, the supervised ML model is leveraged to learn the latent representation to enhance the detection ability. In particular, the latent representation helps to divide the IoT data into two separate regions, including a normal region and an anomalous region. In this context, the new and unknown attack data is likely in the anomalous region as they are expected to have common characteristics. From the ML process perspective, the latent representation is learned by observing the reconstruction error between the input data and the output data and using well-known IoT botnet datasets in the training phase, whereas the learned latent feature is then used in the inference phase to detect abnormal attacks. It is worth noting that advancements in computing hardware enable the learning servers to be equipped with powerful resources for training large models. However, it is possible that a large model is nearly impossible to train on a single machine in special circumstances, such as the very large amount of IoT data and the very long training time. Thus, distributed learning has emerged as a promising approach.

\begin{figure*}[t]
	\centering
	\includegraphics[width=0.975\linewidth]{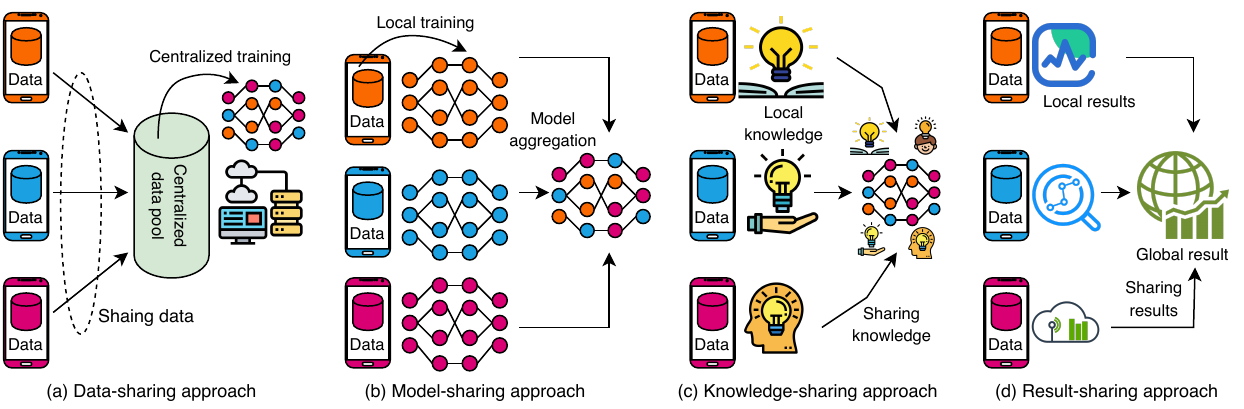}
	\caption{Illustration of main distributed learning approaches.}
	\label{Fig:DtL_classifications}
\end{figure*}

When ML algorithms need to be deployed in a distributed fashion with the involvement of multiple cloud/edge servers and IoT devices, a general solution approach is to share the data, the model, the knowledge, or the result among the clients, which are referred to as the data-sharing approach, the model-sharing approach, the knowledge-sharing approach, the result-sharing approach, respectively \cite{verbraeken2020survey, ma2022trusted}. These four classifications of distributed learning are respectively illustrated in the sub-figures in Fig.~\ref{Fig:DtL_classifications}. In distributed learning, the system typically consists of a centralized server (also known as the parameter server) and multiple distributed devices. In the data-sharing approach (e.g., distributed Q-learning and parallel advantage actor-critic), the server trains the global learning model based on the data shared from distributed devices. Similarly, the model-sharing approach enables distributed devices to train their respective models using their local data and only share the trained model with the server for model aggregation. FL and asynchronous advantage actor-critic are two well-known examples of this approach, in which the aggregated model is distributed from the server to devices to start a new learning round. Unlike these two approaches, which share the data and trained models, the knowledge-sharing approach, such as SL and knowledge distillation, allows devices to extract the knowledge based on their local data and then share it with the learning server. This approach of knowledge sharing is similar to the concept of swarm intelligence, which reflects the collective behavior of distributed devices in IoT systems \cite{pham2021swarm}. Finally, distributed devices in the result-sharing approach only share the outputs of their trained models with the server. An example of this approach is the PATE (i.e., Private Aggregation of Teacher Ensembles) framework proposed in \cite{papernot2017semi}, which models the learning system with multiple teachers with disjoint datasets and one student who, in turn, noisy votes the best teacher without requiring access to trained parameters and the raw data of the teachers.

Moreover, many studies have recognized the importance of the communication aspect in distributed ML, besides the computation aspect in centralized learning. This is because many parts of the learning process are executed at end devices, and the outputs need to be shared with other devices, edge nodes, or clouds in the network. 
Thus, the communication issue should be adequately considered in distributed learning \cite{xu2022edge}. In this regard, several novel solutions have been proposed to leverage wireless communication channels to ease the communication and computation burden in distributed learning. For example, the principle of over-the-air computation (labeled as OAC or AirComp in the literature) can be applied to speed up the model aggregation in FL with digital beamforming \cite{yang2020federated} and analog beamforming \cite{wang2022edge}, and in SL over multi-antenna networks \cite{yang2022over}. When both aspects matter, it is important to scale up conventional DL models over large-scale and distributed IoT systems. A potential solution is to deploy a small device-side DL model on each device and periodically share the knowledge from small models to update the globally large DL model. Taking \cite{he2020group} as an example, small CNNs (e.g., ResNet-4 or ResNet-8) are trained at distributed IoT devices, and a large CNN (e.g., ResNet-56 and ResNet-110) is updated at the server. While the accuracy is almost the same as FedAvg in vanilla FL, the computational complexity is significantly reduced from $9$ to $17$ times on IoT devices.

\color{black}
Different from centralized ML frameworks that train an AI model on a single cloud, distributed ML frameworks distribute the training process across multiple devices and have several advantages. First, distributed ML can be used to train AI models on large datasets that would be too large to fit on a single device. Second, distributed ML can be used to train models on IoT devices that are geographically distributed, thus improving the performance of the model by reducing the training time and inference latency. Third, distributed ML can be used to train models on devices with limited computation power, which can save energy. 
On the other hand, distributed ML also has some disadvantages compared to centralized learning. First, distributed ML can be more complex to set up and manage than centralized ML frameworks as the learning process spans multiple devices. Second, distributed ML can require more communication rounds between the devices, which can slow down the training process and consume more communication resources. Third, spreading data across multiple devices for distributed processing can pose some security risks and privacy threats.
From these aspects, centralized ML frameworks are simpler to set up and manage as the learning process is done on a single device, and less vulnerable to security attacks as powerful security solutions can be employed to protect the centralized data pool.
The choice of which ML framework to be used depends on the specific applications and network scenarios. For training a model on large and geo-distributed datasets or distributed devices with limited computation resources, a distributed ML framework is a good choice. To simplify the training procedure, a centralized ML framework becomes a more suitable solution.
In summary, the key differences between distributed ML and centralized ML are shown in Table~\ref{tab_DisVsCen}.

\begin{table*}[!ht]
\centering
\caption{\textcolor{black}{Comparison of distributed ML frameworks and centralized ML frameworks}}
{\color{black}\begin{tabular}{|l|l|l|}
\hline
Feature & Distributed ML frameworks & Centralized ML frameworks \\ \hline
Data & Distributed across multiple devices & Stored on a single device \\ \hline
Computation & Distributed across multiple devices & Performed on a single device \\ \hline
Communication & Requires communication between devices & Does not require communication between devices \\ \hline
Security & More vulnerable to security attacks & Less vulnerable to security attacks \\ \hline
Complexity & More complex to set up and manage & Simpler to set up and manage \\ \hline
Scalability & Can scale to large datasets & Cannot scale to large datasets \\ \hline
Cost & More expensive & Less expensive \\ \hline
\end{tabular}}
\label{tab_DisVsCen}
\end{table*}
\color{black}

%



\subsection{Federated Learning and Split Learning}
\begin{figure}[t]
	\centering
	\includegraphics[width=0.875\linewidth]{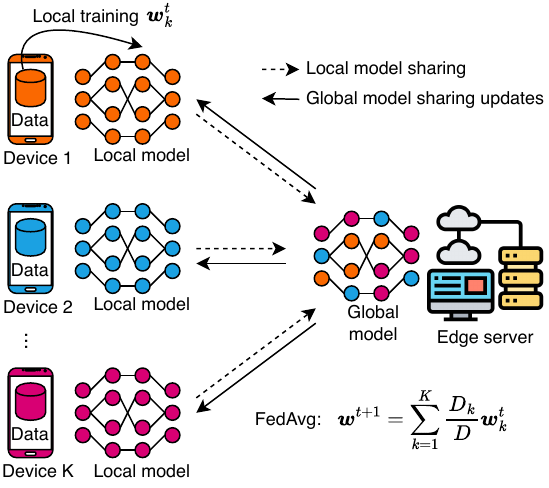}
	\caption{A general architecture of FL.}
	\label{Fig:Federated_learning}
\end{figure}

As a subset of ML that enables multiple devices to train a collaborative model without data sharing, FL has been considered a privacy-preserving solution for IoT \cite{li2020federated}. The general architecture of FL is shown in Fig.~\ref{Fig:Federated_learning}, where a set of $\mathcal{K}=\{1, \dots, K\}$ users (i.e., clients or devices) train the local models using their local data and share the local model updates with the edge server for model aggregation. As aforementioned, FL is a representative approach of model-sharing distributed learning. Each user $k$ has a dataset of $D_{k}$ samples, denoted as $\mathcal{D}_{k} = \{1, \dots, D_{k}\}$, and the total data samples of all users is given as $D = \sum\nolimits_{k = 1}^{K}D_{k}$.

As shown in Fig.~\ref{Fig:Federated_learning}, a global communication round of vanilla FL is composed of three main steps as follows.
\begin{itemize}
    \item \textbf{Model initialization}: The edge server initializes a model weight $\boldsymbol{w}^{0}$ of the global model and broadcasts its value to the selected users in the current communication round. Note that due to data heterogeneity, limited resources, and channel conditions, there may be only a subset of users to be selected to contribute to the FL process. Moreover, the edge server needs to specify the parameters of local models (e.g., local model accuracy and learning rate) and the parameters of the global model (e.g., global model accuracy). 
    
    \item \textbf{Local training}: At communication round $t$, each user $k$ trains a local model using its raw data set $\mathcal{D}_{k}$. The objective of local training of user $k$ is to minimize the loss function $f_{k}(\boldsymbol{w})$ as follows:
    \begin{equation}
        \boldsymbol{w}_{k}^{t} = \argmin f_{k}(\boldsymbol{w}).
    \end{equation}
    Then, the selected users shared the local model parameters with the edge server for model aggregation. Different loss functions may be applied to different learning tasks. For example, the linear regression model over the dataset $(\boldsymbol{x},\boldsymbol{y})$ is identified by the loss function $f(\boldsymbol{w}) = \frac{1}{2}(\boldsymbol{x}^{T}\boldsymbol{w} - \boldsymbol{y})^{2}$.
    
    \item \textbf{Model aggregation}: Once the edge server receives the local model updates from the selected users, the global model parameter can be obtained by minimizing the global loss function as follows:
    \begin{equation}
        f(\boldsymbol{w}) = \sum_{k = 1}^{K}\frac{D_{k}}{D}f_{k}(\boldsymbol{w}).
    \end{equation}
    A popular approach to this problem is Federated Averaging (also known as FedAvg) \cite{mcmahan2017communication}. In particular, the global model parameter can be updated as 
    \begin{equation} \label{Eq:FedAvg}
        \boldsymbol{w}^{t+1} = \sum_{k = 1}^{K}p_{k}\boldsymbol{w}_k^{t},
    \end{equation}
    where $p_{k} = {D_{k}}/{D}$ is the ratio of the data size of the IoT device $k$ to the total data. The FedAvg update in \eqref{Eq:FedAvg} indicates that the edge server takes the weighted average of the local models. In this regard, each user can update the local model parameter multiple times before sending the result to the server for model aggregation. According to \cite{yang2021energy}, the number of local iterations can be specified by the learning hyperparameters and the required accuracy of local models. 
\end{itemize}
The two steps of local training and model aggregation can be executed iteratively until the learning criteria are met. For example, the FL process may stop once the difference of the loss function in the two consecutive iterations is less than a threshold $\varepsilon$, that is, $\vert f(\boldsymbol{w}^{t+1}) - f(\boldsymbol{w}^{t}) \vert \leq \varepsilon$. Similar to the local model iteration, the number of global iterations is dependent on the learning hyperparameters, and the learning accuracy of the global model \cite{yang2021energy}. Furthermore, it is worth noting that the learning task in FL is typically generated by the cloud/edge server. However, in many FL scenarios, FL users are the model owners, and different users have a common interest in collaboratively building an AI model via knowledge sharing. In such cases, the edge server is employed to facilitate the FL process \cite{huang2022collaboration}.  

\textcolor{black}{There are ongoing research efforts and initiatives to make FL more feasible, scalable, and practical. For instance, besides Gboard, Google also developed TensorFlow Federated \cite{bonawitz2019towards}, an open-source framework for implementing FL and distributed ML algorithms. BMW developed an FL framework for autonomous driving, which allows vehicles to exchange their driving models and improve their performance \cite{Li2020May}. Microsoft launched Project Fiddle, a platform for FL experimentation and deployment \cite{BibEntry2022Feb}. Moreover, Microsoft developed Project Florida, a system architecture and software development kit (SDK) to facilitate the deployment of large-scale FL solutions across a heterogeneous ecosystem \cite{Diaz2023Jul}. NVIDIA released Clara Train SDK, featuring FL and making FL applicable with NVIDIA EGX, the edge AI computing platform \cite{BibEntry2022Sep}. NVIDIA also created an FL-enabled healthcare platform, which allows hospitals to train and validate AI models for various tasks, such as cancer detection, brain segmentation, and chest X-ray analysis \cite{BibEntry2023Jun}. By deploying a small-scale wireless network with a couple of desktops, the authors of some recent demos \cite{freitag2021poster, freitag2022experimental, barrachina2023cloud} showed the practical application of FL in wireless environments. However, it is worth noting that the majority of devices are limited in computing capability, storage capacity, and power budget, so the provision of energy sources for FL tasks is important. Therefore, optimization of resource allocation becomes an important research issue, as highlighted in Section~\ref{sec:Introduction}. Moreover, it is expected that advancements in wireless power transfer, energy harvesting, and wireless technologies will further facilitate the training and inference processes in FL at energy-limited IoT devices \cite{pham2021uav, pham2022energy, le2023wirelessly, wang2023overview}.}

\begin{figure}[t]
	\centering
	\includegraphics[width=0.885\linewidth]{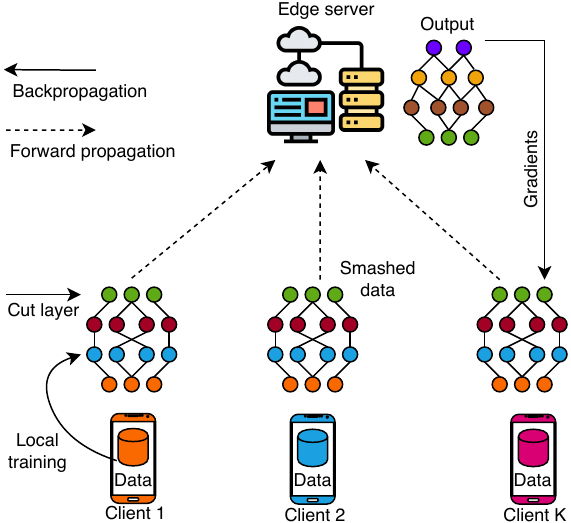}
	\caption{Illustration of split learning.}
	\label{Fig:Split_learning}
\end{figure}

Similar to FL, SL, also known as a split neural network, becomes very effective when clients do not want to share their raw data due to privacy issues \cite{gupta2018distributed}. In SL, a deep neural network (DNN) is split into two parts of neural networks: the first part of layers is trained by clients using their raw data, and the training of the remaining layers is completed on the server. The layer between the two parts is called the cut layer, and its outputs are referred to as smashed data. In general, the training in SL is similar to the training process of conventional deep learning (DL), including forward propagation and gradient backpropagation. Denote by $F_{c}$ ($F_{c}^{b}$) and $F_{s}$ ($F_{s}^{b}$) the function of the deep model (backpropagation) of client and server, respectively, by $L$ the total number of layers, and by $l$ the number of layers that are trained by clients. The client first uses forward propagation of the raw data. At the cut layer, the output  $F_{c}(\mathcal{D}_{k})$ and the corresponding label $\mathcal{L}$ are then sent to the server (e.g., edge and cloud). Taking the smashed data and label from the cut layer, the server continues involving forward propagation of the remaining layers, creating the output $\mathcal{Y}$ of the entire deep network. After that, the server computes the gradient for the final layer and backpropagates it until the cut layer, i.e., $F_{s}^{b}(\mathcal{L}, \mathcal{Y})$. This gradient of the first layer of the server is transmitted back to the client to update its model weights. These steps of one training round in SL are pictorially illustrated in Fig.~\ref{Fig:Split_learning}. 

\textcolor{black}{Since its successful deployment for the Google keyboard, FL has found numerous use cases and applications in mobile networks and distributed IoT systems. For example, FL can offer attractive benefits to current healthcare and Internet-of-Medical-Things (IoMT) systems by effectively overcoming their inherent limitations such as privacy concerns, lack of standard datasets, low performance due to limited healthcare training data, and high costs associated with health data training. In this regard, FL plays a key role in emerging applications in the healthcare sector, from remote health monitoring and medical imaging to COVID-19 detection and electronic health record management \cite{nguyen2022federated}.} Moreover, as an effective distributed learning approach, SL is helpful for many network scenarios due to its advantages. Compared to FL, SL can reduce the training workload on the client side, thus it is suitable for network scenarios when clients (e.g., sensors, drones, tiny IoT devices) are of low computing capabilities. However, this advantage is gained by an increase in the computing workload of the server and the difficulty in specifying the cut layer. In terms of communication efficiency, the work in \cite{singh2019detailed} shows that SL is advantageous when both the number of clients and the model size are large, while FL outperforms with a small number of clients and small model sizes.
In this regard, a hybrid FL-SL architecture may benefit the advantages of both when a set of IoT devices is allowed to implement SL and another set with more powerful computing resources performs FL. Several recent studies \cite{liu2022novel, thapa2022splitfed} show that the hybrid architecture has a higher learning accuracy than vanilla SL and FL-SL while better-preserving data privacy compared to conventional centralized learning. Moreover, a hybrid FL architecture is also promising \cite{huang2022wireless}. Particularly, a set of devices are privacy-sensitive, which are required to perform local training, and the remaining devices are privacy-insensitive, which can decide between two options: local training and sharing, or data offloading and centralized training. 
Furthermore, a variety of research aspects of FL and SL in IoT and mobile edge networks have been reviewed in \cite{nguyen2021federated, khan2021federated, lim2020federated}, and the readers are invited to read these articles and references therein for more details.

\subsection{Multi-Agent Reinforcement Learning}
RL is a central area of ML concerned with how the agents take actions that change the environment and maximize the cumulative reward. At each time, the learning agent perceives the observation from the environment, makes a decision, receives a reward, and transits to a new state. In general, RL is classified into model-based RL and model-free RL. While model-based RL relies on the assumption that the agent knows about the learning environment, the agent in model-free RL has no assumption of prior knowledge of the environment dynamics. From the implementation perspective, RL comprises value-based RL, policy-based RL, and actor-critic RL. 
In value-based RL, the learning agent tries to choose the best action based on the state-value function or the state-action value function. An illustrative example of value-based RL is Q-learning, an off-policy RL method that tries to maximize the expected return of the state-action value function. The core of Q-learning is a Bellman equation as follows:
\begin{align} \label{Eq:Qlearning}
    & Q_{t+1}(\boldsymbol{s}_{t}, \boldsymbol{a}_{t}) =  Q_{t}(\boldsymbol{s}_{t}, \boldsymbol{a}_{t}) \notag\\ & \qquad + \alpha \biggl( \underbrace{R_{t} + \gamma \max_{a}Q_{t}(\boldsymbol{s}_{t+1}, \boldsymbol{a})}_{\text{Predicted Q-value}} - \underbrace{Q_{t}(\boldsymbol{s}_{t}, \boldsymbol{a}_{t})}_{\text{Current Q-value}}\biggl),
\end{align}
where $Q_{t}(\boldsymbol{s}_{t}, \boldsymbol{a}_{t})$ is the Q-value reflecting the accumulated return by taking action $\boldsymbol{a}_{t}$ at the state $\boldsymbol{s}_{t}$, $R_{t}$ is the reward when the learning agents performs the action $\boldsymbol{a}_{t}$ at the state $\boldsymbol{s}_{t}$ and moves to a new state $\boldsymbol{s}_{t+1}$, $\alpha$ is the learning rate, and $\gamma$ is the discount factor. It is seen from~\eqref{Eq:Qlearning} that the notion of Q-learning is to find the difference between the predicted Q-value and the current Q-value of the two consecutive time steps, called temporal difference, which can be controlled by the learning rate $\alpha$. To improve the practicality of Q-learning, a DNN can be used to approximate the Q function. In particular, the optimal Q-value is now parameterized by a learning parameter $\boldsymbol{\theta}$, i.e., $Q_{t}(\boldsymbol{s}_{t}, \boldsymbol{a}_{t}, \boldsymbol{\theta}_{t}) \approx Q_{t}^{*}(\boldsymbol{s}_{t}, \boldsymbol{a}_{t})$. Further, the experience replay and a target network are used to stabilize the network updates in deep Q-learning \cite{mnih2015human}.  
Unlike value-based RL, policy-based RL directly learns a policy that is used to map the state to the action. A stochastic policy $\pi(\boldsymbol{a} \vert \boldsymbol{s}, \boldsymbol{\theta})$ is learned to map the state $\boldsymbol{s}$ to the probability distribution of the action $\boldsymbol{a}$ and the parameter $\boldsymbol{\theta}$ is updated by performing the gradient ascent of the performance measure $J(\boldsymbol{\theta})$ as follows:
\begin{equation}
    J(\boldsymbol{\theta}) = \mathbb {E}_{\pi _{\theta }}\left [{\sum_{t=0}^{\infty } \gamma ^{t} R\left ({\boldsymbol{s}_{t}, \boldsymbol{a}_{t}}\right)}\right].
\end{equation}
As a hybrid method, actor-critic RL combines both value-based RL and policy-based RL by the two components: actor and critic. While the actor aims to decide the action, i.e., policy-based RL, the critic estimates how good the action chosen by the critic is through the value function, i.e., value-based RL. As a result, actor-critic RL makes use of the advantages of both value-based RL and policy-based RL in a single RL approach. 
%

\begin{figure}[t]
	\centering
	\includegraphics[width=0.95\linewidth]{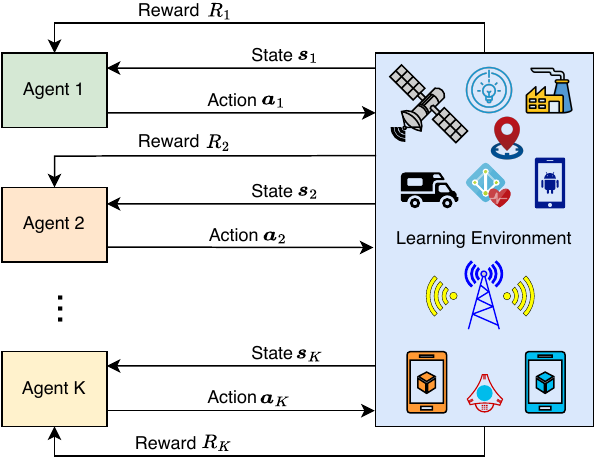}
	\caption{Illustration of the MARL concept.}
	\label{Fig:MARL}
\end{figure}

In practice, most network scenarios involve the participation of multiple users, especially in distributed IoT, where a large number of devices are distributed over different regions, and thus single-agent RL does not \textcolor{black}{well} perform in these networks \cite{guo2022multi}. As a generalization of single-agent RL, MARL concerns the sequential-decision-making problem of multiple agents over a common environment, as shown in Fig.~\ref{Fig:MARL}. Compared to single-agent RL, MARL has multiple benefits, including the speedup due to parallel computation and better performance due to experience sharing \cite{busoniu2008comprehensive}, but also raises several challenges \cite{nguyen2020deep}. First, multiple agents simultaneously interact with and reshape the learning environment, thus leading to a challenge of nonstationarity. Secondly, extending a single-agent system to a multi-agent environment when agents cannot access complete information is a challenging problem. Thirdly, training a multi-agent system is generally complex since the action of an agent is determined by not only its local observation but also the behavior of other agents. \textcolor{black}{Finally, novel learning approaches, such as transfer learning, FL, and meta-learning, have the potential to be an add-on in MARL systems since they can help overcome the challenges of low learning speed and high computational costs \cite{nguyen2022transfer, zhu2023transfer, ding2023multi, liu2023multi}. For instance, the integration of meta-learning and MARL could show a tangible improvement over the baseline schemes in MEC systems and notably the proposed algorithm could achieve a high-quality solution with fewer training steps, better stability, and good generalization \cite{ding2023multi}.}

Markov games, also known as stochastic games in the literature, are a general extension of the Markov Decision Process (MDP) and can be used to model multi-agent systems. A Markov game is represented by a tuple $<\mathcal{S},\mathcal{A},\mathcal{T},\mathcal{R}>$, where $\mathcal{S}$ is the joint state space of all the agents. For a system of $N$ agents as illustrated in Fig.~\ref{Fig:MARL}, $\mathcal{A} = \mathcal{A}_{1} \times \mathcal{A}_{2} \times \dots \times \mathcal{A}_{\textcolor{black}{N}}$ is the joint action space, $\mathcal{T}: \mathcal{S} \times \mathcal{A} \rightarrow p(\mathcal{S})$ is the state transition function, and $\mathcal{R} = \{R_{1}, R_{2}, \dots, R_{N}\}$ is the set of rewards with the reward function $R_{i}: \mathcal{S} \times R_{1} \times \dots \times R_{N}$ depending on the behavior of all the agents. Similar to the MDP objective, each agent in multi-agent systems tries to maximize its cumulative reward under the joint policy $\pi = (\pi_{1}, \pi_{2}, \dots, \pi_{N})$ as follows: 
\begin{equation}
    V_{i}^{\pi}(\boldsymbol{s}) = \mathbb {E}\left[ \sum _{t=0}^{\infty} \gamma^{t} R_{i}\left (\boldsymbol{s}_{t}, \boldsymbol{a}_{t},\boldsymbol{s}_{t+1}\right) \right].
\end{equation}
As such, the value measure of agent $i$ is dependent on the joint policy $\pi$ instead of its own policy $\pi_{i}$ in single-agent systems. Therefore, the concept of Nash equilibrium in game theory is applied in Markov games to find the optimal solution. More specifically, the joint policy $\pi^{*} = (\pi_{1}^{*}, \pi_{2}^{*}, \dots, \pi_{N}^{*})$ is said to constitute an Nash equilibrium if 
\begin{equation}
    V_{i}^{\pi_{i}^{*}, \pi_{-i}^{*}}(\boldsymbol{s}) \geq V_{i}^{\pi_{i}, \pi_{-i}^{*}}(\boldsymbol{s}), \forall s \in \mathcal{S} \text{ and } \pi_{i},
\end{equation}
where $\pi_{-i}$ denotes the set of policies of other agents except $\pi_{i}$. This definition can also be extended to the general case of $\epsilon$-Nash equilibrium as follows:  
\begin{equation}
    V_{i}^{\pi_{i}^{*}, \pi_{-i}^{*}}(\boldsymbol{s}) \geq V_{i}^{\pi_{i}, \pi_{-i}^{*}}(\boldsymbol{s}) - \epsilon, \forall s \in \mathcal{S} \text{ and } \pi_{i}.
\end{equation}
Here, $\pi_{i}^{*}$ is referred to as the best response of agent $i$ to the policies $\pi_{-i}^{*}$ of other agents.  Therefore, the principle under the design of most MARL algorithms is to find such a Nash equilibrium policy. Further, stochastic games can address different settings of multi-agent systems, including fully cooperative, fully competitive, and mixed settings \cite{zhang2021multi}. While the agents share a common reward function in fully cooperative multi-agent systems, the fully competitive setting and mixed setting can be modeled by well-known stochastic games, called two-player zero-sum stochastic games and general sum games, respectively. 

\subsection{Distributed Inference}
Once completed, the trained model may be used to run the inference stage on new data. Distributed learning is typically performed on large-scale datasets, which are spanned over a large number of distributed devices with disjoint datasets. AI techniques we have considered so far are mostly focused on the training phase and ignore the importance of the inference phase. However, inference is also an important step in distributed learning when the trained ML model needs to be deployed at distributed IoT devices at the network edge for inference purposes, such as object classification (i.e., classification) and price prediction (i.e., regression) \cite{gao2022wide}. Similar to the classification of distributed learning approaches, as shown in Fig.~\ref{Fig:DtL_classifications}, distributed inference can be categorized as cooperative inference, model inference, knowledge inference, and result inference. 
It is worth noting that inference can be completed at different tiers in the hierarchical cloud-edge-IoT architecture. According to the classification in \cite{zhou2019edge}, there are seven levels of inference, including cloud inference, cloud-training cloud-edge co-inference, cloud-training in-edge inference, cloud-training on-device inference, cloud-edge co-training and co-inference, edge inference, and on-device inference. The first four levels require training on the cloud, while the inference can be carried out at either individual tiers (e.g., on-device) or their cooperation (e.g., cloud-edge). On the contrary, both cloud and edge are involved in the training and inference phases in the fifth level. In the last two levels, both phases are done within the edge and the on-device manner only. Conventional inference methods, such as cloud inference and cloud-training on-device inference, require the sharing of raw data from IoT devices to the cloud for centralized training and inference.
As such, distributed inference is highly promising in protecting data privacy and easing the communication burden between the cloud and IoT devices. 

\begin{figure}[t]
	\centering
	\includegraphics[width=0.95\linewidth]{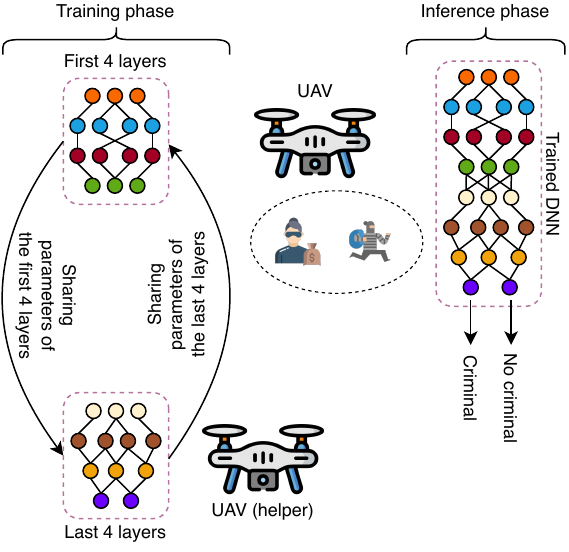}
	\caption{Illustration and example of cooperative inference.}
	\label{Fig:Cooperative_inference}
\end{figure}

In cooperative inference, a large neural network is divided into multiple parts, which are trained by different devices (e.g., IoT devices, edge clouds, and centralized clouds). For example, a UAV can be deployed to take images for surveillance activities and train a DNN to detect criminals on the ground. However, the UAV may not be able to train the whole DNN due to its limited computing and storage capabilities. Therefore, the UAV may ask for help from other UAVs in the swarm to train parts of the DNN. Suppose that the DNN is composed of $8$ layers; the UAV will train the first four layers while the remaining layers are trained by its helper (i.e., another UAV in the swarm), as illustrated in Fig.~\ref{Fig:Cooperative_inference}. Thus, the UAV shares the learning parameters with its helper once the training is finished. Then, the helper trains the remaining layers and updates gradients to the UAV. These processes are repeated until the learning criteria are met. Finally, the UAV can use the trained model in the inference phase to generate the output, i.e., whether or not there are criminals under its coverage area. 
In knowledge inference, IoT devices extract knowledge from their local data and observations and share it with the cloud/edge server to make the inference output. Similarly, in model inference and result inference, each IoT device trains the whole DNN before sharing necessary information with the cloud/edge server. While the parameters of local models are shared for model aggregation at the server, only local results (based on local models) are shared with the server to make the final inference. Compared with cooperative inference, where IoT devices only train a part of the DNN, knowledge inference, model inference, and result inference cause the great burden of high computing workload on IoT devices as each one needs to fully train large DNNs to extract knowledge, to achieve model parameters, and to generate local results, respectively. For example, ResNet-152 has $58.5$ and $117.6$ million parameters with 2D-CNN and three-dimensional (3D)-CNN, respectively. Storing a large number of parameters of such a large deep model in small IoT devices is cumbersome. \textcolor{black}{As a result, it is necessary to investigate innovative approaches to reduce the computation burden on IoT devices, such as sparse regularization, model compression, structural matrix designing, pruning, and binarization \cite{shi2020communication}.}

\subsection{Summary}
This section presents recent advancements in distributed learning techniques, including FL, SL, MARL, and distributed inference. 
However, these learning techniques can be fused in various ways to remarkably improve learning performance compared with that obtained by conventional and individual approaches. \textcolor{black}{For example, the integration of FL and RL is an efficient approach for solving complex and challenging problems in decentralized IoT systems}. Here, the principle of FL is considered to facilitate the implementation of centralized RL algorithms \cite{zhang2022federated, wu2022distributed, yin2022decentralized}. Taking \cite{yin2022decentralized} as an example, each base station in a multi-cell network is treated as a learning agent, which finds resource allocation actions via an RL technique (e.g., Q-learning and deep deterministic policy gradient (DDPG)). Then, the agent exchanges the learned parameters with its neighbor base stations in a fashion that preserves data privacy and facilitates decentralized implementation. Alternatively, RL is helpful to address optimization problems of distributed inference, such as classification latency \cite{baccour2022rl} and latency-energy tradeoff \cite{wang2022decentralized}. All in all, distributed learning is important in optimizing IoT services and applications with superior performance, as comprehensively reviewed in the following two sections.

\section{Distributed Learning for IoT Services}
\label{Sec:IoTservices}
This section extensively reviews distributed learning for IoT services, including data sharing and computation offloading, localization, MCS, and security and privacy.








\subsection{Distributed Learning for IoT Data Sharing and Computation Offloading}

\begin{figure}[t]
	\centering
	\includegraphics[width=1.0\linewidth]{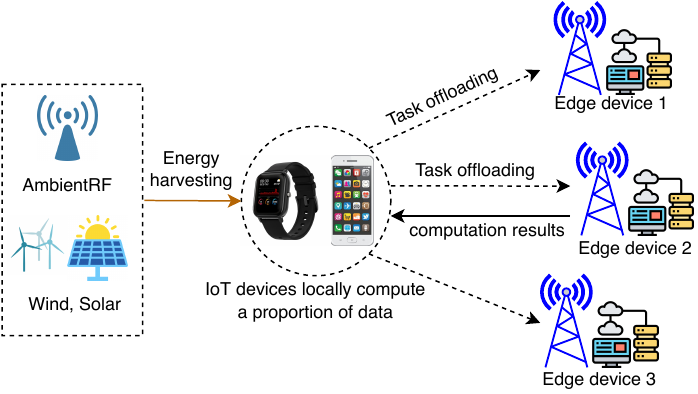}
	\caption{An MEC system with energy harvesting-powered IoT devices \cite{min2019learning}.}
	\label{Fig:MEC_EH_FL}
\end{figure}

In recent years, the rapid increase in the massive number of connected IoT devices has brought many challenges in enhancing the quality of service of emerging applications, e.g., smart cities, smart healthcare, and smart grid, through data sharing. Due to the limited computation resources of IoT devices, computation task offloading has thus received much attention. The computation tasks from a distributed device are offloaded to one or multiple edge/fog/cloud servers to reduce the execution latency and energy consumption by making optimal offloading and resource allocation decisions. In this regard, distributed learning models (e.g., FL and MARL) have been realized as a powerful approach to efficiently maintain the data sharing and computation offloading in IoT systems.
%
%
%
%
%
%
In \cite{min2019learning}, an RL-based computation offloading framework is proposed for IoT devices with limited energy constraints, aiming to optimize the offloading policy without knowledge models of the MEC, computational delay, and energy consumption. 
The MEC network architecture of a MEC system with an IoT device and multiple edge servers is illustrated in Fig.~\ref{Fig:MEC_EH_FL}.
Each IoT device, e.g., a smartphone and a smartwatch, is equipped with electricity storage elements and EH components, such as an RF energy harvesting module. A deep RL-based offloading (DRLO) algorithm is also introduced in \cite{min2019learning} to compress the state space dimension and thus improve the convergence speed of the offloading process. A quality function denoted by $Q$ is updated for each state-action pool by the DRLO algorithm involved in a dynamic computation offloading process as follows:
\begin{equation}
Q\left ( \boldsymbol{s}_{t}, \boldsymbol{a} \right)= \mathbb{E}_{\boldsymbol{a}'}\left [ R_{t} +\gamma \max_{\boldsymbol{a}' \in \boldsymbol{A}} Q\left ( {\boldsymbol{s}}', \boldsymbol{a}'\right ) | \boldsymbol{s}_{t},\boldsymbol{a} \right ],
\end{equation}
where the reward is defined as a function of task offloading ratio, task drop, energy consumption, and completion time. The numerical results show that the proposed learning-based computational offloading model can achieve significant $58.3\%$ lower energy consumption and $26.7\%$ less computation delay than the baseline binary MEC offloading scheme. However, such an offloading scheme in \cite{min2019learning} may not be applicable for large-scale distributed IoT systems, which raise the issues of low-quality data and data privacy preservation. Moreover, the agent in traditional single-agent RL approaches typically ignores the actions learned by other agents, making it vulnerable to the environment dynamics. 

\begin{figure}[t]
	\centering
	\includegraphics[width=1.0\linewidth]{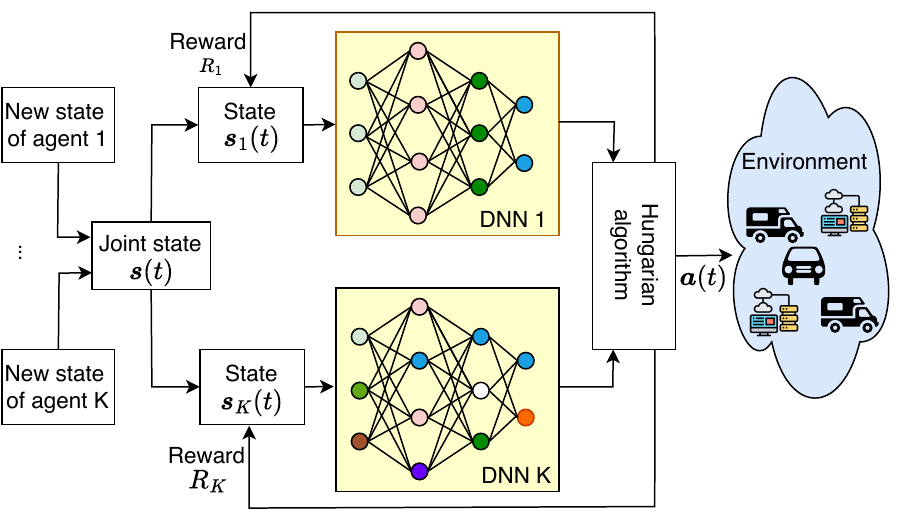}
	\caption{The integration of MARL and Hungarian algorithm for solving the computation offloading problem.}
	\label{Fig:MARL_CO}
\end{figure}

Recent studies have adopted the MARL technique to address these issues of computation offloading and data sharing in MEC systems \cite{tang2020decentralized, jiang2021distributed, zhu2021federated, alam2022multi, liu2022blockchain, chen2022joint, sun2022joint, zhou2022multiagent}. For instance, the work in \cite{tang2020decentralized} studies the decentralized computation offloading problem in energy harvesting-enabled IoT fog system to allow the IoT devices to make their own optimal offloading decision based on its local observed system state, e.g., the number of its remaining tasks, battery energy level, and available computing resources of the fog node. Specifically, an optimization problem is formulated as a decentralized, partially observable MDP, in which the main objective is to maximize the utility of all IoT devices in the network given strict delay requirements. In a high-mobility IoV environment, ignoring other agents’ actions during the training process reduces the benefits of collaborative interactions. Therefore, the work~\cite{alam2022multi} develops three-tier vehicular edge computing networks to enable vehicles to perform tasks locally, offload the task to the other cloudlet, or migrate the task to the centralized cloud. In such a case, the computation offloading problem can be formulated as a weighted bipartite graph, which can be solved optimally using the Hungarian algorithm, but it is very costly. To tackle this hurdle, \cite{alam2022multi} proposes an algorithm combining the MARL technique and the Hungarian algorithm, as shown in Fig.~\ref{Fig:MARL_CO}. In particular, the algorithm is implemented in a fashion of centralized learning and distributed inference. Aside from that, based on the $Q$ values learned by DNN agents, the cost of assigning the task from the vehicle $k$ to the cloudlet connected with the RSU $j$ when it associates with RSU $r_{i}$ is calculated as follows:
\begin{equation}
    c_{k,r_{i},j} = \frac{\exp(\alpha Q(\boldsymbol{s}_{k}(t), a_{k,r_{i},j}))}{\sum\nolimits_{o \in \mathcal{O}}\exp(\alpha Q(\boldsymbol{s}_{k}(t), a_{k,r_{o},j}))},
\end{equation}
where $\mathcal{O}$ denotes the size of the action space of the vehicle $k$. Then, the cost matrix is used by the Hungarian algorithm for cooperative resources to obtain the best reward for each vehicle.
As expected, the proposed MARL-assisted Hungarian algorithm is not only superior to the single-agent RL counterpart but also outperforms the classical MARL scheme and several baseline methods. The work~\cite{liu2022blockchain} proposes a computation task offloading approach based on blockchain and distributed RL. Besides a consortium blockchain to securely share data among IoT devices, evaluate and repair low-quality data, and achieve a fast consensus mechanism for task offloading, a distributed DDPG strategy is designed to maximize the system's utility. The distributed RL framework is composed of multiple distributed actors, which choose the actions of offloading decisions, computing resources, and repairing factors to maximize the following cumulative reward function
\begin{equation}
    R(\boldsymbol{s}_{t}, \boldsymbol{a}_{t}) = \sum\limits_{t = 0}^{\infty}\gamma^{t}\sum\limits_{k = 1}^{K} \left( \alpha \log(1 + \beta - T_{k}) + q_{k}  \right),
\end{equation}
where $\alpha$ and $\beta$ are system parameters, $T_{k}$ is the task completion time, and $q_{k}$ is the data quality of the IoT device $k$. It is shown that the distributed DDPG scheme needs more learning rounds to converge but offers a better cumulative reward than the scheme without considering data quality. Taking advantage of both game theory and RL, several solutions for computation offloading and resource allocation are derived based on both techniques, such as joint coalitional game and MARL for user association and data collection in \cite{chen2022joint} and for service caching and computation offloading in \cite{sun2022joint}, and joint non-cooperative game and MARL for task offloading in WiFi-5G coexisting networks in \cite{zhou2022multiagent}. The key result from these works is that the game-theoretic MARL approach can provide superior performance to conventional MARL methods or game-based schemes.

\textcolor{black}{To address bandwidth and privacy concerns in IoT data sharing, Chiu \textit{et al.}~\cite{chiu2020semisupervised} propose an edge learning system based on semi-supervised learning and FL. The system uses semi-supervised learning at the edge devices and periodically uploads the training results to the cloud server. To prevent weight divergence, a novel FedSwap operation is proposed to share a few data during federated training. The proposed FedSwap method achieves higher accuracy than FedAvg for image classification and object detection in video analysis applications.}
Also, aiming at dealing with the non-IID issue in FL, the work in \cite{zhao2022federated} analyzes the expected weight divergence between the FL model with non-IID data and the standard centralized learning with IID data. The theoretical proof shows that the weight divergence mainly depends on the data distribution divergence (denoted by $\Bar{\mathcal{L}}_{k}$ for IoT device $k$). Specifically, the weight divergence is inversely proportional to the sum of $\Bar{\mathcal{L}}_{k}$ of $K$ IoT devices. Inspired by this, a new weight scheme is proposed for improving the standard FedAvg scheme in \eqref{Eq:FedAvg} as follows: 
\begin{equation} \label{Eq:weightdivergence}
    \boldsymbol{w}^{t+1} = \sum_{k = 1}^{K} \frac {1 / \bar {\mathcal {L}}_{k}^{2}}{\sum _{k = 1}^{K} (1 / \bar {\mathcal {L}}^{2}_{k})} \boldsymbol{w}_k^{t}.
\end{equation}
Herein, the proportion weight $p_{k}$ is determined by the data size, i.e., $p_{k} = ({1 / \bar {\mathcal {L}}_{k}^{2}}) / {\sum _{k = 1}^{K} (1 / \bar {\mathcal {L}}^{2}_{k})}$. It is worth noting that the weight $p_{k}$ reflects not only the data size but also the data distribution divergence. In addition, a data sharing and user selection scheme is also proposed in \cite{zhao2022federated}, which allows IoT devices to train the local models with the improved training data consisting of local data and on-server data. Experimental results based on the MINST dataset show that the aggregation scheme with \eqref{Eq:weightdivergence} provides much better learning accuracy than FedAvg. For example, when employing the support vector machine (SVM) and CNN models, the proposed scheme offers learning accuracy up to $84.25$\% and $94.25$\%, respectively.

{\color{black}The research in \cite{lu2020blockchain} studies a collaborative data-sharing model in IIoT, where the proposed solution is to combine FL with permissioned blockchain to protect data privacy and prevent data leakage. Numerical results executed with two benchmark real-world datasets for data categorization tasks show that the proposed architecture can achieve high efficiency and enhanced security in data sharing. In another work \cite{ren2019federated}, multi-agent DRL (MADRL) is applied at multiple edge nodes to optimally determine whether IoT devices should offload their computation tasks. To reduce transmission costs among the network devices and cope with privacy issues, some randomly chosen IoT devices perform the training process using their local data and transmit the updated model parameters of the DRL agent back to the edge node for model aggregation. This approach can achieve superior performance compared to the conventional centralized training methods. More recently, \cite{bayerlein2021multi} formulates and solves a multi-UAV path planning problem for data harvesting missions from distributed IoT sensor nodes. This approach aims to identify efficient trajectories that maximize the collected data while satisfying safety and navigation constraints, thereby enabling optimally allocating data collection tasks to multiple UAVs and providing effective cooperation between them by learning and adapting to large complex environments and state spaces.}
\begin{figure*}[t]
	\centering
	\includegraphics[width=0.85\linewidth]{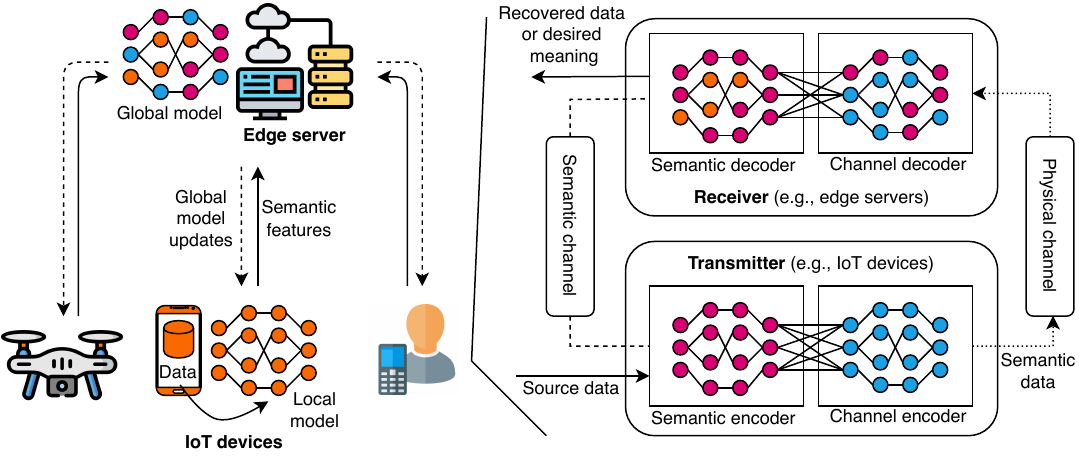}
	\caption{Illustration of FL-enabled SemCom systems.}
	\label{Fig:SemCom}
\end{figure*}

Distributed learning has also been integrated with emerging 6G technologies to increase the effectiveness of data transmission in many wireless scenarios. Seen as a key enabler of 6G IoT, semantic communication (SemCom) allows IoT devices to only transmit relevant data to the receiver instead of sending the entire data as in conventional communication systems \cite{yang2022semantic}. Thus, SemCom reflects the principle of many distributed learning techniques, such as MARL and FL, showing the full potential of their integration. Fig.~\ref{Fig:SemCom} illustrates the procedure of FL-enabled SemCom systems. Concretely, each IoT device trains a local model to extract semantic features from the source data (e.g., text \cite{xie2021deep} and audio \cite{tong2021federated}). Unlike FL, in FL-SemCom, the edge server receives the extracted semantic features to recover the meaning of the source data and update the global model. For text data, \cite{xie2021lite} shows the superiority of FL-SemCom over the conventional SemCom method, e.g., the model size is reduced from $12.3$ MB to $1.28$ MB while the bilingual evaluation understudy (BLEU) score maintains at around $89.7$\%. For audio data, \cite{tong2021federated} shows that SemCom with FL can reduce the mean square error (MSE) by around $100$ times compared to the conventional coding scheme using pulse code modulation (PCM), low-density parity-check (LDPC), and 64-QAM modulation. There is also an interplay between FL and MARL, by which they are applied to increase the performance of each other \cite{gunduz2022beyond}. It is shown in \cite{he2022learning} that the task-oriented communication protocol can be modeled as a MARL problem in wireless IoT networks with limited bandwidth resources and noisy channels. Reversely, SemCom is used to evaluate the relatedness among agents in heterogeneous IoT environments \cite{lotfi2022semantic}. Notably, the semantic-aware MARL shows a rewarding improvement of $83$\% and $75$\% compared to the baseline schemes only considering the structural and semantic similarities, respectively.

\subsection{Distributed Learning for IoT Localization}
Localization services are becoming increasingly popular in daily life with the support of \textcolor{black}{IoT} \cite{nessa2020survey}. However, various challenges exist for the prevalent deployment of IoT-assisted localization services, such as the accuracy of physical location, robustness, and privacy preservation of the location-based services caused by the unpredictable radio propagation characteristics and broadcasting nature of the dynamic (wireless) channels. To tackle these challenges, many studies have recently proposed using decentralized learning approaches (e.g., MARL, FL, and federated distillation) as alternative ways of providing efficient and robust localization services.
For example, in WiFi-based indoor localization services, the work \cite{liu2019floc} develops a fingerprinting-based localization system called FLoc to improve the risk of privacy leakage by using a crowdsourcing-based model updating mechanism. In the considered system, FL is used to update localization models. More specifically, every distributed device manages its local localization mode and sends the encrypted updating parameters to a server, where a global model is synthesized and then sent back to all local devices for the next updating round.
Similarly, the authors in \cite{ciftler2020federated} exploit the usefulness of FL for fingerprint-based localization services to improve the accuracy of the conventional radio signal strength (RSS) approach without affecting the individual information of distributed users. To do this, the received signal strength measurements from received beacons will be trained locally to generate the learning model. At the same time, the centralized server uses the previous local models to train a global multilayer perceptron model and then feeds that model back to the distributed users. Throughout real-world experimental data, the results demonstrate that the proposed method can localize users within $4.98$ meters accuracy in a $390$ meters to $270$ meters building and can especially narrow localization accuracy by $1.8$ meters when used as a booster for centralized learning.
In \cite{li2020pseudo}, the authors suggest a centralized indoor localization model using pseudo-label data (CRNP) combined FL and decentralized approaches to offload the burden of labeled fingerprint data collection and the network cost on the central server. Specifically, the CRNP first collects and merges the labeled and unlabeled fingerprint data to lessen the cost of data collection and speed up the localization performance. Later, both approaches are operated to prevent users’ privacy leakage from distributed training models of local location data.
In another aspect, the authors in \cite{yin2020fedloc} develop a FedLoc framework based on cooperative localization and location data processing to elegantly address the privacy issue in cooperation among numerous mobile IoT users. Driven by this, two potential wireless network infrastructures have been introduced to meet the fundamental requirements of FL-based models. In the first kind, cloud-based architecture, the entire network is divided into several non-overlapping sub-areas, each composed of a series of collaborative mobile terminals. In which core network components with high-speed computing, cache/storage, and communication entity will rely on the local update of mobile users to compute a global parameter one, giving a new position prediction for online communications. For the second kind, edge-based architectures, a trusted third-party edge node with sufficient storage and strong computation capability is added to the first one to construct a locally global learning model. In this way, the entire network can be built in a hierarchy, contributing to minimizing latency at each level.
\begin{figure}[t]
	\centering
	\includegraphics[width=0.995\linewidth]{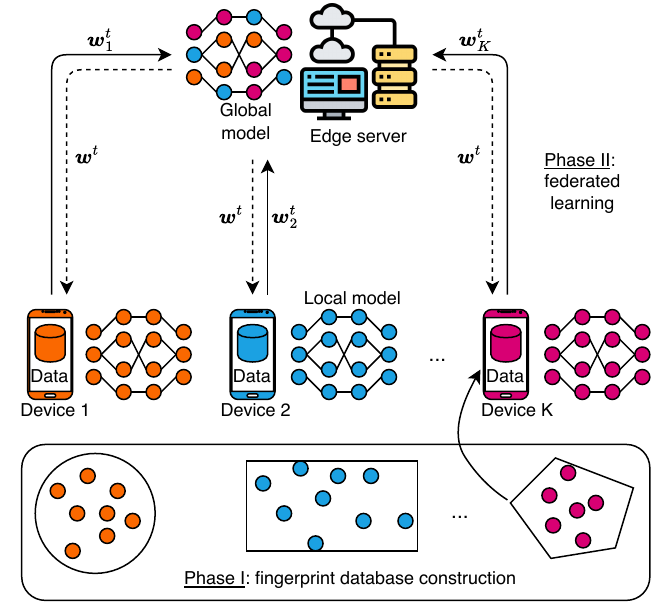}
	\caption{Illustrative use of FL for IoT localization \cite{cheng2022federated}.}
	\label{Fig:Localization}
\end{figure}

In practical fingerprint-based localization applications, database heterogeneity due to volatile client behaviours is often a major factor in performance degradation. To handle such problems, the work in \cite{cheng2022federated}, therefore, has released a novel heterogeneous FedLoc algorithm with the area of convex hull-based aggregation, as shown in Fig.~\ref{Fig:Localization}. In this model, the FedLoc scheme includes two phases: phase I of fingerprint database construction and phase II of FL. In the working principle, each distributed IoT node moves around the area of interest to collect the RSS fingerprint data. Then, it optimizes the learning model based on the collected local dataset before uploading the updated model. At the central server, the proportion weight of the IoT device $k$ is chosen as
\begin{equation} \label{Eq:localization}
    p_{k} = \frac{S_{C_{k}}}{\sum\nolimits_{k = 1}^{K}S_{C_{k}}},
\end{equation}
where $S_{C_{k}}$ denotes the area of the convex hull characterized by the local fingerprint data. 
Based on \eqref{Eq:localization}, the server generates the new global model and transmits it back to the distributed nodes for executing the next learning round. Compared to the FedLoc framework in \cite{yin2020fedloc}, the proposed heterogeneous FedLoc algorithms can obtain a significant location prediction gain of around $20$\%.
Considering the networks with limited communication resources, the work in \cite{etiabi2022federated} evolves a federated distillation framework for regression tasks data. In this framework, the server (called coordinator/teacher) is in charge of
supervising the clients (workers/ students) as well as defining the global training model. The clients focus on two missions: 1) train local models using its private datasets and 2) periodically upload the local average estimations per each target segment to the server. Afterward, the communication system produces a bridge connection to return the estimated global parameters from the server to the clients to refresh their respective loss functions for the next local training phase. From this framework, the number of transmitted bits for the communication load can be reduced by approximately $98\%$ compared to the conventional FL method. In addition, due to the reasonable savings of transmission power energy per communication round, the proposed framework can be also envisaged as a powerful approach for battery-powered IoT systems with very limited bandwidth.
Different from \cite{liu2019floc,ciftler2020federated,li2020pseudo,yin2020fedloc,cheng2022federated,etiabi2022federated}, the authors in \cite{kong2022fedvcp} recently propose an FL-enabled outdoor localization framework to provide high-precision positioning correction for dynamic IoT environments. Taking into account specific applications, the authors also develop an add-on FL-integrated transfer learning approach to reduce individual location prediction errors in cooperative IoV systems. Experimental datasets with sensor-rich vehicles show that the proposed framework is not affected by changes in the scale of the dataset but also offers a convergence rate lower than 10 epochs, much faster than using conventional DNN methods.

{\color{black}In the past, several centralized ML localization approaches have been introduced for IoT networks; however, most of them have high computing complexity for networks with many IoT devices. To overcome this problem, \cite{jia2022distributed} proposes a distributed and online localization approach with MARL to model the dynamic localization of portable IoT devices by exploiting Q-learning for each non-anchor node as an intelligent agent that can be localized by interacting with its surrounding nodes. Such the proposed approach offers a more accurate localization with a smaller mean localization error than two other distributed localization algorithms \cite{jia2013distributed} while having an acceptable computation for resource-constrained devices. In another work, \cite{alagha2022target} develops an efficient MADRL framework for target localization in IoT-aided manufacturing environments, which uses CNN models optimized by a proximal policy optimization algorithm to trigger an actor-critic structure model under multiple agent constraints. The framework also uses a team-based reward mechanism to encourage agent cooperation and resource management, and a centralized learning and decentralized execution procedure to handle scalability. The simulated results show a better performance than other target localization methods in different complex conditions. Compared with target localization in a two-dimensional plan, identifying and tracking UAVs in a 3D space is more challenging.} In~\cite{chen2020autonomous}, to obtain high localization accuracy and energy efficiency, an enhanced MARL framework with Q-learning is proposed for automatic flight planning decisions, in which the searching time constraint is considered in an action selection mechanism. For dealing with the unknown path loss of UAVs, the authors of this work adopted Gaussian process regression to estimate the location of reference points without paying attention to channel conditions. As calculated from reference points, the constrained reward and summarized reward values are used by UAVs to update the shareable and independent Q-tables. Relying on the simulation results, the proposed method achieved the shortest searching time versus the number of UAVs, the packet loss probability, and the acceleration constant compared with a single-agent Q-learning approach~\cite{chowdhury2019rss} and a regular MARL framework~\cite{wang2019improved}.

\subsection{Distributed Learning for IoT Mobile Crowdsensing}
As IoT and mobile devices are more powerful in storage, sensing, and computing capabilities, IoT can sense and collect a large amount of data, laying the foundation of integrated sensing, communication, and computing 6G IoT systems \cite{le2023wirelessly, xu2022edge, zhu2022pushing}. The crowdsensing systems can employ the traditional centralized machine learning approaches to train and process the collected sensing data. However, the massive volume of data from the sensing devices could consume significant communication resources (e.g., bandwidth and power budget), affecting applications' QoS targets. Moreover, since the centralized server directly accesses whole user data, the user privacy concerns could not be protected from leakage \cite{liu2018survey}.
To this end, several studies have recently assessed the distributed learning approach as a promising tool to leverage the learning and training process for MCS.
\textcolor{black}{For example, the work in \cite{hu2021blockchain} investigates a new FL-based framework for MCS that integrates crowdsensing tasks and subsequent data analysis. The main objective of the design in \cite{hu2021blockchain} is to diminish the communication cost and the requirement of computing and storage capabilities for requesters and to protect the workers' privacy at the edge servers. First, by applying a mechanism design based on an incentive-compatible game rule where selected mobile devices collect the sensed data at specific locations and forward it to the nearest edge server, therefore, the local privacy of the submitted data can be achieved at the edge servers. Then, a consortium blockchain-based FL is developed to secure the distributed learning process. Lastly, a cooperation-enforcing control scheme is deployed to evoke full compensation from the requesters.}
In \cite{chen2020intelligentcrowd}, considering the impacts of stochastic sensing environments in multi-device MCS scenarios, the authors propose an online IntelligentCrowd algorithm based on the MARL technique to make the best real-time sensing decisions for each participant. To glean the maximal payoffs of distributed IoT participants, the proposed sensing policy defines the incentive function of each participant as follows:
\begin{equation} \label{Eq:crowdsensing}
    U_{k}^{t}(x_{k}^{t},q_{k}^{t})= \frac{x_{k}^{t} q_{k}^{t}}{\sum_{k = 1}^{K}x_{k}^{t} q_{k}^{t}}R_t - c_k x_{k}^{t},
\end{equation}
\textcolor{black}{where $x_{k}^{t}$ $(x_{k}^{t}>0)$ denotes an effort level that each user $k$ has to choose to take part in an MCS task before sending the sensed data to the service provider in time slot $t$, $R_{t}$ is the total reward budget of the service provider, $q_{k}^{t}$ is an indicator expressing the quality of collected data, and $c_{k}$ is a cost coefficient. Then, the MARL-based IntelligentCrowd scheme learns to optimally find the effort level variables $x_{k}^{t}$ so that the reward function is maximized as in \eqref{Eq:crowdsensing}.}
The algorithm's performance is shown to be superior to the model predictive control and single-agent RL schemes under different stochastic environments.
%
\textcolor{black}{Alike, Chen \textit{et al.}~\cite{chen2021federated} propose a novel FL-based risk-aware decision framework to mitigate fake tasks from MCS. Each detection device has a local ML model and dataset, and the aggregation module gathers predictions from devices to minimize loss. The proposed framework achieves $100\%$ detection accuracy with small datasets and $8\%$ improvement over traditional methods.}

\begin{figure}[t]
	\centering
	\includegraphics[width=0.975\linewidth]{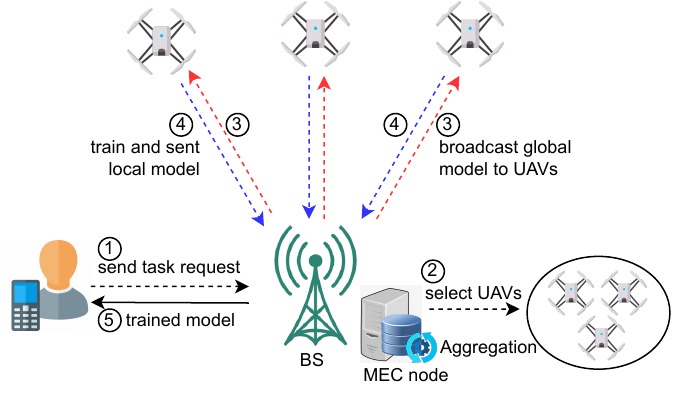}
	\caption{\textcolor{black}{Illustration of the FL framework for MCS assisted by UAVs \cite{wang2020learning}.}}
	\label{Fig:Learning_Air_FL}
\end{figure}

\textcolor{black}{FL and AI can potentially solve many challenging tasks in MCS, such as autonomous crowdsensing with UAVs, as depicted in Fig.~\ref{Fig:Learning_Air_FL}. However, security and privacy threats are two critical challenges. In~\cite{wang2020learning}, Wang \textit{et al.} proposed a secure FL scheme for collaborative AI model training in UAV-assisted crowdsensing. The scheme uses a blockchain-based collaborative learning architecture and local differential privacy to guarantee secure local model updates, contribution verification, and privacy preservation. It also introduces a two-tier RL-based incentive mechanism to enhance the UAVs' motivation to share high-quality models. Three critical challenges in incorporating FL into MCS systems to offset privacy concerns: (1) adding extra cloud servers as the sensing platforms to assist data aggregation cannot be a secure solution when a colluding attack exists in multi-cloud server configurations; (2) sharing the same private key to execute synthetic model decryption can lead to high information leakage relatively since each participant with this key in hand can decrypt any ciphertexts, producing high decryption costs; and (3) providing a truthful and fair incentive mechanism is necessary to offset resource consumption as well as stimulate participation. To address these challenges, Zhao \textit{et al.}~\cite{zhao2022crowdfl} developed a privacy-preserving crowdsensing system called CrowdFL. CrowdFL allows participants to process sensing data locally according to the FL paradigm and uses the threshold Paillier cryptosystem to aggregate participants' training models. This approach provides a better privacy-preserving solution and reduces the participant's decryption cost. CrowdFL also introduces a hybrid incentive mechanism to promote the participant's data sharing by rewarding a payoff to each participant based on their amount of sensing data.}

\begin{figure}[t]
	\centering
	\includegraphics[width=0.95\linewidth]{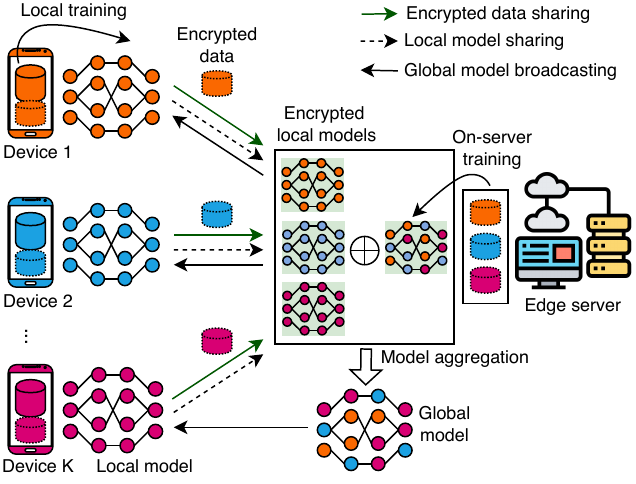}
	\caption{Illustration of the MPOA framework for privacy preservation and straggling mitigation \cite{saputra2022federated}.}
	\label{Fig:MPOA}
\end{figure}

With the same objective, the work in \cite{zhang2021fedsky} builds a novel privacy-preserving scheme called FedSky. Considering the participants' dynamics and heterogeneous natures, the designed scheme only selects qualified participants to improve the efficiency of the model training process while ensuring secure aggregate model updates based on an additive homomorphic encryption technique. \textcolor{black}{Various security analyses and experiments via an image classification task report that FedSky takes about $0.2$ hours to complete one training round, and only needs ten and one hundred rounds to achieve 96\% target accuracy in the IID and Non-IID settings, respectively. The fact is that as the distribution of each local dataset is significantly different from the global distribution, the local models are updated towards the local optima and thus far from the global optima. As a result, non-IID datasets can seriously impact the accuracy of FL-based models \cite{Li22NonIIDSilos,karimireddy20aStochastic}.}
Unlike studies in \cite{zhao2022crowdfl, zhang2021fedsky}, very recent work in \cite{saputra2022federated} divides the local data into two parts: one is trained locally on IoT devices and the remainder is encrypted for remote training at the edge server, as shown in Fig.~\ref{Fig:MPOA}. With this mechanism, the proposed framework, namely MPOA, can not only address the straggling issues in FL but also improve the learning performance, e.g., a reduction of $49\%$ in training time and an increase of $4.6$ times in model accuracy compared to the baseline FL method.

\textcolor{black}{Another FL-based approach is studied in \cite{pandey2020crowdsourcing}, where a novel crowdsourcing framework is designed to target a high-quality global model with communication efficiency during parameter exchange. Specifically, the framework uses an incentive mechanism to encourage participants to share their data, and a two-stage Stackelberg game to find the equilibrium where both the participants and the servers benefit. Numerical simulations show that the proposed framework can achieve up to $22\%$ utility gain over a heuristic approach.}
In addition, in \cite{bonawitz2019towards}, a system design for FL-assisted crowdsensing is built in the domain of mobile phones, where each mobile phone trains a DNN using TensorFlow. The weighted factors are then combined with the FedAvg algorithm, in the cloud server, to create a global model, which is sent back to the phones for inference in the next iteration. For securing the individual devices' updates against threats within the centralized server, the secure Aggregation protocol \cite{bonawitz2017practical} is implemented as a privacy preservation to the FL approach. Moreover, the challenges and open problems for the future developments of the proposed design are clearly addressed.
Also, the work in \cite{liu2020boosting} studies a new FL framework for privacy-preserved MCS, called FEDXGB, based on the original extreme gradient boosting scheme (XGBoost) \cite{chen2016xgboost}. FEDXGB comprises a secure gradient aggregation algorithm that takes into account the advantages of both homomorphic encryption and secret sharing so that the central server has to conduct the aggregation operation, thus ensuring robustness against user dropout. Comprehensive experimental results conducted using two standard datasets indicate that when compared with the original XGBoost, FEDXGB obtains negligible accuracy loss and a significant reduction in computation and communication cost for secure aggregation.

\subsection{Distributed Learning for IoT Security and Privacy}
\textcolor{black}{B5G/6G networks are essential for IoT services and applications, but security and privacy are critical issues, especially when deploying ML-based systems~\cite{sharma2020security}. This is because traditional ML architectures require users to upload their data to a central server, which compromises privacy~\cite{van2022joint, rawal2022identifying}. To address these challenges, researchers have proposed a communication-efficient and privacy-preserving deep learning framework~\cite{du2020approximate}. This framework uses a novel private approximate mechanism to achieve differential privacy with lower noise, which improves model performance. It also proposes a new gradient sparsification method to reduce communication costs by reducing the number of exchanged parameters between local users and the centralized server. Experiments show that the proposed framework can reduce communication costs by up to $2\%$ and improve accuracy by up to $16\%$ compared to existing work~\cite{shokri2015privacy}.}

\textcolor{black}{Recently, there have been many studies investigating FL to improve accuracy and protect user privacy in emerging IoT systems.}
\textcolor{black}{For example, Wu \textit{et al.}~\cite{ wu2020ddlpf} propose a decentralized deep learning model that uses FL, meta-learning, and blockchain techniques. This model significantly improves classification accuracy, security, data privacy, and total computation time compared to traditional learning approaches.
In~\cite{ lu2020privacy}, Lu \textit{et al.} designed an FL-based learning method for edge computing networks that provides privacy protection for multiple edge nodes without sacrificing training accuracy. They develop a novel gradient compression algorithm to reduce gradient exchanges and communication costs, as well as an asynchronous FL with a dual-weights correction method to address performance degradation caused by practical constraints of edge nodes.
Yin \textit{et al.}~\cite{yin2021privacy} propose a hybrid privacy-preserving FL framework for data sharing in IoT networks that relies on an advanced function encryption algorithm with a sparse difference matrix and a local Bayesian differential privacy mechanism. The former method ensures privacy protection for distributed clients, while the latter adjusts the user's privacy budget depending on the distribution of datasets to improve service quality.}

\begin{figure}[t]
	\centering
	\includegraphics[width=1.0\linewidth]{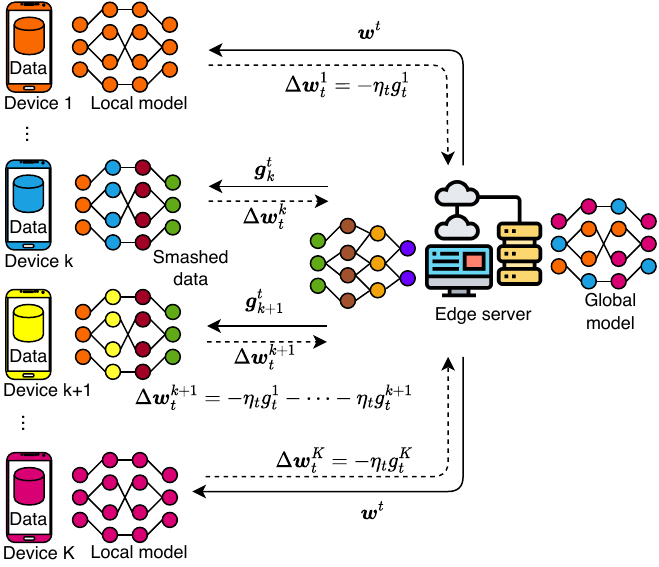}
	\caption{Illustration of the hybrid SL-FL (HSFL) architecture \cite{liu2022wireless} with $K$ IoT devices, where the two representative devices $1$ and $K$ perform FL, and devices $k$ and $k+1$ perform SL.}
	\label{Fig:HSFL}
\end{figure}
{\color{black}The work in \cite{liu2022wireless} discusses that deploying SL or FL alone in heterogeneous IoT systems is inefficient. As such, it proposes a hybrid architecture of SL and FL (HSFL) to exploit the advantage of both learning paradigms, as shown in Fig.~\ref{Fig:HSFL}. Considering the efficiency, FL typically requires less communication overhead than SL. However, SL is more suitable for IoT devices with limited computing resources, as they only need to train a part of the model. Combining these together, the model aggregation in HSFL is updated as follows:
\begin{equation} \label{Eq:HSFL}
\boldsymbol{w}_{t+1} = \boldsymbol{w}_t - \sum\limits_{k = 1}^{{K}_{FL}} p_{k}\eta_{t}\boldsymbol{g}_{t}^{k} - \sum\limits_{k = 1}^{K_{SL}} p_{k}\eta_{t}\boldsymbol{g}_{t}^{k} - \sum\limits_{k = 2}^{K_{SL}} p_{k} \Delta \boldsymbol{g}_{n},
\end{equation}
where $p{k} = D_{k}/D$ is the data size ratio, $\eta_{t}$ is the learning step at round $t$, $K_{FL}$ and $K_{SL}$ denote the numbers of FL devices and SL devices, respectively, and $\boldsymbol{g}_{t}^{k}$ denotes the gradient of the IoT device $k$. Different from FedAvg, the model aggregation in HSFL is based on two sets of IoT devices (FL set and SL set), and the last term in \eqref{Eq:HSFL} indicates that IoT devices in HSFL receive more local model updates when SL is employed. 
For example, a decentralized learning approach in \cite{chen2022decentralized} enables each IoT device to fully train the local model but only share the intermediate results with neighbor devices, which will calculate the intermediate gradients and send them back to update the bottom layers of the local model.}

\begin{figure}[t]
	\centering
	\includegraphics[width=1.0\linewidth]{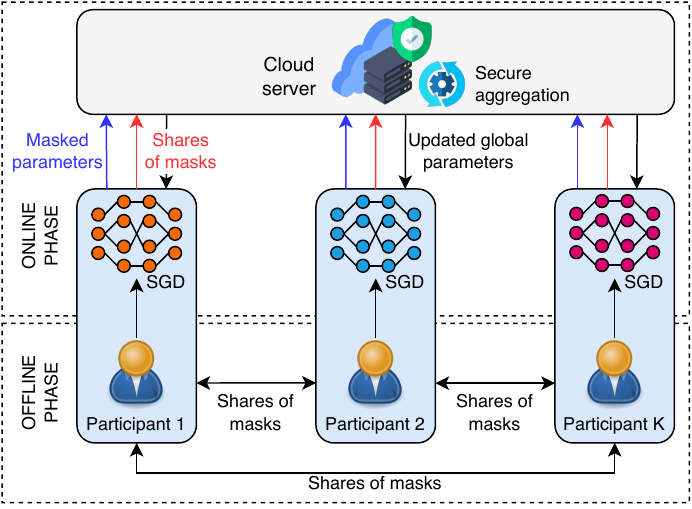}
	\caption{An illustrative system model of a lightweight privacy-preserving FL scheme for IoT devices in \cite{wei2021lightweight}.}
	\label{Fig:Lightweight_Privacy_FL}
\end{figure}

\textcolor{black}{The work in \cite{zheng2020preserving} reviews the fundamentals and compares the performance of two popular privacy-preserving learning models, namely local differential privacy and FL. The authors show that FL can achieve better clarification performance with privacy protection than local differential privacy for a moderate number of users with high computing resources. 
Zhang \textit{et al.}~\cite{zhang2019deeppar} propose two asynchronous FL-based deep learning schemes, DeepPAR and DeepDPA, based on proxy re-encryption technique and group dynamic key management, respectively. DeepPAR protects the input information of each individual participant, while DeepDPA guarantees backward secrecy of group participants in an efficient and lightweight manner.}
%
\textcolor{black}{Moreover, Wei \textit{et al.}~\cite{wei2021lightweight} propose a novel lightweight FL scheme for resource-limited IoT devices. The scheme protects individual local data while enabling the devices to train a high-accuracy global model. As shown in Fig.~\ref{Fig:Lightweight_Privacy_FL}, the FL process consists of two phases: offline and online. In the offline phase, each participant randomly selects a mask and shares it with others, meanwhile, each participant trains their local data with the additive masking scheme in the online phase to protect their parameters.}
%
For IoT systems to prevent user privacy leakage and cope with the impact of non-IID local datasets, Zhang \textit{et al.}~\cite{zhang2021privacy} introduce an asynchronous grouped FL framework with two key components: (i) an adaptive privacy budget allocation protocol based on the Renyi Differential Privacy mechanism to adjust the privacy budget and (ii) an asynchronous weight-based grouped update algorithm to combat against the poisoning attack in non-IID dataset. As a result, the proposed framework can guarantee the user's privacy while improving the utility and robustness of IoT systems under various cyberattack conditions.

\subsection{Summary and Lessons}
This section has reviewed the use of distributed learning for a wide range of IoT services in future 6G networks, including data sharing and computation offloading, localization, MCS, security, and privacy. Table~\ref{Table:Summary_IoTServices} summarizes the technical aspects and the contributions of representative references on IoT services based on distributed learning.

In practical offloading applications, the conventional approach based on cloud computing requires an enormous amount of data to be transmitted and aggregated at the central data centers, which causes high communication costs and transmission delays. As an improvement, edge computing is a promising computing paradigm for offloading tasks from massive IoT devices that can reduce a significant amount of data over the network and thus lessen the latency. However, in both cloud and edge computing approaches, privacy concerns remain a challenging design issue due to transferring raw data and processing it at a single central point. In this context, FL allows distributed devices to jointly train a shared model without sending the raw data to the cloud/edge server, which reduces communication costs and achieves higher user privacy. On the other hand, in real-time applications with dynamic workloads, computational offloading decisions may need complicated resource management schemes. \textcolor{black}{Optimal offloading decisions can be made by combining FL and multi-agent DRL, where FL plays the role of a coordinator to drive the local training process at the DRL agents, thus effectively dealing with the resource allocation optimization problem}. Especially, the long-term benefits of local computation and energy consumption at the agents should be considered to gain sustainable learning procedures and low-cost system operation.

Traditional ML approaches may be propitious for IoT-based localization services in some perspectives, such as providing robustness and scalability, improving accuracy, and reducing complexity. Nevertheless, distributed clients are required to send a large amount of raw data to the server for model training, which induces high communication costs and weak privacy issues. In this context, FL has been introduced as a game-changing technology to reduce the communication load and strengthen the client's data privacy in IoT-assisted localization services. Some advanced FL frameworks have allowed transferring the local model outputs from clients to the cloud server instead of the whole models having larger sizes, thus obtaining communication cost efficiency and saving network bandwidth. To increase localization accuracy, some FL-based cooperative learning mechanisms are beneficial to exploit the relative correlations between different portable IoT devices. Besides, using multiple sensory data from different sources is another way to improve localization accuracy, but it should require some advanced learning models for fused data, including time series data, high-dimensional data, and structured and unstructured data. Some semi-supervised learning models can be considered to learn unlabeled data due to the difficulties of notating localization data in complex systems. Compared with indoor localization, outdoor localization is more challenging, especially with resource-constrained IoT devices like UAVs in 3D spaces, with many factors and agents that can be handled by MARL frameworks using shareable and independent Q-learning models.

\begin{table*}[hb!]
    \renewcommand{\arraystretch}{1.1875}
	\caption{\textcolor{black}{Summary of the literature on distributed learning for IoT services.}}
	\label{Table:Summary_IoTServices}
	\centering
	\begin{tabular}{|c|c|p{2.25cm}|p{2.5cm}|p{9.25cm}|}
		\hline 
		\textbf{Service} & \textbf{Ref.} & \textbf{Use case}  & \textbf{Learning Approach}  & \multirow{1}{*}{\textbf{Technical Contributions}}  \\ \hline 
		
		\multirow{1}{*}{\rotatebox{90}{\textbf{~\makecell{Data Sharing and \\Computation Offloading}~}}}
        &
		\cite{min2019learning}
        &
        Efficient offloading policy for energy harvesting-enabled IoT networks
		&
		RL-based offloading scheme
		& 
		A computation offloading framework to select optimal offloading policy for energy harvesting-enabled IoT devices.
		\\ \cline{2-5}		
		
		
		&
		\cite{tang2020decentralized}
            &
            Efficient offloading policy for IoT fog systems
		& 
		A low-complexity learning-based offloading algorithm
		& 
		A decentralized computation offloading problem allows IoT devices to make their own optimal offloading decision depending on their local observed state.
		\\ \cline{2-5}

            &
		\cite{alam2022multi}
            &
            Offloading policy in three-tier vehicular edge computing
		& 
		MARL-based computation offloading
		& 
		The MARL technique is employed to generate the cost matrix of cooperative resources, which is used by the Hungarian algorithm to achieve the best performance.		
		\\ \cline{2-5}		
            
            &
		\cite{liu2022blockchain}
		&
            Efficient offloading policy
            &
		Blockchain and distributed RL-based computation offloading
		& 
		Blockchain enables IoT devices to share and evaluate the quality of data. Meanwhile, a distributed RL approach is developed to maximize task completion time.
		\\ \cline{2-5}		
		
		&
		\cite{chiu2020semisupervised}
            &
            Video data analysis for AIoT service platform
		& 
		Semi-supervised learning and FL approach
		& 
		Edge devices implemented a semi-supervised learning scheme periodically upload the training results to the cloud server following the concept of the FL scheme.
		\\ \cline{2-5}		

        &
		\cite{zhao2022federated}
		& 
            Network edge intelligence
            &
		A new proportion weight scheme in FL
            &
		The relationship between the weight divergence and data divergence is investigated. Then, a new weight design is proposed to improve FedAvg with non-IID data. 
		\\ \cline{2-5}		
		
		&
		\cite{lu2020blockchain}
            &
            Data sharing in industrial IoT
		&
		FL model with permissioned blockchain
		&
		The integration of FL and blockchain enables secure and intelligent data sharing among decentralized multiple parties in IIoT.
		\\ \cline{2-5}		
		
		&
		\cite{ren2019federated} 
		& 
            Efficient offloading policy for edge computing supported IoT
            &
		Multi-agent DRL and FL 
		& 
		Multi-agent DRL built at edge nodes to determine optimal offloading decisions. FL trains DRL agents for optimal communication and computing resource allocation.
		\\ \cline{2-5}		
		
		&
		\cite{bayerlein2021multi}
		& 
            Multi-UAV path planning
             &           
		MARL approach
		& 
		A multi-UAV path planning problem subject to flying time and collision avoidance constraints that maximizes the data harvested from distributed IoT sensor nodes.
		\\ \hline
		
		\multirow{1}{*}{\rotatebox{90}{\textbf{~\makecell{IoT \\Localization}~}}}
		&
		\cite{ciftler2020federated}
		& 
            RSS-based localization
            &
		FL approach
		& 
		FL-based indoor localization scheme to improve the accuracy of the RSS-based localization method while protecting the privacy preservation of distributed users.
		\\ \cline{2-5}		
		{}
		&
		\cite{li2020pseudo}
		& 
            RSS-based indoor localization in sensor networks
            &            
		FL and centralized model
		& 
		An integrated FL and a centralized indoor localization model using pseudo-label data to improve indoor localization accuracy while protecting local data privacy.
		\\ \cline{2-5}		
		{}
		&
		\cite{yin2020fedloc}
		& 
            Cooperative localization and location data processing
            &
		A data-driven FL-based learning model
		&
		An FL-based cooperative localization and location data processing framework in cooperation between mobile IoT users targeting the privacy-preserved accurate location services.
		\\ \cline{2-5}		
		{}
		&
		\cite{cheng2022federated}
		& 
            Fingerprint-based localization
            &
		A heterogeneous FL-based localization algorithm
		& 
		FL-based localization algorithm to enhance the localization accuracy taking into account the fingerprint database heterogeneity.
		\\ \cline{2-5}		
		{}
		&
		\cite{etiabi2022federated}
		&
            Indoor localization system
            &
		Federated distillation model
		&
		An indoor IoT localization algorithm based on federated distillation to improve communication efficiency and scalability capability in large-scale IoT networks.	
		\\ \cline{2-5}		
		{}
		&
		\cite{liu2019floc}
		&
            Fingerprinting-based indoor localization
            &
		FL approach
		&
		A fingerprinting-based localization system called FLoc, where FL is used to update the localization model.
		\\ \hline	
	\end{tabular}
\end{table*}

\begin{table*}[t!]
\ContinuedFloat
	\renewcommand{\arraystretch}{1.1875}
    \caption{\textcolor{black}{Summary of the literature on distributed learning for IoT services (cont.).}}
    \centering
	\begin{tabular}{|c|c|p{2.25cm}|p{2.5cm}|p{9.25cm}|}
		\hline 
		\textbf{Service} & \textbf{Ref.} & \textbf{Use case}  & \textbf{Learning Approach}  & \multirow{1}{*}{\textbf{Technical Contributions}} \\ \hline 

		\multirow{1}{*}{\rotatebox{90}{\textbf{~\makecell{IoT Mobile \\Crowdsensing}~}}}
		&
		\cite{hu2021blockchain}
		& 
            Privacy-preserving MCS
            &
		FL approach
		& 
		A FL-based framework for MCS to diminish the communication and computing/storage requirements for requestors while protecting the workers' privacy.
		\\ \cline{2-5}
		{}
		&
		\cite{chen2020intelligentcrowd}
		& 
            Multi-device MCS
            &
		MARL approach
		&
		An online crowdsensing policy that makes the best real-time sensing decisions for each distributed participant.
		\\ \cline{2-5}
		{}
		&
		\cite{chen2021federated}
		& 
            Illegitimate task detection in crowdsensing platforms
            &
		FL approach
		&
		An FL-based system for illegitimate task detection caused by the distributed nature of MCS to minimize the prediction loss.
		\\ \cline{2-5}
		{}
		&
		\cite{wang2020learning}
		& 
            UAV-assisted crowdsensing
            &
		FL and RL-based incentive mechanism
		&
		A secure FL framework is proposed for collaborative AI model training to cope with potential security and privacy threats.
		\\ \cline{2-5}
		{}
		&
		\cite{zhao2022crowdfl}
		&
            Privacy-preserving MCS
            &
		FL approach
		& 
		An FL-based privacy-preserving MCS system that allows participants to process sensing data locally and upload encrypted training models to the centralized server.
		\\ \cline{2-5}
		{}
		&
		\cite{zhang2021fedsky}
		&
            MCS systems
            &
		FL approach
		& 
		The FL scheme optimizes the global model by only selecting qualified participants and protects user privacy by securely aggregating model updates based on the additive homomorphic encryption system.
		\\ \cline{2-5}
		{}
		&
		\cite{saputra2022federated}
		&
            Data privacy preservation and straggling mitigation
            &
		FL approach
		& 
		A multi-principal one-agent contract-based framework is developed to solve the data-sharing problem while information asymmetry is considered.	
		\\ \cline{2-5}
		{}
		&
		\cite{pandey2020crowdsourcing}
		&
            Communication-efficient crowdsourcing
            &
		FL approach
		& 
		A novel crowdsourcing framework that targets a high-quality global model with communication efficiency during parameter exchange. A two-stage Stackelberg game is adopted to identify the equilibrium where the participants achieve their utility maximization.	
		\\ \cline{2-5}
		{}
		&
		\cite{bonawitz2019towards}
		&
            Smartphones-based crowdsensing
            &
		FL approach
		& 
		A scalable system design for FL-assisted crowdsensing is built in the domain of mobile phones.
		\\ \cline{2-5}
		{}
		&
		\cite{liu2020boosting}
		&
            Privacy-preserving MCS
            &
		FL approach
		& 
		An FL framework is developed based on the original extreme gradient boosting scheme to ensure privacy preservation and robustness against user dropout.
		\\ \hline
		
        \multirow{1}{*}{\rotatebox{90}{\textbf{~\makecell{IoT Security \\and Privacy}~}}}
		&
		\cite{du2020approximate}
		& 
            Privacy-preserving IoT systems
            &
		Deep learning approach
		& 
		A communication-efficient and privacy-preserving learning model is designed to guarantee users' privacy with high resource utilization efficiency.
		\\ \cline{2-5}
		{}
		&
		\cite{wu2020ddlpf}
		& 
            Secure and privacy-preserved IoT
            &
		FL, meta-learning, and blockchain techniques
		&
		A decentralized deep learning paradigm to tackle technical challenges such as timely response, privacy preservation, security, and biased data distribution amongst local IoT sensing datasets.
		\\ \cline{2-5}
		{}
		&	
		\cite{lu2020privacy}
		& 
            Privacy-protecting in edge computing
            &
		FL-based asynchronous learning
		& 
		An FL-based learning method that provides multiple edge nodes with privacy protection in a freer learning environment without reducing the accuracy of training.
		\\ \cline{2-5}
		{}
		&
		\cite{yin2021privacy}
		& 
            Multiparty data sharing in social IoT
            &
		FL approach
		& 
		A hybrid privacy-preserving FL method for data sharing in IoT to preserve the privacy of distributed clients and computing and storage resource utilization.
		\\ \cline{2-5}
		{}
		&
		\cite{liu2022wireless}
		&
            Communication efficiency in UAV networks
            &
		Hybrid SL-FL approach
		& 
		A hybrid architecture, namely HSFL, is proposed by leveraging the learning features of both FL and SL. HSFL can achieve as good performance as vanilla SL under IID data while always outperforming FL under both IID and non-IID data. 
		\\ \cline{2-5}
            {}	
            &
		\cite{zheng2020preserving}
		& 
            Privacy preservation in mobile networks
            &
		FL approach
		& 
		A performance evaluation of two popular privacy-preserving learning models, namely local differential privacy and FL.
		\\ \cline{2-5}
		{}
		&
		\cite{zhang2019deeppar}
		&
            Privacy-preserving industrial IoT
            &
		Asynchronous federated deep learning
		& 
		Two asynchronous FL-based learning schemes, namely DeepPAR and DeepDPA, are developed to protect the data of each participant and guarantee the backward secrecy of group participants in an efficient and lightweight manner.
		\\ \cline{2-5}
		{}
		&
		\cite{wei2021lightweight}
		& 
            Privacy-preserved large-scale IoT
              &
		FL approach
		& 
		A lightweight FL scheme enables the massively distributed devices to train a high-accuracy global model while guaranteeing the privacy of individual local data.
		\\ \cline{2-5}
		{}
		&
		\cite{zhang2021privacy}		
		& 
            Privacy-preserving IoT systems
            &
		Asynchronous grouped FL approach
		& 
		An adaptive privacy budget allocation protocol adaptively adjusts the privacy budget to achieve an efficient local model, while an asynchronous grouped update algorithm defends against the poisoning attack in the non-IID local dataset.
		\\ \hline
	\end{tabular}
\end{table*}%

Privacy leakage and incentive mechanism design are major challenges to the widespread deployment of MCS, especially when conventional machine learning is employed to process the aggregated sensing data at a centralized server. As a practical solution, FL allows the server to only gather training models from distributed participants without sharing their local sensing data, which shows the effectiveness of FL for sensing data in terms of saving communication costs and privacy preservation \cite{zhao2022crowdfl}. However, the FL-based crowdsensing design should not employ extra servers to assist data aggregation, significantly increasing the communication burden between participants and multiple servers and collusion attacks among them. Furthermore, the secure aggregation algorithms only allow each participant and the server to hold one partially private key so that any participant cannot decrypt any cyphertexts. This may effectively prevent privacy key leakage and reduce the decryption cost. In addition, due to the nature of heterogeneous IoT devices with different computational capabilities and data resources, the set of participants should be carefully chosen based on, e.g., the computational power of the devices and/or the local data size, to improve the local computational time and the server's waiting time so as to enhance the system efficiency \cite{zhang2021fedsky}. On the other hand, MARL is a powerful solution to encourage distributed users to participate in crowdsensing tasks and enhance the truthfulness and fairness in crowdsensing, which could significantly consume their computing and communication resources. According to a specific incentive mechanism, each participant would make optimal sensing decisions under the uncertainties of sensing quality and payoffs based on their amount of sensing data.

\section{Distributed Learning for IoT Applications}
\label{Sec:IoTapplications}
In future network systems, distributed learning is expected to play a critical role in offering efficient and sustainable wireless connections to a wide range of IoT applications. 
\color{black}
Notably, FL and SL can be exploited to overcome the challenges of limited computation capacity, storage, and energy in IoT devices in the following ways~\cite{imteaj2022survey}:
\begin{itemize}
    \item Lightweight ML/DL models: ML and DL models should be designed to be lightweight and cost-efficient. This can be done by using simple architecture models, or by using techniques such as quantization and pruning to reduce the size and complexity of the model.
    \item Transfer learning: In this technique, a model trained on a large dataset is used as a starting point for training a model on a smaller dataset, thus reducing the amount of computation and data required to train the model.
    \item Compression techniques: Some advanced compression techniques can be applied to reduce the size of the data that needs to be transferred between devices and the server. This can help to reduce the bandwidth requirements and the energy consumption of the devices.
    \item Edge computing: As a distributed computing paradigm where computation is performed closer to the data source, edge computing can significantly reduce the amount of data that needs to be transferred between devices and the server, and it can also improve the latency of the system.
    \item Secure protocols: To protect the privacy and security of the data that is shared between devices and the server, some secure protocols can be integrated into FL and SL frameworks.
\end{itemize}
\color{black}
Therefore, this section concentrates on providing an overview of the functionality/applicability of distributed learning for smart healthcare, smart grid, autonomous vehicles, aerial IoT networks, and smart industries.  

\subsection{Distributed Learning for IoT-enabled Healthcare}
Smart healthcare has become an integral part of our daily life, which integrates innovative technologies to connect people with institutions and deliver intelligent healthcare services. In fact, healthcare data is often fragmented owing to the complex and diverse nature of healthcare and medical systems and processes, as well as the variety of data structures collected from different IoT devices. Clearly, this exposes some challenges in modeling and learning complex patterns for an increasing number of IoMT devices with a variety of healthcare and medical applications when applying traditional ML frameworks with centralized learning performed at central servers. In this context, many distributed learning frameworks have been developed for IoMT in future 6G networks.

Several recent smart healthcare applications and services are built on distributed computing frameworks ~\cite{jeon2019privacy,qian2020wearable,yan2021concurrent,kasyap2021privacy}, including data acquisition, storage and processing with AI/ML/DL algorithms. 
For a triangle goal, including privacy preservation, network bandwidth utilization reduction, and computing efficiency increment, a distributed DL framework is proposed in~\cite{jeon2019privacy}. 
\textcolor{black}{In the framework, a single central server plays the following roles: weight initialization (initializes the weights for the deeper layers of a CNN model while the geo-distributed health centers initialize the weights for the first layer), gradient coordination and aggregation (coordinates the gradient updates from the health centers and aggregates them to compute the updates to the model parameters), and broadcasting of the updates (broadcasts the updates to the model parameters to all of the health centers). The proposed framework is distinguished from the traditional FL framework in terms of model sharing (sharing a partial model with the health centers instead of the whole model for data privacy improvement) and communication overhead (requiring more communication overhead for coordination and aggregation activities).}
\textcolor{black}{Similarly, a distributed framework with CNN-based models~\cite{qian2020wearable} is developed} for self-health monitoring services to address the geographic fragmentation of data sources and other regulatory constraints in the healthcare domain like sensitive data privacy. In this system, smartphones and IoMT devices train a shared consensus AI model collaboratively with owned private data to guarantee patients' data privacy, while a centralized cloud server solely focuses on coordinating whole local models to reduce the computation on smartphones. As expected, the proposed system achieves competitive performance against other centralized ML ones when executed with the healthcare dataset of fall detection using smartphones, smartwatches, and shimmer devices. \textcolor{black}{Furthermore, a decentralized learning framework in~\cite{qian2020wearable} is introduced to address} some common privacy problems in healthcare systems, such as inference, poisoning, and Sybil attacks, in which local ML/DL models are partially trained on edge IoMT devices, and the remainder of model learning is done at intermediate gateways as hospitals in a collaborative healthcare framework. In the meanwhile, there exists a central server that takes the role of global model aggregation and computing process delineation. Compared with a centralized learning framework~\cite{lu2020blockchain}, 
the proposed decentralized framework achieves better by $35\%$ of model inference detection precision when benchmarked on the MedNIST dataset for the medical image classification application despite classification accuracy is less than $1.8\%$.

Many secure healthcare and medical systems apply FL to obtain flexible and privacy-preserving data processing, in which ML models for automatic decision-making are trained by a distributed mechanism~\cite{kumar2021federated, antunes2022federated, cheng2022aafl}. 
For instance, an innovative healthcare data analysis IoMT-based system with a deep FL (DFL) framework is introduced in \cite{elayan2022sustainability} to monitor and analyze healthcare data in real-time while preserving patients' privacy without sacrificing model learning efficiency. In this decentralized learning framework, IoMT devices complete the local training process with their own acquired data, and the models' changes are summarized and sent to a cloud server for FedAvg-based model aggregation. The system is generally developed for medical image processing tasks, such as the skin disease detection task, and evaluated on the Dermatology Atlas dataset. In particular, the proposed DFL framework outperforms the regular centralized learning framework to reach the area under the curve up to $97\%$ while using a lower operational cost. 
\textcolor{black}{Similarly, a DFL framework is introduced to analyze histopathology images} having IID and non-IID distributions with some specific security techniques~\cite{adnan2022federated}. 
Each client in the FL framework is responsible for several processing steps locally, including visual feature extraction with DenseNet~\cite{gao2017densely} and local model training. 
Remarkably, to deal with non-IID data, memory-based exchangeable models \cite{kalra2020learning} are trained with the multiple-instance learning algorithm over a differentially private stochastic gradient descent mechanism and subsequently aggregated by the FedAvg algorithm at the central server. Based on the performance evaluation of the medical image dataset of non-small cell lung cancer ($2580$ images with more than two TB of data), the proposed FL framework classifies lung adenocarcinoma and lung squamous cell carcinoma more accurately than the centralized learning mechanism for both IID and non-IID data distributions.
In another work ~\cite{baghersalimi2022personalized}, a standard FL framework is investigated for epileptic seizure detection as one among many mobile-aided personal services, in which each IoMT device trans a DNN model relied on its local electrocardiography (EEG) data. Moreover, to verify the practicability and realistic efficiency, this learning framework is implemented on the NVIDIA Jetson nano developer kit and evaluated on the EPILEPSIAE dataset. 
\textcolor{black}{Furthermore, to effectively deal with two problems of massive healthcare sensor data and initially unknown data category}, a scalable and transferable FL framework is developed in \cite{sun2022scalable} for healthcare monitoring and analysis with different modalities of healthcare sensory data, such as EEG, electrocardiogram (ECG), photomicrography (PPG), and electrooculography (EOG). Especially, this framework applies the transfer learning technique with a parameter initialization scheme to accelerate the training process and learning convergence, besides new parameter protection terms added in the loss function to prevent catastrophic forgetting. 
\textcolor{black}{Being different from the aforementioned approaches that process sensory data, the proposed FL framework in~\cite{xu2021federated} has been extensively exploited for electronic health records (EHR)} based healthcare informatics analysis to renovate fragmented healthcare data sources and securely protect sensitive patients' data. Moreover, further details on FL for IoT healthcare are also discussed in a recent review \cite{nguyen2022federated}.
\textcolor{black}{It is recognized that the DFL framework can outperform the centralized learning framework in some specific scenarios, due to its advantages in reducing overfitting and enhancing model generalization.
    \begin{itemize}
        \item Overfitting reduction: DFL trains a model on small batches of local data from each device, which helps to prevent overfitting. This is because the model is not exposed to the entire dataset, which can help to prevent it from memorizing the training data and making inaccurate predictions on new data.
        \item Model generalization enhancement: DFL trains a model on diversified data collected from a variety of devices. This helps to improve the model's generalization ability, which is the ability to make accurate predictions on new data that is not part of the training dataset. This is because the model is exposed to a wider variety of data, which helps it to learn the underlying patterns in the data.
    \end{itemize}}
\textcolor{black}{Besides being trained on data from a variety of devices to improve its generalization ability for skin detection and histopathology image analysis, DFL can also deal with non-IID data effectively by using memory-based exchangeable models. Additionally, a scalable and transferable FL framework is recommended for real-time medical and healthcare monitoring services due to its ability to handle massive healthcare sensor data and initially unknown data categories.}

\begin{figure}[t]
	\centering
	\includegraphics[width=0.95\linewidth]{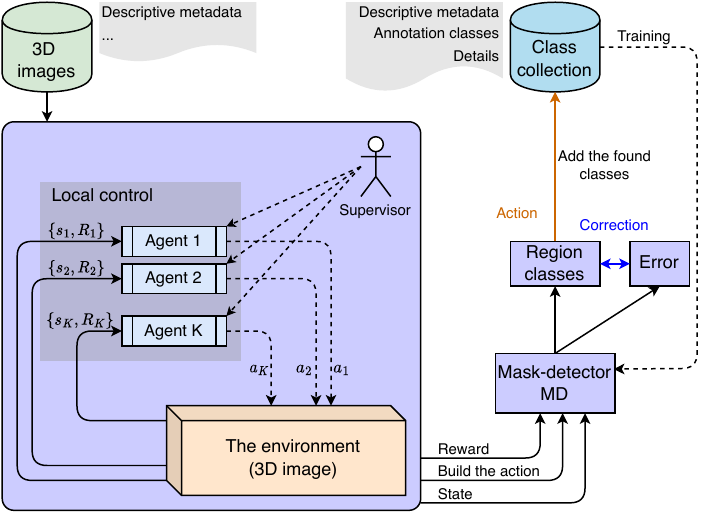}
	\caption{The architecture of mask extraction for image-based COVID-19 diagnosis using multi-agent DRL in~\cite{allioui2022multi}.}
	\label{Fig:healthcare_marl}
\end{figure}

\textcolor{black}{Compared to single-agent RL algorithms, MARL algorithms are more effective in solving large-scale optimization and decision-making problems~\cite{wen2021dtde}. This makes them well-suited for healthcare and medical applications on decentralized systems, such as the multiple landmark detection approach proposed in~\cite{vlontzos2019multiple}. This approach uses a MARL algorithm to train multiple agents simultaneously, allowing them to share their knowledge and collaborate to detect different landmarks. In experiments on MRI landmark detection datasets, the collaborative RL model achieved significant reductions in computation cost and training time, as well as higher accuracy, compared to training multiple agents separately.}
%
\textcolor{black}{Being similar to~\cite{vlontzos2019multiple} in exploiting MARL in healthcare and medical concerns, Liao \textit{et al.}~\cite{liao2020iteratively} proposed IteR-MRL, an interactive medical image segmentation update method that uses MARL to solve two problems: the dynamic process for successive interactions and segmentation map binarization with prediction uncertainty loss. Each voxel is treated as an agent, and a relative cross-entropy gain-based reward is used to enhance the efficiency of segmentation exploration. Segmentation probability is also used as a part of the RL state to preserve the prediction uncertainty and provide finer granularity and dense spaces. IteR-MRL shows superior segmentation accuracy compared to state-of-the-art methods while offering robustness and generalization with different initial segmentation and datasets.}
{\color{black}The work in \cite{allioui2022multi} introduces a new mask extraction method based on MADRL algorithms for COVID-19 CT image segmentation, as drawn in Fig.~\ref{Fig:healthcare_marl}. This approach uses DQN-LSTM networks to produce masks and pick superior visual features, and adopts MDP-based learning to recommend the optimal mask extraction approach to specialists in order to get a more accurate clinical diagnosis and faster diagnostic time. Experiments based on a united CT image dataset of different classes show that the method can reach over $97\%$ of dice score and $83\%$ of F1-score with the mean absolute error of $0.86\%$, verifying its efficiency in making lung CT masks for COVID-19 diagnosis.}

\textcolor{black}{It is not hard to realize that several institutions are currently using FL to develop AI models for improving the accuracy of different healthcare and medical applications and services. For instance, Owkin is a healthcare company that is exploiting FL to train ML models for various purposes, such as drug discovery (deliver novel drug targets and optimize drug positioning), drug development (increase the probability of success of clinical trials), and diagnostics (pre-screen for biomarkers and predict outcomes)~\cite{xu2021federated} Additionally, many hospitals and medical centers are deploying FL in their healthcare and medical systems to effectively address many challenging tasks~\cite{li2020federated}, such as improving accuracy of AI models for diagnosing and treating diseases in Mount Sinai hospital, developing new AI modes for detecting and treating cancer in Stanford University Medical center, and designing a new deep network architecture to optimize the performance of medical scans in Massachusetts General hospital. In addition, in a paper published by Nature Medicine \cite{dayan2021federated}, 20 institutes around the world have collaborated to realize a training AI model using federated learning that predicts the future oxygen demands of symptomatic patients with COVID-19 infections. Also, coordination between industry and academia has built MedPerf \cite{karargyris2023federated}, an open benchmarking platform that enables measuring the performance of medical AI models in large-scale heterogeneous data.}

\subsection{Distributed Learning for IoT-enabled Smart Grid}

In modern IoT networks, smart grids play a core role in automatic power supply with the ability to monitor and control every customer and node, thereby maintaining a two-way flow of information and electricity between power plants and appliances and all nodes. Thanks to the advances in AI-aided automated control systems, it is now capable of delivering real-time applications and services via numerous tasks, such as modeling different virtual power plants (VPPs), predicting individual users' power demands, designing decentralized control mechanisms with cycle optimization, etc, which enables the distributed smart grid to be safer and reliable~\cite{ramchurn2012putting}. Besides, the advanced evolution of sensing technologies, communication infrastructures, and diverse computing platforms has facilitated the modernization of the smart grid, where distributed intelligence can effectively deal with many obstacles related to security and privacy issues instead of centralized decision-making in IoT-based smart grid systems \cite{tariq2021vulnerability}. 
For instance, to protect the important power grid infrastructure and the integrated distributed energy resources (DER) from potential threats and malicious cyberattacks, a distributed learning framework for holistic attack-resilient has been introduced in \cite{qi2016cybersecurity}. Concretely, a cooperatively distributed learning mechanism is produced in an energy management system, in which several commonly statistical ML models (such as SVM, decision trees, naive Bayes network, and self-organizing) play the role of analyzing relevant data (e.g., smart inverters, smart meters, and supervisory control and data acquisition measurements) to identify anomalies and learn different DER cyberattack patterns. 
\textcolor{black}{As being similar to~\cite{qi2016cybersecurity} in taking into consideration DERs, an agnostic ML model using multiple step-wise linear regression algorithms\cite{sondermeijer2019regression} is developed} to process with a local dataset selectively collected from the central dataset. For the prediction of nearly optimal inverter actions from local measurements of multiple controllable DERs, a decentralized data-driven optimal power flow method is introduced in \cite{dobbe2020toward} to automatically adapt voltage and power flow in intelligent electric networks. Specifically, the distributed ML model wishes to achieve multiple predefined objectives regarding distribution grid operations
\begin{align} 
\begin{split}
    f_o := & \alpha \sum _{(m,n)\in\varepsilon } r_{mn}\ell_{mn}+\beta \sum _{n\in\mathcal N}\left ( y_n-y_{\mathrm{ref}} \right )^2\\
    & +\gamma \left ( P^2_{01}+Q_{01}^2 \right ),
\end{split}
\end{align}
where $\alpha,\beta,\gamma$ denote hyperparameters, $y_n$ is the squared voltage magnitude, $y_{\mathrm{ref}}$ is the reference voltage throughout the network, $r_{mn}$ and $\ell_{mn}$ are the resistance and squared current magnitude on the branch $\varepsilon$ from node $m$ to node $n$ of the set $\mathcal N$. $P$ and $Q$ denote the real and reactive power flowing out of a node. At each learning iteration, a Bayesian information criterion-based feature selection mechanism is adopted to selectively offer the most efficient feature subset that can yield the optimal DERs' power injection. Aiming to provide secure and economic (e.g., reduce power outage loss) energy delivery services to electricity consumers, the work in \cite{amini2020distributed} 
evolves a distributed ML paradigm based on local resilience management systems (RMS). In which, the RMS agents (e.g., autonomous microgrid or smart home) share their local data (i.e., microgrid operational data) with neighboring RMS and adopt collaborative learning to coordinate decisions in a distributed fashion, e.g., minimizing power loss by leveraging objective function-based optimal resource sharing policy as a measure of resilience. This leads to an optimization problem of RMS by maximizing the control signal for resilience enhancement under the constraint of power load and battery 
\begin{equation}
\max_{a\left ( \mathcal W \right )} \mathcal{F}^{RMS} \left ( x \left ( \mathcal W \right ) \right ),
\end{equation}
where $\mathcal{F}^{RMS}(\cdot)$ refers to the function of RMS, $\mathcal{W}$ is the control signal of the demand response and distributed storage agents, and $x$ is the decision variable input of load and battery.

\textcolor{black}{For training AI models in smart grids, FL has shown its capability to preserve privacy and scalability. For instance, Liu \textit{et al.}~\cite{liu2022federated} proposed a federated reinforcement learning (FRL) approach for decentralized voltage control in distributed energy networks. FRL outperforms standard centralized learning algorithms in terms of privacy, scalability, and communication costs. 
Consequently, a novel privacy-preserving FL framework, namely FedDetect~\cite{wen2022feddetect}, is designed to detect energy thefts securely and automatically.
FedDetect trains the same TCN-based model on local data at distributed detection stations. The model parameters are encrypted using Lifted elliptic curve ElGamal's homomorphic encryption and Camenisch-Lysyanskaya signature scheme before being sent to the control center, enhancing security and privacy. Experiments on a real dataset~\cite{yao2019energy} showed that FedDetect is comparable in efficiency to existing centralized DL-based systems while enhancing security and privacy significantly. FedDetect also provides more accurate energy theft detection than other FLs using different AI models, albeit with slightly higher memory usage and compute consumption.}

\begin{figure*}[t]
	\centering
	\includegraphics[width=0.80\linewidth]{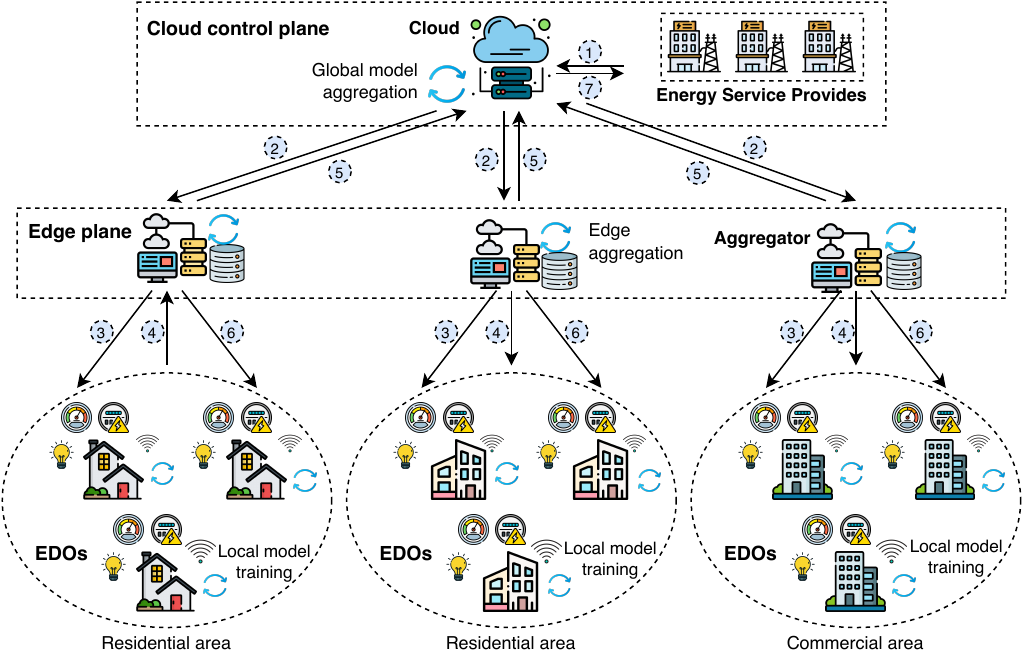}
	\caption{The architecture of edge-cloud integrated FL in~\cite{su2022secure} for energy data analysis in the smart grid with the following notations: (1) Request an FL task to the cloud, (2) Broadcast the aggregated global model to participating aggregators, (3) Forward the global model to EDOs, (4) Train all local models with local energy data and upload their updates to the corresponding aggregators, (5) Do the edge aggregation and send to the cloud for a global aggregation, (6) Reward to EDOs based on their contributions, and (7) Repeat steps (2)-(6) and then send the trained model to the EPS.}
	\label{Fig:smartgrid_fl}
\end{figure*}

In recent years, FL's preeminence in privacy and security protection for Industrial IoT-integrated smart grid networks has been profoundly demonstrated in the early detection and classification of harmful threats by ML/DL models \cite{abdel2022privacy,singh2021privacy}. To detect faults and anomalies in fog-assisted smart grids, the work in \cite{abdel2022privacy} builds an innovative federated semi-supervised class-rebalanced learning framework, namely Fed-SCR. There are two main modules in Fed-SCR. First, a lightweight generative network dealt with unbalanced data learned at fog nodes. Second, a geometric median-based aggregation scheme handles noisy gradients caused by unreliable fog nodes participating in training rounds. Especially, by reserving a period of training iterations before synchronizing with the central cloud, the aggregation scheme is capable of reducing the total number of communication rounds and total communication overhead significantly. As expected, Fed-SCR obtains outstanding detection accuracy over some existing cutting-edge semi-supervised models, viz., Fed-AE in \cite{zanjani2021adversarial} and Fed-GAN in \cite{zhang2021detecting}, in terms of both IID and non-IID data. 
FL for smart grid systems has the potential to improve privacy and efficiency, but it can cause overload on resource-constrained IoT devices. To address this, Singh \textit{et al.}~\cite{singh2021privacy} propose a real-time privacy-preserving FL-based electricity usage data analysis solution for distributed and serverless smart grid systems. Their framework generates a global model in the serverless cloud by aggregating locally trained models from home networks. They also use blockchain-enabled DEW servers to preserve the privacy of electricity usage information and local model training. However, there are still some security and efficiency concerns associated with implementing FL-based IoT services in smart grids, such as weak-learning shared local models, non-IID data distribution, and unpredictable communication delays. To address these concerns, Su \textit{et al.}~\cite{su2022secure} propose an edge-cloud-assisted FL framework for communication efficiency and privacy-preserving energy data sharing in smart grids. Their framework clusters neighboring energy users to lighten the heavy data traffic to clouds. They also formulate rewards for energy data owners (EDOs) and energy services providers (ESPs) to address the optimization problems of payoff functions separately. This ensures that all participants contribute to the learning process and are fairly compensated.
\textcolor{black}{In a nutshell, compared to the traditional 3-tier FL, the edge-cloud integrated FL offered three innovative things: the introduction of an edge aggregator to reduce the communication latency and bandwidth requirements, execution of local aggregation before sending model updates to the cloud for a better global aggregation, and integration of a reward mechanism to encourage the participation of data owners.}

\textcolor{black}{MARL has been successfully applied to various decentralized smart grid services, including energy control and management, and cooperative controls in power grids.
For example, Prasad \textit{et al.}~\cite{prasad2019multiagent} develop a novel MARL framework for efficient energy control and management in multiple energy-shared buildings to achieve a near-zero energy community. Each agent learns the energy consumption and generation data using the DQN algorithm to approximate Q-values representing a set of actions in a particular state to generate rewards. 
Another MARL-aided solution for decentralized smart grid services is the combination with LSTM networks for cooperative controls in power grids~\cite{chen2022powernet}. This combination results in three advantages: (i) it speeds up the training process and increases scalability; (ii) it encourages collaboration among neighboring agents with a learning-based communication protocol; and (iii) it addresses system uncertainty and learning disturbance.
Compared to the centralized approach with proximal policy optimization~\cite{schulman2017proximal}, PowerNet achieved a remarkable improvement in convergence speed and average training reward.
To minimize the voltage bias while maintaining the photovoltaic minimum active power drawdown, the work in~\cite{cao2021data} proposes an innovative model-free centralized training and decentralized execution (CT\&DE) MADRL framework. The term model-free is achieved through two steps: system identification with a sparse pseudo-Gaussian process to formulate the relations between power injections and voltage magnitudes, and voltage regulation with a smart meter-based state estimation process. The proposed framework can achieve the same voltage offset reduction performance as the centralized multi-agent soft actor-critic framework~\cite{iqbal2019actor} by evaluating 342-node low voltage network test systems without prior knowledge of exact physical models.}
Besides, MARL can be extensively applied to various services and applications in distributed smart grid systems, e.g., decentralized cooperative control in multiple IoT-based energy storage systems \cite{zhu2020decentralized}, energy management in microgrids \cite{kofinas2018fuzzy}, and voltage stabilization in intelligent inverter-based distributed generators \cite{tomin2021management}.

\textcolor{black}{
In closing, FL and MARL are two promising ML techniques that can be applied to a variety of decentralized smart grid applications. FL can be developed for privacy-preserving electricity usage data analysis, fault and anomaly detection with unbalanced data, and edge-cloud-assisted data storage and sharing; while MARL should be used to solve complex optimization problems such as energy control and management in multiple energy-shared infrastructures and voltage stabilization.
Notably, some important requirements should be taken into consideration when developing a distributed learning framework for IoT-enabled smart grid applications. First, to protect sensitive data collected from sensors in a smart grid from unauthorized access, distributed learning systems should only share necessary data with selective clients~\cite{su2022secure}. In addition, to meet the stringent latency requirement of responding to changes in power usage in real-time, these frameworks should conserve computational resources by distributing the training procedure across multiple clients and efficiently handle communication overhead to minimize the effect of limited communication resources~\cite{abdel2022privacy}. Finally, the frameworks should be easy to maintain, flexible for various specific applications, and scalable for the rapid growth of the smart grid.}

\subsection{Distributed Learning for IoT-enabled Autonomous Vehicle}

Vehicles are becoming increasingly autonomous thanks to recent advances in computing hardware and AI technologies and significant industry investment \cite{sangdeh2022cf4fl}. Therefore, the integration of vehicle-to-everything (V2X) networks with ML and distributed decision-making is needed to provide a comprehensive smart-connected vehicle solution. Following that, intelligent driverless vehicles can make highly accurate and reliable predictions in numerous practical and complicated scenarios by using distributed ML algorithms to perform collaborative multi-vehicle planning and control functions. 
In \cite{barbieri2022decentralized}, a novel distributed ML framework with a consensus-driven mechanism for road object classification is proposed for autopilot systems of automated vehicles to connect to a V2X network. In the framework, vehicles cooperatively learn the PointNet model \cite{qi2017pointnet}, which is developed for visual-based object classification and semantic segmentation based on a global dataset of Lidar point clouds acquired by networked vehicles. Notably, decentralized deep model parameter sharing and adaptation can be made over an average consensus mechanism to optimize learning efficiency by accommodating model parameters with mixed weights. To alleviate traffic congestion in V2X networks, the authors in \cite{lin2020distributed} have designed a new distributed software-defined IoV (SDIoV) architecture with edge intelligent computing to support vehicles with real-time routing and to improve decision-making accuracy. Based on each task requirement, an LSTM network gives a reasonable resource cost to predict the routing resource cost while the edge vehicles and devices cooperate in an asynchronous and parallel mechanism to minimize latency.
\textcolor{black}{Remarkably, the work in \cite{ma2021joint} not only proposes a joint scheduling} and resource allocation method for the platooning network of connected autonomous vehicles (CAVs) but also an efficiency-oriented distributed learning framework to obtain a comfortable balance between accuracy and complexity. To satisfy the learning service heterogeneity of every participating vehicle, the framework works on a two-phase Markovian stochastic process \cite{perel2010queues}. First, all clients join learning rounds with the same opportunity, which is totally different from classical distributed learning. Second, the efficiency-oriented global aggregation algorithm aims to provide a learning efficiency model for each vehicle relying on the constraint of learning latency, thereby avoiding the deviation in scheduling and resource allocation processes. From the mentioned above, it is obvious that distributed learning is becoming an efficient solution to autonomous driving tasks, such as routing, scheduling, and resource allocation, for performance improvement while presenting low processing latency and privacy protection. 

\begin{figure}[t]
	\centering
	\includegraphics[width=0.975\linewidth]{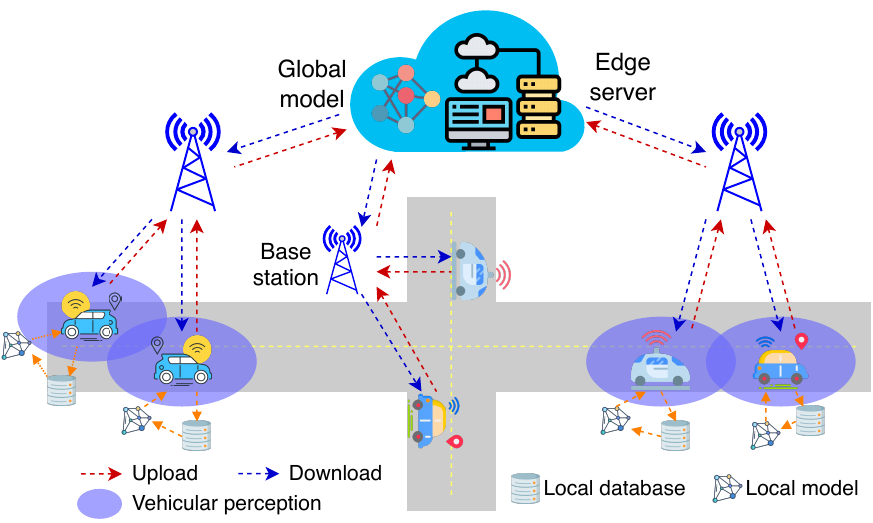}
	\caption{Overall FL process in IoV networks.}
	\label{Fig:vehicle_fl}
\end{figure}

In recent years, the evolution of FL has been impressively shown to IoV and CAVs' intelligent transportation systems (ITS) in order to against cyberattacks in learning-communication rounds between clients as vehicles and the central server \cite{manias2021making, sangdeh2022cf4fl}. An illustration of the overall IoVs' FL process can be described in Fig. \ref{Fig:vehicle_fl}. Due to its scalability and high availability, the work in \cite{ye2020federated} has exploited FL for several IoV applications (e.g., moving object detection and scene classification for autonomous driving). In which, a suggested selective model aggregation mechanism endeavors to mitigate the effect of image caliber's heterogeneous nature and computational ability among vehicular clients, such as different data quality, amount of data, communication condition, and computation capacity. Besides, a model selection procedure is also proposed to reformulate the information asymmetry problem (central server without regard to local client data uploads) into a two-dimensional contract theory problem, thereby readily solved by greedy algorithms. \textcolor{black}{In the comparison with FedAvg~\cite{mcmahan2017communication}, the proposed selective aggregation algorithm~\cite{ye2020federated} achieves} better than $2.42-6.28\%$ in terms of scene classification accuracy when evaluating the MNIST and Belgium datasets. \textcolor{black}{As being different from~\cite{ye2020federated} that focuses on enhancing the learning efficiency of AI/ML models, the work~\cite{liu2022distributedjsac} aims to} minimize the resource utility of AI-aided vehicular applications and services by emerging intelligent content caching in IoV networks.
By collaboratively learning deep models among CAVs in a group, the study in \cite{zeng2022federated} demonstrates the feasibility of the FL framework to deal with the performance degradation of centralized learning-based autonomous control in harsh road conditions and traffic dynamics. In addition, to solve several challenging issues related to the mobility of CAVs, wireless channel impairments, and imbalanced and non-IID data, a dynamic federated proximal (DFP) algorithm has also been regarded as an effective alternative. Unlike FedAvg, the DFP algorithm proposed a $L_2$ regularizer to ensure that the trained model parameters of local CAVs will converge to the parameters of the model received at the central server. The CAV's model parameters at the communication round $t$ are computed as follows:
\begin{equation}
f\left ( \boldsymbol{w}_{t+1,I} \right ) = \boldsymbol{w}_t+\eta _t \sum_{i=0}^{I-1}\left (\triangledown f_n \left ( \boldsymbol{w}_{t,i},\xi _i \right ) + \gamma _t \left ( \boldsymbol{w}_{t,i} - \boldsymbol{w}_t\right ) \right ),
\end{equation}
where $I$ is the number of learning iterations, $\eta$ is the learning rate, $\gamma$ is the $L_2$ regularizer coefficient, and $\xi _i$ refers to the gradient of a local data sample. The DFP method obtains a superiority in learning efficiency over FedAvg and FedProx in \cite{li2020federatedFedProx} when applied to the BDD datasets (imbalanced and non-IID data collected by various practical traffic scenarios), e.g., faster-learning convergence with lower loss. 
Meanwhile, in \cite{yu2021mobility}, a proactive edge caching approach that leverages FL and a context-aware adversarial autoencoder (C-AAE) model has been put forward to tackle three key challenges of high mobility, privacy concerns, and scalability in IoV networks. 
First, the global C-AAE model at the round $t$ is updated at the roadside unit (RSU) by aggregating the parameters of locally trained models with a position-based weighted averaging algorithm: 
\begin{equation}
\boldsymbol{w}_{t+1} \leftarrow \boldsymbol{w}_t -\eta \sum_{k=1}^{K} \delta_k \frac{d_k}{d}\boldsymbol{w}^k_{t+1},
\end{equation}
where $K$ is the total selected vehicles, $\eta$ is the fixed learning rate, and $\delta$ is the relative position of vehicle. Second, a designed mobility-aware vehicle selection based on various factors (e.g., channel condition, sufficient local data for training, and the standing time of the vehicle) is in charge of the optimization of caching resource utilization in IoV networks. Overall, the proposed FL framework demonstrated a desirable cache hit ratio and achieved faster convergence in learning compared to FedAvg. For instance, it reached a cache hit ratio of $16\%$ ten times quicker with $27$ vehicles. On another front, the work in \cite{liang2022semi} recommended a semi-synchronous FL framework, called Semi-SynFed, to address the high mobility and uncertainty of CAVs in IoV networks. In Semi-SynFed, flexible mechanisms assess their compute capacity, network bandwidth, and learning load in order to select an appropriate client for deep model learning, whereas a dynamic waiting time scheme is in charge of adjusting the server's timeout at each communication round. Furthermore, an elastic parameter aggregation plays an important role in balancing the time consumption and available computation resources. Following that, the global model $w$ at round $t$ can be written as follows:
\begin{equation}
\boldsymbol{w}_{t+1} = \boldsymbol{w}_t+\sum_{k=1}^{K} \alpha^k_t \triangledown \boldsymbol{w}^k_t,
\end{equation}
$\triangledown w^k_t$ refers to the gradient of client $k$ at round $t$ while $\alpha^k_t$ implies the mixing hyper-parameter involving the multiplication of the proportion of training samples and the linear function of staleness. From a learning convergence perspective, Semi-SynFed is faster than FedAvg but slower than SAFA in \cite{wu2021safa}.

On another front, the use of MARL has increased significantly in IoV networks and CAVs' autopilot systems to effectively handle different operational and driving issues caused by complicated multi-agent scenarios \cite{gao2022video}. For example, leveraging the advance of  DRL in complex interaction and highly non-stationary environments, Palanisamy \textit{et al.} in~\cite{palanisamy2020multiagent} develop a multi-agent learning platform, namely MACAD-Gym, based on an extensible set of simulation environments (e.g., joint action, join observation, state transition, and reward) to apply to different connected autonomous driving systems with unlimited operational design and multi-agent settings. To optimize long-term cumulative rewards, each agent in MACAD-Gym platforms endeavors action for a vehicle actor enforced by its local state.

Due to fast channel variations in highly mobile vehicular environments, inaccurate collection of instantaneous channel state information hinders the centralized frequency spectrum management and sharing in IoV networks, including vehicle-to-vehicle (V2V) and vehicle-to-infrastructure (V2I) links. To address the above-mentioned problems, the work in \cite{liang2019spectrum} evolves a MARL platform, where a fingerprint-based DQN model can associate each agent's policy change with the training iteration number and the exploration rate (or the probability of random action selection). To maximize the sum V2I capacity $C$ and increase the successful V2V payload delivery probability $L$, the reward $R$ at each time step $t$ of the $m$-th V2I link and the $k$-th V2V link is designed as follows:
\begin{equation}
R_{t+1}=\lambda_c \sum_{m} C^c_m\left [ m,t \right ] + \lambda_d \sum_k L_k\left ( t \right ),
\end{equation}
where $\lambda_c$ and $\lambda_d$ refer to positive weights to balance the objectives of V2V and V2I.

MARL methods have been widely introduced to ensure safe and efficient lane exchange in autonomous vehicles. However, these can be frustrating for drivers due to the complex and unpredictable nature of driving behaviors. Inspired by this hurdle, the work in \cite{zhou2022multi} develops a lane change decision-making method for multiple CAVs cooperating in a mixed traffic environment as a MARL problem. After rebuilding the actor-critic network in \cite{mnih2016asynchronous} with multiple actors, inter-agent communication will be set up to enhance the scalability and stability of the learning procedure. Most importantly, a multi-objective reward with multiple factors and constraints (including safety $r_s$, headway $r_d$, speed $r_v$,  and driving comfort $r_c$) is designed according to $R = w_s r_s + w_d r_d + w_v r_v - w_c r_c$, where $w_s, w_d, w_v$, and $w_c$ are the weighted hyper-parameters for dynamic traffic environment setups. For Internet of Electric Vehicles (IoEV) networks, the application of MARL has demonstrated substantial enhancement in the charging experience despite the challenges posed by the incomplete charging infrastructure and the unpredictable, imbalanced charging demands prevalent in the actual world. 
\textcolor{black}{Also evolving MARL, Zhang \textit{et al.}~\cite{zhang2021intelligent} develop multi-agent spatiotemporal RL (MASTER)}, an advanced learning framework that manipulates various long-term factors to intelligently achieve accessible charging stations publicly.
Besides, CT\&DE methods are also used in MASTER frameworks to ensure agents comply with learning policies while adapting well to non-stationary environmental conditions. Towards multiple-objective optimization (viz., charging wait time, charging price, charging failure rate, and charging station recommendation), a dynamic gradient re-weighting scheme is further developed to regulate the optimization process by transitioning from a centralized attentive critic to a multi-critic system, thereby facilitating diverse training stages. 
\textcolor{black}{Indeed, the use cases of distributed ML and MARL in ITS are diverse, in which distributed ML can be applied to improve the accuracy of models and reduce the amount of data needed for training, while MARL can be exploited to solve complex optimization problems. Some important distributed ML-based use cases in ITS include road object classification, distributed software-defined IoV architecture for routing, joint scheduling and resource allocation for platooning networks, FL for autonomous driving tasks, MARL for lane change decision-making, and multi-agent spatiotemporal RL for charging station recommendation.}

\subsection{Distributed Learning for Aerial IoT Networks}
As a key component of future 6G networks, aerial IoT (AIoT) networks are expected to fully fill the shortfall of previous network generations in terrestrial networks. AIoT networks define their hierarchical manner with four layers, including terrestrial IoT devices, low-altitude platforms (LAPs), high-altitude platforms (HAPs), and satellite communications \cite{dao2021survey}. With the advent of distributed learning paradigms, AIoT networks are potentially capable of delivering more advanced and intelligent wireless services to distributed IoT devices on a global scale  \cite{wang2022incorporating, liu2022wireless}. 

\textcolor{black}{A distributed DL-based offloading decision method is proposed in~\cite{li2021aerial} to enhance offloading decision-making and resource allocation while minimizing the weighted sum of latency and energy consumption in hybrid cloud-edge-ground layers.
Li et al.~\cite{li2021aerial} propose a distributed DL-based offloading decision method to enhance offloading decision-making and resource allocation while minimizing the weighted sum of latency and energy consumption in hybrid cloud-edge-ground layers. The proposed method outperforms existing approaches, including local learning, low-orbit (LEO) learning, and cloud-based learning, in terms of total weighted consumption.
Similar to \cite{li2021aerial} in optimizing the offloading procedure in dynamic IoT services to obtain reliable connectivity with UAV base stations, the work in \cite{zhao2021predictive} introduces a distributed learning framework called PRUDENT for multi-access edge computing (MEC).
PREDENT predicts link usage more accurately by two times than FedAvg while being more lightweight than its variant without clustering by six times.
For IoT-oriented satellite-terrestrial relay networks, Zhao et al.~\cite{zhao2021distributed} intelligently leverage a distributed Q-learning (DQL) approach to effectively tackle network congestion caused by massive IoT devices and multiple relays in IoT-oriented satellite-terrestrial relay networks. The proposed methods significantly reduce the complexity of onboard processing and signaling overhead in satellites compared to classical access control approaches.}

On another front, FL shows its feasibility in enabling multiple UAVs to collaboratively train ML models with locally raw data, thus efficiently addressing the serious concerns of conventional cloud-centric AI/ML-based UAV networks (e.g., weak privacy preservation, unacceptable latency, and resource burden). By leveraging the advanced features of the inexact stochastic parallel random walk alternating direction method of multipliers (ISPW-ADMM), the FL framework proposed in \cite{qu2021decentralized} demonstrates significantly improved communication efficiency for specific on-board missions in UAV swarm systems, such as noise-robust beamforming design. Precisely, the local extreme learning machine (ELM) model is deployed at each UAV to robustly learn beamforming patterns from locally collected noisy/imperfect channel state information (CSI) data when traveling a specific region/area, and the global model can be obtained by gradually converging all local models over stochastic gradients and first-order moment. As such, ISPW-ADMM achieves superior beamforming efficiency compared to state-of-the-art decentralized optimization methods (viz., D-ADMM~\cite{mota2013dadmm} and PW-ADMM~\cite{ye2020decentralized}) even if CSI is imperfect. However, the fact is that applying AI/ML to intelligent aerial services and applications typically lacks continuous connections between UAV swarms and ground base stations in centralized systems. Faced with such challenges, the work in \cite{zeng2020federated} presented a novel distributed FL framework to resolve the problem of joint power allocation and scheduling of multiple collaborative UAVs in an intra-swarm network, in which the impact of many wireless factors, including fading, transmission delay, and UAV antenna angle deviation on the learning convergence should be investigated and optimized for UAV-aided flying networks. To ensure the successful updating of the global FL model and local models, the transmission latency of the uplink $T_{iL}$ and downlink $T_{Li}$ are taken into account as additive constraints of the model aggregation function. Concretely, the global FL model is updated at communication round $t$ as follows:
\begin{equation}
\boldsymbol{w}^t = \frac{\sum_{k=1}^{K} D_k \boldsymbol{w}_k^t C_{k}^{t}}{\sum_{k=1}^{K} D_k C_{k}^{t}},
\end{equation}
with 
\begin{equation}
C_{k}^{t}=\left\{\begin{matrix}
1 & \mathrm{with}~\mathbb{P}\left ( T_{kL}^{t} \leq T_u(\beta ), T_{kL}^{t}\leq T_d(\beta )\right ), \\ 
0 & \mathrm{otherwise},
\end{matrix}\right.
\end{equation}
where $K$ is the number of UAVs, $D_{k}$ is the number of collected samples, $\mathbb{P}$ is the probability term of an objective function, and $T_u$ and $T_d$ correspond to the transmission delay thresholds of uplink and downlink. Based on the simulation results, the proposed method significantly improved the convergence performance of joint task design compared to a single design of either power allocation or power scheduling. Meanwhile, the work in \cite{pham2022energy} focuses on practical scenarios where terrestrial communications are disrupted due to malfunction/overload and IoT devices having limited batteries. Subsequently, a sustainable UAV-aided FL framework, namely energy-efficient FL (E2FL), is developed towards a joint algorithm for multiple objectives, for example, UAV arrangement, transmission time, power control, model learning efficiency, bandwidth allocation, and computing resources. Compared to \cite{pham2021uav}, E2FL demonstrated superior performance on multiple evaluation criteria (e.g., global model accuracy, model size, total energy consumption, and learning convergence). Remarkably, the studies in \cite{pham2022energy, pham2021uav} show the prospective integration of aerial edge computing \cite{pham2022aerialiotj},  aerial FL \cite{pham2022aerial}, wireless powered FL \cite{wu2022swipt}, which is seen to be a key enabler for many vertical applications as well as an ideal complement to existing terrestrial computing infrastructures. 

Recently, AI/ML algorithms have modernized IoT with edge computing and remote sensing technologies in terrestrial-aerial-space networks, which has been used in many services and applications for supervision and forecasting, such as air quality index, hazardous zone detection, and earthquake wave prediction \cite{liu2022wireless}. As stated in \cite{chhikara2021federated}, decentralized FL frameworks deliver more accurate remote sensing analytics at edge devices with limited resources while preserving the privacy of informative data compared to the traditional centralized cloud-based framework. 
\textcolor{black}{Indeed, the work in \cite{fadlullah2021smart} exploits FL to AIoT networks to accurately detect wildfires and estimate their spread in real-time. In this framework, a novel asynchronous weight updating scheme is applied to mitigate communication overhead by scheduling the global model aggregation with shallow or deep model parameters. This approach saves limited storage capacity and reduces bandwidth overload.
For accurate, fine-grained 3D air quality index (AQI) prediction, a newly FL-based aerial-ground sensing framework is developed~\cite{liu2021federated}. This framework alleviates computationally constrained devices like UAV swarms by taking advantage of a lightweight deep CNN model. It builds two collaborative FL streams to aggregate at the central server: 1) a region-level CNN-based AQI scale for a mobile vision-based aerial sensing stream (i.e., UAV swarm) and 2) small-scale accurate LSTM-based spatial-temporal for a sensor-based ground sensing stream (i.e., wireless sensor networks).}

\begin{equation}
\boldsymbol{w}_{t+1}=\argmin\frac{1}{D} \sum_{\ell \in L} D_\ell \mathcal L_\ell\left ( \boldsymbol{w} \right ),
\end{equation}
with $D=\sum _{\ell \in L} D_\ell$, where $D_\ell$ is the size of the local dataset for edge learning, $L$ refers to the set participating points (UAV or ground stations), and $\mathcal L_\ell$ refers to as the local loss calculated for the $\ell$-th points. Similarly, \cite{chhikara2021federatedIoT} combines sensory data and aerial panoramic images to give forecasting indexes of air quality and monitoring, where the multi-modality data collected by UAV swarms is learned collaboratively over an FL framework based on CNN-LSTM models to gain two objectives of improving prediction accuracy and preserving data privacy. Moreover, the use of FL for UAV swarms in \cite{lim2021towards} has shown promise in predicting traffic and enabling automated parking, offering active competitive opportunities for service providers with the Internet of Drones and Drones-as-a-Service towards modern aerial IoV (AIoV) services.

\begin{figure*}[t]
	\centering
	\includegraphics[width=0.985\linewidth]{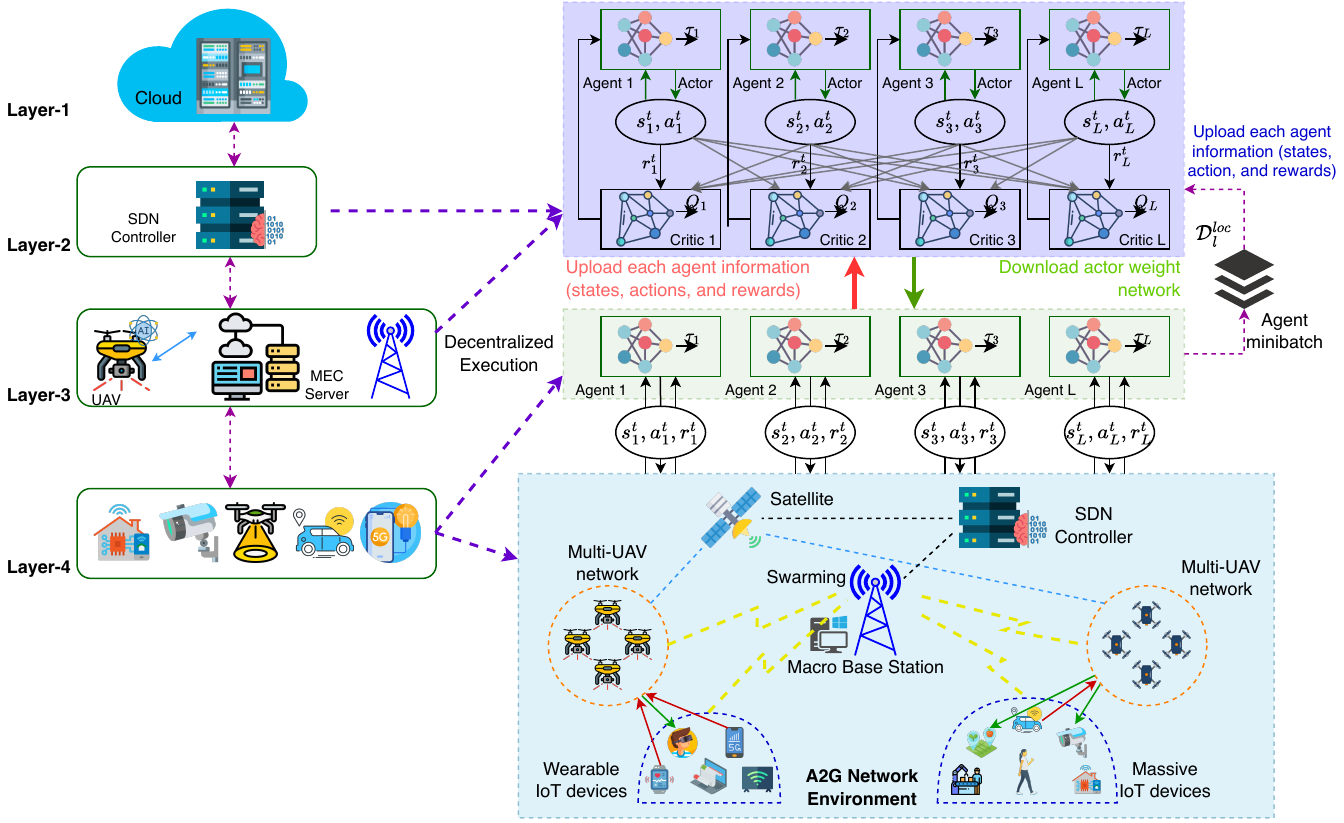}
	\caption{An overall architecture of MA-DDPG for AIoT networks~\cite{seid2021multiagent}.}
	\label{Fig:aerial_marl}
\end{figure*}

Nevertheless, the deployment of distributed learning in AIoT networks with multiple agents in realistic environments has remained challenging, especially when updating agents independently with their own policies over a non-stationary Markov process cannot guarantee learning convergence. The use of MARL techniques becomes beneficial to deal with AIoT problems thanks to its capability of distributed resource management for computation offloading with efficient resource allocation among UAVs \cite{li2022deep}, thus strongly supporting communications between aerial base stations and ground users. For instance, the work \cite{seid2021multiagent} applies MADRL to solve a joint problem of network congestion in MEC and UAV networks using stochastic game theory to offload costs in long-term computation and resource allocation. As shown in Fig.~\ref{Fig:aerial_marl}, agents in the AIoT network can increase the long-term reward through a multi-agent DDPG (MA-DDPG) method to optimize their learning policy, therefore reducing training cost with the CT\&DE technique. The main idea of MA-DDPG is learning a centralized action-value function. Each agent $l$ is updated with the global trainable weight information based on minimizing the loss of an actor-critic network as follows:
\begin{equation}
\mathcal L\left ( \boldsymbol{\theta} _l \right )=\frac{1}{H} \sum_{j} \left ( y^j-Q_l^\tau \left ( \boldsymbol{s}^j, \boldsymbol{a}_1^j,\dots,\boldsymbol{a}_L^j \right ) \right )^2,
\end{equation}
where $y^j = R^j_l + \gamma Q^\tau{}'_l \left ( S{}'^j, \boldsymbol{a}'_1,\dots,\boldsymbol{a}'_L \right )|_{\boldsymbol{a}'_k=\tau{}'_k\left ( \boldsymbol{s}^j_k \right )}$, $H$ refers to as the tuple of state $\boldsymbol{s}$, action $\boldsymbol{a}$, reward $R$, and next state $\boldsymbol{s}'$, $\gamma$ denotes the QoS threshold. Compared to the state-of-art of DDPG~\cite{lillicrap2015continuous}, Dueling-DQN~\cite{seid2021collaborative}, and DQN~\cite{seid2021collaborative}, MA-DDPG shows outstanding improvement in terms of average reward and consuming lower average cost when taking into account the diversity of numerous distinct agents, data sizes, and server's computation capacities. With the same in mind, the work in \cite{zhang2020uav} has also shown MA-DDPG benefits in optimizing UAVs' trajectories and flexibly allocating the transmit power at UAVs' transmitters and jammers. In order to further improve the distributed learning convergence and efficiency in complicated multi-agent environments, \cite{zhang2020uav} also proposes a continuous action attention MA-DDPG (CAA-MADDPG) by adding an attention layer network to capture attention weights associated with actions and observations of other agents. As a result, CAA-MADDPG achieves superior performance over DDPG~\cite{lillicrap2015continuous} and MA-DDPG~\cite{seid2021multiagent} in terms of reward gain and secure rate for different transmitter-jammer patterns. Besides, the work in \cite{peng2021multiagent} displays MA-DDPG's reliability in managing spectrum, computing, and catching resources, thus cooperatively supporting heterogeneous and delay-sensitive AIoV services and applications. 
\textcolor{black}{On the other hand, \cite{peng2021multiagent} demonstrates} the potential for task offloading and resource management in satellite-terrestrial IoV networks. In \cite{yin2021resource}, the problem of resource allocation and trajectory design in UAV-aided IoT networks can be well solved by employing MADRL with parameterized DQNs to handle the mixture space of actions and a QMIX mechanism \cite{rashid2018qmix} to aggregate UAV's local critics.

\subsection{Distributed Learning for IoT-enabled Smart Industry}
With numerous advancements in high-tech, we have witnessed an incredible revolution in smart industries from 4.0 to 5.0. In the fourth industrial revolution, starting in 2011, we saw the emergence of digitalization, spearheaded by many innovative technologies, such as IoT, cognitive computing, and the combination of AI, especially DL, and big data. Subsequently, a variety of frameworks have arisen to serve a wide range of applications of smart industries. While Industry 4.0 is accelerated by adopting the interconnectivity of centralized technologies, the next inevitable wave of revolution, a.k.a., Industry 5.0, asks for the full integration of intelligent systems featured by mass customization, collaboration, and cyber-physical cognitive computing. Importantly, these systems can leverage the available computing resources of IoT devices to learn a massive amount of industrial data using distributed AI-based frameworks, hence enabling systems to achieve reliability, privacy, and efficiency in dealing with multiple agents in smart industries.

In fundamental, a distributed ML framework for smart industries should fulfill some key requirements, including distributed control function, self-organization, communication between smart IoT devices, and decision-making capability in real-time. As such, deploying distributed learning frameworks is necessary to make an interactive learning mechanism in various manners for multiple distributed control purposes. For smart factories, \cite{banadaki2019cyber} provides a distributed learning solution to collaboratively train IoT-connected devices to obtain the highest learning efficiency while maintaining the scalability of manufacturing systems and the specialization of applications. For aerospace industries, especially the application of aerofoil self-noise prediction, the work \cite{ren2018distributed} has succeeded in combining a distributed ML framework and a distributed optimization algorithm at multiple edge devices using fuzzy logic models to optimize the global cost function. Additionally, the framework has two key advantages, including speeding up learning convergence with a small amount of communication and computation usage and preventing leakage of raw and sensitive information during data exchange between devices. For robot-aided production line commission and re-configuration, the work in \cite{kotriwala2020supporting} demonstrates the potential of distributed ML in assisting developers in programming robots to simplify certain critical tasks of localizing workpieces, estimating movements, and optimizing control parameters. The main feature of this framework is to generate a central node that is in a chance of triggering multiple simulated robots (a.k.a., edge nodes, which are initialized with the same model but with different parameters initialization) to satisfy diverse requirements from customers and improve productivity.

\begin{figure*}[t]
	\centering
	\includegraphics[width=0.925\linewidth]{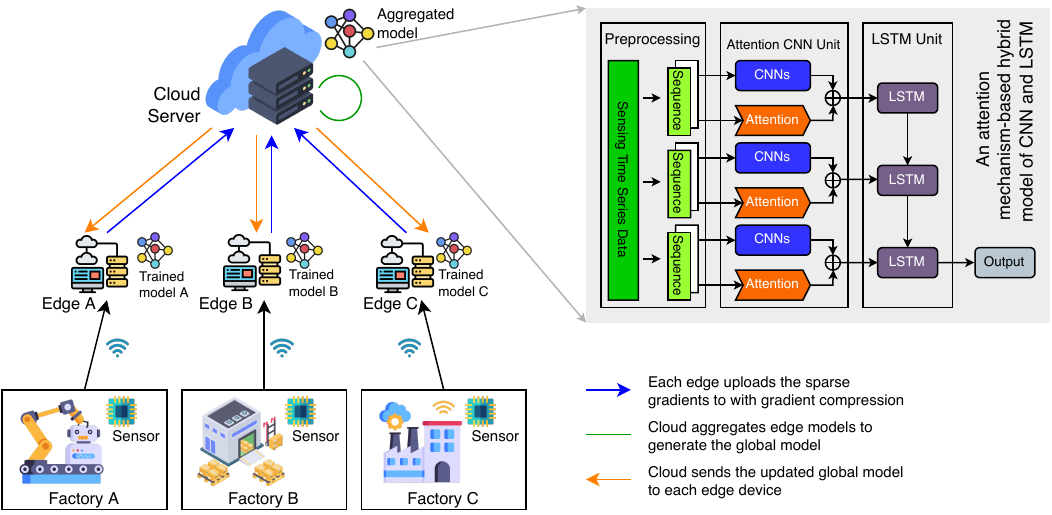}
	\caption{An overall architecture of FL framework for anomaly detection in smart industries~\cite{liu2021deep}.}
	\label{Fig:industry_fl}
\end{figure*}

In smart industries, the most pivotal factor is data privacy concerns, which limit the capability of centralized ML methods in wireless environments. However, in response to this challenge, FL with distributed learning mechanisms has emerged as a promising alternative to deliver higher industrial data preservation and generates incredibly competitive learning efficiency compared to classical centralized learning approaches. In recent years, FL has frequently innovated and reformed to address more specialized tasks. For example, \cite{zhao2022afl} proposes an adaptively federated multi-task learning (AFL) in order to customize the specialized function of industrial equipment and specify learning models to distinct tasks. The core idea of this framework is to create sparse shareable structures that can iteratively prune, which combined with some add-on auxiliary layers in a tailored task routing network to carry out learning tasks by each subnet. Consequently, the model sharing between different edge nodes is effectively improved. Meanwhile, the study in \cite{deng2022federated} utilizes FL to overcome challenges associated with restricted data acquisition in aircraft manufacturing industries, such as complex implementation, time-consuming, and high cost. However, different data structures and formats from distributed enterprises pose an intractable problem in performing learning processes for some particular tasks, e.g., tool wear estimation and smart process monitoring. To this end, employing a domain adaptation method inspired by transfer learning can effectively allow the refinement of irrelevant features while keeping domain-invariant attributes. 
In \cite{huong2021detecting}, FL plays the role of helping edge devices detect anomalies without waiting for a response from a centralized server whenever there are cyberattacks in IoT-based manufacturing control systems. To complete this mission, each edge node should deploy a hybrid model consisting of a variational autoencoder and LSTM to give suitable responses from input time-series data across multi-scale time frames. Unlike \cite{huong2021detecting}, the work in \cite{liu2021deep} focuses on an attention mechanism-based hybrid model of CNN and LSTM to improve the anomaly detection accuracy, in which CNN is to capture fine-grained relevant features without vanishing gradient descend and LSTM is to analyze time-series data at multiple signal resolutions. Besides, a novel gradient compression algorithm based on Top-\textit{k} selection is developed for FL to reduce the edge-cloud communication cost. In addition to time-series data, FL remains exploited image data in many manufacturing tasks, such as defect detection and predictive maintenance. For high-accuracy defect detection in additive manufacturing, \cite{mehta2022federated} applies FL to devastate deficiency involving data availability and privacy. The layer-wise images are collected from the laser powder bed fusion procedure at each edge device in order to train the learning model with the U-Net architecture \cite{ronneberger2015u}. The numerical results argued that the data diversity across edge devices improves learning efficiency and detection accuracy of FL without computational cost increment, and transfer learning can improve the model's generalization to be suitable with heterogeneous machinery systems.

The inevitable deterioration of machines during prolonged usage renders periodic maintenance actions indispensable to ensure the reliability of manufacturing systems and minimize production costs. Driven by this fact, the work in \cite{wang2016multi} dissects the applicability of a distributed MARL in handling open and dynamic manufacturing issues in resource-constrained flowline systems. By thoroughly scrutinizing the interrelatedness between the individual decisions taken by each agent and the overall optimization objective of multiple agents, such MARL has shown immense potential in conjunction with heuristic algorithms to enhance learning efficiency and convergence rate in dealing with intricate flowline systems. 
In a serial production line with multi-level actions,  the work in \cite{su2022deep} builds a deep MARL framework-assisted CT\&DE paradigm to transform a multi-agent maintenance decision-making problem into a decentralized partially observable Markov decision process (Dec-POMDP) in which each module/machine was modeled as a cooperative agent. Consequently, the proposed framework can reach two objectives: leveraging a reward function to measure system-level production loss and specializing in a value-decomposition multi-agent actor-critic algorithm. For smart factories containing heterogeneous manufacturing units, solving scheduling issues with high-dimensional data collected by IIoT devices are an impediment. 
%
\textcolor{black}{In this context, Zhou \textit{et al.}~\cite{zhou2021multi} propose a MARL-based approach for online scheduling in manufacturing. The approach uses multiple AI schedulers to make decisions cooperatively under uncertainties. Each AI scheduler has four deep neural networks to infer online schedules from real-world data and the shareable policies of others. The approach minimizes makespan and balances workload, and it can also deal with unforeseen manufacturing events. MARL has also been shown to be effective for distributed learning in collaborative robotics tasks.}
Similarly, the work in \cite{yu2021optimizing} presents a novel DQN-MARL approach to optimize task scheduling in human-robot collaboration (HRC). The task scheduling problem was mapped into a Markov game model under multi-agent constraints to capture correlated equilibrium policies among agents in a dynamic real-time environment. Each DQN agent was trained by perceiving all the remaining agents' actions and rewards, enabling them to discover dynamic task scheduling policies that are constantly updated in response to environmental changes. The proposed approach was evaluated in scenarios with different numbers of agents and task spaces. The results showed that the DQN-MARL approach achieved better task scheduling performance than the naive Nash-Q learning, dynamic programming, and DQN-based single-agent RL approaches.
\textcolor{black}{In a nutshell, distributed learning with deep models is a promising technology for a variety of smart industry services and applications. It can help to improve efficiency, reduce costs, improve quality, and protect privacy in a diversified range of use cases, including smart factories, aerospace industries, robot-aided production line commission and reconfiguration, data privacy, anomaly detection, and predictive maintenance. However, some practical issues should be taken into consideration, such as the large number of devices and sensors, which can result in a large number of access requests and a massive amount of data for real-time processing.}

\textcolor{black}{Interestingly, the requirements for designing and developing an effective distributed learning framework for IoT-enabled smart factories are fairly similar to those for smart grids, but there are some key differences~\cite{wang2022federated}. Specifically, smart factories typically have more computational resources than smart grids, so they may not need to rely as heavily on distributed learning to conserve resources. However, they may still benefit from distributed learning to improve the efficiency of the training process and to make better use of the data collected from a variety of sensors. Additionally, smart factories do not have as strict latency requirements as smart grids, but they still need to be able to make decisions quickly. Finally, smart factories often collect more data from different sensors (i.e., a large volume and high diversity of data types) than smart grids, so the distributed learning framework needs to be able to handle large datasets efficiently~\cite{mazzocca2023trulaas}.}

\textcolor{black}{Several companies are developing and implementing FL in their factories. For example, Bosch is using FL to develop AI models that can predict when machines are likely to fail. This information can then be used to schedule preventive maintenance, which can help to reduce downtime and improve production efficiency. To improve the efficiency of the supply chain, Siemens develops FL models to predict demand for products and optimize inventory levels, thus reducing costs and improving customer satisfaction. Toyota exploits FL models to detect defects in vehicles before they are shipped to customers, consequently cutting warranty costs and increasing customer satisfaction. Similarly, Volkswagen is using FL to enhance the efficiency of its production lines by optimizing the scheduling of production tasks, resulting in boosting production throughput besides operating cost reduction. The development and implementation of FL in these companies is still in its early stages. However, the potential benefits are significant. FL can help companies to improve the quality and efficiency of their manufacturing operations, while also reducing costs and improving customer satisfaction.}

\textcolor{black}{In general, distributed ML is a promising technology, rather than centralized ML, for various IoT-enabled applications. For example, distributed ML is well-suited for IoT-enable healthcare and medical applications and services (e.g., predicting patient outcomes, detecting diseases, and developing personalized treatment plans) thanks to it enables the development of accurate and privacy-preserving ML models. In the smart grid domain wherein efficiency and reliability are stringently required in energy demand prediction, outage prevention, and energy distribution optimization, distributed ML is more applicable than centralized ML. In many scenarios of IoT-enabled autonomous vehicles, distributed ML is essential to deploy ML models that should make real-time decisions based on data from a variety of sensors quickly and efficiently.
It can be realized that distributed ML can offer scalability, robustness, and privacy preservation benefits. However, it is important to note that distributed ML can be more complex to implement and manage than centralized ML. For more information, an overall comparison between distributed ML and centralized ML in various IoT-enabled domains is given in Table~\ref{tab_DisVsCen2}.}
\begin{table*}[!ht]
\centering
\caption{\textcolor{black}{Comparison of distributed ML frameworks and centralized ML frameworks}}
{\color{black}
\begin{tabular}{|p{3.5cm}|p{6cm}|p{6cm}|}
\hline
Domain & Distributed ML & Centralized ML \\ \hline
IoT-enabled healthcare & More scalable and robust, more privacy-preserving & Easier to implement and manage \\ \hline
IoT-enabled smart grid & Can improve energy efficiency and reduce costs, can help to detect and prevent outages & Can be more vulnerable to single points of failure and privacy concerns \\ \hline
IoT-enabled autonomous vehicles & Can improve safety and efficiency, can help to develop new features and services & Can be more vulnerable to single points of failure and privacy concerns \\ \hline
Aerial IoT networks & Can improve coverage and reliability, can help to develop new applications such as drone delivery & Can be more complex to implement and manage \\ \hline
IoT-enabled smart industry & Can improve productivity and reduce costs, can help to develop new products and services & Can be more vulnerable to single points of failure and privacy concerns \\ \hline
\end{tabular}}
\label{tab_DisVsCen2}
\end{table*}
\subsection{Summary and Lessons}


\textcolor{black}{In general, distributed learning has several advantages over centralized learning for IoT-enabled applications (such as healthcare, smart grids, autonomous vehicles, aerial networks, and smart industry). First, it can be used to train models on large datasets that would be too large to fit on a single device. Second, it can be used to train models on devices distributed geographically, while still protecting the privacy of the data. This can further improve the performance of ML models, as the data is closer to the source and there is less latency. Third, it can be used to train models on devices with limited computation power, which can save energy.
For example, in aerial networks, distributed learning can be exploited to train models using the data collected from multiple modality sensors distributed across multiple drones. This can be used to improve the safety and efficiency of drone operations (trajectory prediction and collision avoidance) and drone-based applications.
On the contrary, centralized learning trains models with sensory drone data at a central server, where a powerful computing capacity of the server is required. This can be a bottleneck in terms of latency and bandwidth, and it can also pose a security risk, as the data is centralized in a single location.
Distributed learning is becoming increasingly popular for IoT-enabled applications, as it can overcome the challenges of limited computation power, storage, and energy on IoT devices. However, centralized learning is still an alternative for applications that require real-time training or security.}


In healthcare and medical domains, preserving patients' data privacy collected by hospitals, health centers, and medical institutions is of utmost priority whenever a distributed approach is contemplated for a specific application. In this regard, FL has demonstrated superiority in safeguarding data privacy against cyberattacks and utilizing the computing resources of IoMT devices to collaboratively learn ML/DL models while not compromising the overall performance as compared to centralized learning models. In order to achieve a more efficient diagnosis, a vast amount of healthcare and medical data needs to be collected from multiple sources with varying formats and structures (e.g., time-series sensory data, images, and electronic medical records), which may be noisy and contain outliers. Therefore, DL models with single architectures (e.g., RNN, LSTM, and CNN) and hybrid architectures (e.g., RNN+DNN, CNN+LSTM) are more advantageous than traditional ML models, especially being modernized by advanced structural connection to improve accuracy while maintaining an acceptable complexity for resource-constrained devices.

Due to some natural difficulties in data collection and labelling, building a fully annotated dataset in the smart grid domain for pattern recognition using supervised distributed learning is unachievable in practice. In this context, semi-supervised learning and unsupervised learning become two potential solutions and receive significant interest in a variety of smart grid applications. Another challenging issue in smart grids is that different energy delivery systems use non-identical hardware and software (e.g., distributed energy allocation and virtual power plants). This may cause data bias (e.g., sampling bias, measurement bias, exclusion bias, racial bias, and association bias). To avoid some learning mistakes and failures derived from data bias, some data normalization and standardization techniques should be applied in combination with the expert knowledge of data scientists.

Currently, numerous autonomous driving and aerial network applications have exploited the computing resources of edge IoT devices to implement distributed learning. In these applications, FL has shown the primary advantage of data privacy, while MARL  effectively handles many challenging driving and controlling tasks in complicated multi-agent scenarios. FL allows for collaborative learning of AI models among CAVs and UAVs in a group or swarm; however, specialized supportive algorithms are required to address dynamic mobility, wireless channel impairments, and imbalanced and non-IID data, such as dynamic federated proximal \cite{zeng2022federated} and networking-based client selection \cite{liang2022semi}. Furthermore, conventional CT\&DE strategies should be renovated to federated training and decentralized execution to address the scalability problem in large-scale CAV and UAV networks. Therefore, designing an efficient reward algorithm subject to the constraints of multiple associated agents is always an essential task in any MARL framework. For instance, the MA-DDPG framework offers an efficient long-term accumulative reward algorithm for discovering optimal learning policies that maximize average reward and reduce training costs.

In smart industries, dealing with multi-task and multi-agent learning issues is arduous due to the complicated manufacturing environments involving collaborative robotics. Additionally, the capabilities of function customization of industrial equipment and learning adaptation of AI models should be revisited when developing distributed learning for IIoT-based manufacturing control and production line systems. A federated multi-task learning framework can be a promising solution, where each subnet is responsible for a single task, learned at selected edge nodes, and managed by a task routing network. To effectively handle the multi-agent learning problem in manufacturing environments, where each machine in a system or each module in a machine is modeled as a cooperative agent, a specialized model is needed to drive underlying learning policies, thus allowing to uncovering of the collaborative relations between local decisions made by different agents and synthesize all agents for an optimal objective.

In this section, we have provided an extensive review of the use of distributed learning for IoT applications: healthcare, smart grid, autonomous vehicle, aerial network, and smart industry. Thereafter, several cutting-edge frameworks of distributed ML, FL, and MARL are introduced to address numerous challenging problems in a wide spectrum of applications, from medical image analysis and energy theft detection to publicly accessible charging station recommendation and manufacturing defect detection. Additionally, we have further delivered the representative references with their technical contributions in Table~\ref{Table:Summary_IoTApplications}.

\section{Challenges and Future Directions}
\label{Sec:challenges}
As presented in the previous sections, distributed learning demonstrates its potential in numerous IoT services and applications. Besides its potential, distributed learning also faces several critical challenges that need to be considered in future network systems. In this section, we discuss key challenges of security and privacy, communication efficiency, resource allocation, the next Internet revolution, and specifications and standardizations. Furthermore, we highlight several promising solutions and research directions.

\subsection{Security and Privacy}
The development and use of distributed learning are motivated by increasing computational resources at end devices and privacy concerns of a large amount of sensitive IoT data. However, distributed learning also raises critical issues in terms of security and privacy. It is because IoT devices collect not only much sensitive information, such as names, bank accounts, and habits but also are aware of many scandalous security and data breaches. It is worth highlighting the difference between security and privacy in the future IoT. While security refers to protecting IoT data from bad things (e.g., malicious access and denial of data), privacy refers to preventing personal information from unintentional disclosure. For example, an adversary can steal local model updates from IoT devices in FL, which are then used by the adversary to create its global model and identify private data from IoT devices. The adversary can also inject a small amount of poisoned data into the training phase of distributed learning models, thus leading to an incorrect inference result.
Due to the nature of distributed learning and 6G IoT, there are many issues of privacy (e.g., passive attack and model inference) and security (e.g., poisoning attack, evasion attack, model inversion attack, membership inference attack, model stealing) \cite{ma2022trusted}. Therefore, numerous studies have been investigated to develop security and privacy solutions for the future IoT. For example, the work in \cite{zhang2022robust} improves privacy in FL through a robust FL framework. In particular, two game-theoretic approaches are designed to motivate the participation of IoT devices in the training stage. These approaches make use of differential privacy to limit the impact of adversary IoT devices on the aggregated FL model. Another study in \cite{huang2021multi} leverages MA-DRL to solve two optimization problems in an IRS-aided relay network with buffering, including one to maximize the secrecy rate with a delay constraint and another to maximize the throughput with a secrecy constraint.

There have been various effective solutions to come up with privacy and security issues in 6G IoT. However, there is a huge demand for better solutions that can cope with various aspects of complexity and deployment because many IoT services and applications will emerge in the future. Firstly, since 6G IoT will be realized with new key-enabling technologies, it is vital to revisit existing privacy and security solutions once these technologies are deployed in the future IoT. These technologies (e.g., intelligent surfaces, compressive sensing, quantum information, THz communications, and semantic communications \cite{de2021survey}), when combined with distributed learning, can help us achieve 6G requirements but also introduce additional security and privacy issues. Secondly, the convergence of blockchain, edge computing, quantum computing, and distributed learning has shown the potential to improve IoT security. For example, blockchain can be leveraged to design serverless FL frameworks and solutions \cite{nguyen2021federated_BC}, thus increasing the scalability of IoT systems. However, despite significant advantages, blockchain security and privacy issues may significantly impact the performance of IoT services and applications. Therefore, more research is needed to exploit blockchain characteristics to protect data privacy and avoid security issues in the future IoT. Thirdly, scalability and complexity are significant issues as the communication overhead grows with the number of devices in massive IoT, and the model complexity increases with the amount of training data on IoT devices. Therefore, lightweight frameworks and scalable solutions should be investigated in the future by considering the resource availability of IoT devices and the requirements of IoT services and applications. As an example, the work in \cite{nguyen2022hcfl} proposes to use an autoencoder to design a communication-efficient FL framework and shows that the distortion rate is reduced quadratically as the number of IoT clients increases. Accordingly, the proposed scheme in \cite{nguyen2022hcfl}, namely high-compression FL (HCFL), is highly scalable and applicable to massive IoT. Furthermore, by exploiting the power of quantum computing, the security issues and data privacy in distributed learning can be effectively solved, as demonstrated by quantum FL \cite{huang2022quantum}, quantum SL \cite{yun2022quantum}, and quantum RL \cite{narottama2021quantum}, thus stimulating future research in quantum distributed learning for the future IoT. 

\begin{table*}[!b]
    \renewcommand{\arraystretch}{1.05}
	\caption{\textcolor{black}{Summary of the literature on distributed learning for IoT applications.}}
	\label{Table:Summary_IoTApplications}
	\centering
	\begin{tabular}{|c|c|p{2.5cm}|p{3.5cm}|p{8.25cm}|}
    \hline
		\textbf{Service} & \textbf{Ref.} & \textbf{Use case}  & \textbf{Learning Approach}  & \multirow{1}{*}{\textbf{Technical Contributions}} \\ \hline 
	
	\multirow{1}{*}{\rotatebox{90}{\textbf{~\makecell{Smart \\Healthcare}~}}}
	& \cite{qian2020wearable}
        & Self-health monitoring
	& Distributed hierarchical DL approach
	& Smartphones and IoMT train neural networks with their own data. A shared consensus AI model is to guarantee data privacy.
		\\ \cline{2-5}

	& \cite{kasyap2021privacy}
         & Healthcare data privacy preservation
    & Intermediate decentralized learning approach
	& Local AI models are partially trained on edge IoMT devices and completely done at intermediate gateways.
		\\ \cline{2-5}
	
	& \cite{elayan2022sustainability}
         &  Real-time healthcare data monitoring
	& Deep FL approach.
	& Local training is done on edge IoMT devices, and models' changes are synthesized for FedAvg-based model aggregation on the cloud server.
		\\ \cline{2-5}
	
	& \cite{adnan2022federated}
         &  Histopathology images analysis
	& Differentially private FL approach
	& 
    Multiple-instance learning algorithm over a differentially private stochastic gradient descent mechanism for non-IID data. 
		\\ \cline{2-5}
	
	& \cite{sun2022scalable}
    & Healthcare monitoring and analysis services    
	& Scalable and transferable FL approach
	& 
    Transfer learning with parameter initialization to accelerate the training process and learning convergence of unknown data.
		\\ \cline{2-5}
    
	& \cite{vlontzos2019multiple}
         &  Multiple landmark detection
	& MARL approach
	& Multiple agents are configured with implicit inter-communication based on a DQN model.
		\\ \cline{2-5}
    
	& \cite{liao2020iteratively}
         &  Interactive 3D medical image segmentation
	& MARL approach
	& A relative cross-entropy gain-based reward in a progressive model update process to address prediction uncertainty.
		\\ \cline{2-5}
    
	& \cite{allioui2022multi}
         &  COVID-19 CT image segmentation
	& MARL approach
	& 
    An MDP-based learning mechanism is adopted for the DQN model to provide a more accurate clinical diagnosis and save diagnostic time.
	\\ \hline
	
	\multirow{1}{*}{\rotatebox{90}{\textbf{~\makecell{Smart \\Grid}~}}}
	&  \cite{amini2020distributed}
         & Power system resilience
	& Distributed ML approach
    & A collaborative learning is adopted for each RMS and its neighbors, and an optimal resource-sharing policy is formulated under the constraint of power load and battery.
		\\ \cline{2-5}
	
	& \cite{qi2016cybersecurity}
         &  Power grid infrastructure protection
	& Regular distributed learning approach
	& A cooperatively distributed learning mechanism with different statistical ML models.
		\\ \cline{2-5}
	
	& \cite{dobbe2020toward}
         &  Predicting inverter actions in multiple controllable DERs
	& Decentralized data-driven learning approach
    & An agnostic multiple step-wise linear regression algorithm with Bayesian information criterion is to select the relevant feature subsets.
		\\ \cline{2-5}
	
	& \cite{liu2022federated}
         &  Decentralized voltage control in distributed energy networks
	& FL approach
    & Perform the federated training $\&$ decentralized execution framework instead of centralized training $\&$ decentralized execution.
		\\ \cline{2-5}
	
	& \cite{wen2022feddetect}
         &  Energy theft detection in IoT-based smart grids
	& Privacy-preserving FL approach
	& Distributed detection stations collaboratively train the same TCN with local encrypted data.
		\\ \cline{2-5}
	
	& \cite{abdel2022privacy}
         &  Faults and anomalies detection
	& Privacy-preserving federated semi-supervised class-rebalanced learning
	& A lightweight generative network to deal with unbalanced data and a geometric median-based aggregation scheme to handle noisy gradients.
		\\ \cline{2-5}
	
	& \cite{singh2021privacy}
         &  Electricity usage data analysis in serverless smart grid systems
	& Privacy-preserving FL approach
	& All local models are aggregated to generate the global model in a serverless cloud with blockchain-enabled data privacy preservation.
		\\ \cline{2-5}
	
	& \cite{su2022secure}
         &  Privacy-preserving energy data sharing
	& Edge-cloud-assisted FL approach
	& An aggregator, as a group of edge computing nodes geographically distributed in the edge plan, lightens heavy data traffic to the cloud.
		\\ \cline{2-5}
	
	& \cite{prasad2019multiagent}
         &  Efficient energy control and management
	& MADRL approach
	& Each agent learns the energy consumption and generation data using the DQN algorithm to approximate Q-values.
		\\ \cline{2-5}
	
	& \cite{chen2022powernet}
         &  Inverter-based secondary voltage control in IoT-based DGs
	& MADRL approach
	& PowerNet, an on-policy MARL algorithm with LSTM networks, is featured by a novel spatial discount factor, an innovative learning-based communication protocol, and an action smoothing factor
		\\ \cline{2-5}
	
	& \cite{cao2021data}
         &  Minimization of voltage deviation
	& Model-free MADRL approach
	& A system identification technique with sparse pseudo-Gaussian to formulate the relations between power injections and voltage magnitudes.
	\\ \hline	
	\end{tabular}
\end{table*}%

\begin{table*}[!htbp] 
\ContinuedFloat
\caption{\textcolor{black}{Summary of the literature on distributed learning for IoT applications (cont.).}}
\centering
	\begin{tabular}{|c|c|p{3.0cm}|p{3.5cm}|p{7.95cm}|}
    \hline
		\textbf{Service} & \textbf{Ref.} & \textbf{Use case}  & \textbf{Learning Approach}  & \multirow{1}{*}{\textbf{Technical Contributions}} \\ \hline 
	\multirow{1}{*}{\rotatebox{90}{\textbf{~\makecell{Autonomous \\Vehicle}~}}}
	& \cite{barbieri2022decentralized}
        &   Road object classification
	& Distributed ML approach with a consensus-driven mechanism
	& Decentralized deep model parameter sharing and adaptation are performed over an average consensus mechanism with mixing weights.
		\\ \cline{2-5}
	{}
	& \cite{lin2020distributed}
         &  Accuracy enhancement for SDIoV-based vehicle routing
	& Distributed ML approach
	& LSTM networks estimate overhead to predict the resource cost of routing. CAVs cooperatively minimize the overall system latency.
		\\ \cline{2-5}
	{}
	& \cite{ma2021joint}
         &  Joint scheduling and resource allocation for the platooning network of CAVs
	& Distributed ML-based methods
	& A distributed framework is specialized by a two-phase Markovian stochastic process to facilitate learning service heterogeneity. 
		\\ \cline{2-5}
	{}
	& \cite{ye2020federated}
         &  Different applications and services in IoV networks
	& Advanced FL approach
    & A model selection procedure for information asymmetry is studied with a greedy algorithm under multiple complicated constraints.
		\\ \cline{2-5}
	{}
	& \cite{zeng2022federated}
         &   Performance enhancement in harsh road conditions and traffic dynamics
	& Dynamic FL framework
	& The $L_2$ regularizer in DFP compels the trained model parameters of local CAVs to converge to the parameters of the received model.
		\\ \cline{2-5}
	{}
	& \cite{yu2021mobility}
         &  Intelligent mobility-aware proactive edge caching of CAVs in IoV networks
	& FL approach
	& The global C-AAE model is updated by aggregating the parameters of local models with a position-based weighted averaging algorithm.
		\\ \cline{2-5}
	{}
	& \cite{liang2022semi}
         &  High mobility of CAVs in IoV networks
	& Semi-synchronous FL approach
	& A flexible mechanism to select appropriate clients and an elastic parameter aggregation for balancing time consumption and available computation resources.
		\\ \cline{2-5}
	{}
	& \cite{palanisamy2020multiagent}
         &  Complex interaction between CAVs in a highly non-stationary environment
	& MADRL approach
	& Extending MADRL with many constraints: joint action, joint observation, state transition, and reward, to adapt to different driving systems. 
		\\ \cline{2-5}
	{}
	& \cite{zhou2022multi}
         &  Lane-changing decision-making of multiple cooperative CAVs in mixed-traffic environments
	& MADRL approach
	& An actor-critic network extends a certain communication among many participating agents to enhance scalability and stability.
	\\ \hline
	
	\multirow{1}{*}{\rotatebox{90}{\textbf{~\makecell{Aerial IoT \\Networks}~}}}
	&  \cite{zhao2021predictive}
         &  IoT services and connectivity over MEC for UAV base
	&  Distributed learning framework
    & A proactive mobility management method has a trajectory similarity measurement and a similarity search scheduling to enhance the accuracy of distributed mobility and traffic flow predictions.
		\\ \cline{2-5}
	{}
	& \cite{zhao2021distributed}
         &  IoT-oriented satellite-terrestrial relay
	& DQL method
	& The joint relay selection and access control problem is addressed by a DQL algorithm. 
		\\ \cline{2-5}
	{}
	& \cite{zeng2020federated} 
         &  Joint power allocation and scheduling of collaborative UAV networks
	& Distributed FL
    & The transmission delays of uplink and downlink are additionally considered in model aggregation to update global FL and local models.
		\\ \cline{2-5}
	{}
	& \cite{fadlullah2021smart} 
         &  Real-time forest fire detection using AIoT networks
	& FL approach
	& In a novel asynchronous weight updating scheme, UAVs send partly shallow trainable parameters of local models for aggregation, thus saving limited storage capacity and reducing bandwidth overload. 
		\\ \cline{2-5}
	{}
    & \cite{liu2021federated} 
            & Accurate fine-grained 3D AQI prediction
    & FL-based aerial-ground sensing approach
    & Two FL streams are deployed collaboratively: a mobile vision-based aerial sensing stream for region-level AQI scale and a sensor-based ground sensing stream for small-scale spatial-temporal AQI inference.
		\\ \cline{2-5}
	{}
    & \cite{seid2021multiagent}
    & Computing task offloading and resource allocation in AIoT networks
    & MADRL approach
    & The joint MEC-UAV network congestion problem is formulated as an extension of the Markov decision process regarding the stochastic game to be solved by MADRL.
		\\ \cline{2-5}
	{}
    & \cite{zhang2020uav}
    &  Highly secure capacity in A2G channel links
    & MADRL approach with CT\&DE
    & CAA-MADDPG is with an attention layer network to capture attention weights associated with actions and observations of other agents in complicated multi-agent environments.
	\\ \hline
	
	\multirow{1}{*}{\rotatebox{90}{\textbf{~\makecell{IoT-enabled Smart Industry}~}}}
	& \cite{ren2018distributed}
         &  Aerofoil self-noise prediction in the aerospace industry
    & A distributed ML approach
    & A distributed optimization algorithm to cooperatively learn the prediction patterns at multiple edge devices.
		\\ \cline{2-5}
	{}
	&\cite{zhao2022afl}
         &  Function customization and model separation for different tasks
    & FL framework
    & An adaptively federated multi-task learning with a sparse shareable structure over an iterative pruning network. 
		\\ \cline{2-5}
	{}
	&\cite{huong2021detecting}
         & Cyberattack detection in IIoT-based manufacturing control systems
    & FL framework
    & Edge devices deploy a hybrid autoencoder-LSTM model to detect anomalies from time-series data over multi-scale time frames.
		\\ \cline{2-5}
	{}
    &\cite{liu2021deep}
            & Anomaly detection in manufacturing control systems.
    & FL framework
    & Edge devices deploy an attention mechanism-based hybrid CNN-LSTM model and a Top-\textit{k} selection-based gradient compression algorithm.
		\\ \cline{2-5}
	{}
    & \cite{wang2016multi}
            &   Optimal maintenance in resource-constrained flowline systems
    & Distributed MARL approach
    &  A set of maintenance policies is created for modeling correlative relations between the local decision made by each agent and the overall multi-agent optimization goal.
		\\ \cline{2-5}
	{}
    & \cite{su2022deep}
            &   Cost-efficient preventive maintenance
    & Deep MARL framework with CT\&DE
    & Dec-POMDP is specialized by a production loss-based reward function and a value-decomposition multi-agent actor-critic algorithm.
		\\ \cline{2-5}
	{}
    & \cite{yu2021optimizing}
            &   Optimal task scheduling in human-robot collaboration for manufacturing processes
    & DQN-MARL approach
    & The task scheduling problem is formulated into a Markov game model under multi-agent constraints to discover correlated equilibrium policies among agents in a dynamic real-time environment
	\\ \hline
	
	
	\end{tabular}
\end{table*}%

\subsection{Communication Efficiency}
In distributed learning, massive IoT devices share limited communication resources. Thus, limited resources of resource-poor IoT devices make the communication aspect in distributed learning much more important than the computing aspect (e.g., training and inference). It is reasonable since the training phase in distributed learning may require multiple communication rounds between the server and IoT devices in order to update necessary information, such as local data, model updates, and knowledge sharing. Moreover, many IoT devices may increase the model accuracy but can result in a communication bottleneck and high communication/computation costs. The limitation in network resources (e.g., bandwidth) can be relieved by adopting emerging 6G technologies (e.g., THz communications), which can offer a wide range of unused bandwidth. However, the advantage of abundant bandwidth resources is sacrificed by an increase in the IoT device complexity and a shorter communication distance.
Furthermore, the communication bottleneck can be caused by the statistical heterogeneity in distributed learning, where different IoT devices may have different data distributions, model properties, and learning environments \cite{liu2022fedbcd}. In FL, non-independent and identically distributed (non-IID) data is a major cause of model difference, and the fluctuation in data size may lead to communication inefficiency due to the straggling effect, i.e., different IoT devices may have different training and communication times. Therefore, it is vital to develop communication-efficient schemes in distributed learning to enable IoT applications and IoT services in future networks effectively \cite{park2021communication}.

Due to its importance, there has been a surge of research on improving communication efficiency in distributed learning. Based on the entire process of distributed learning frameworks (e.g., FL and SL), future works in this direction can focus on local training, IoT device selection, information sharing reduction, and information sharing compression. For example, \cite{liu2022joint} focuses on communication-efficient FL via joint model pruning and IoT device selection to reduce communication overhead and global loss. Based on the close-formed solutions of the model pruning ratio and wireless resource allocation, and a device selection mechanism, the proposed algorithm is evaluated on the real datasets, CIFAR10 and CIFAR100, and is demonstrated to have superior performance to alternative schemes, such as no-pruning and baseline device selection. The work in \cite{krouka2022communication} proposes a communication-efficient MARL solution using the alternating direction method of multipliers (ADMM) and an analog transmission scheme. The main notion of this work is to integrate wireless channel conditions in designing a distributed learning framework where the RL agents share the models via analog transmissions. In SL, a communication-efficient scheme is investigated in \cite{koda2020communication} to predict the received power of mmWave signals, showing a respective reduction of $93$\% and $2.8$\% in communication latency and data privacy over a baseline scheme. 
Although, as explained, communication efficiency plays a vital role in distributed learning, it is expected that more effort will be taken to carry out communication-efficient distributed learning and its applicability to the future IoT.

\subsection{Resource Allocation}
Due to limited computing and communication resources, massive connectivity in 6G IoT, and network dynamics of IoT communication environments, optimizing resource allocation can help improve the performance of distributed learning. On the one hand, it is necessary to jointly consider communication metrics in conventional IoT systems, such as bit error rate, energy, spectral efficiencies, outage probability, and network throughput. On the other hand, resource allocation schemes need to consider computing aspects from the learning perspective, such as computation efficiency, training model, and completion time. Further, unlike prior works on resource allocation frameworks in conventional IoT systems, solutions with distributed learning are required to address new challenges in future communication networks and distributed IoT systems. For example, in view of the scalability issue in multi-agent IoT systems, especially when the centralized resource allocation methods are designed, new solutions should be aware of that issue in network systems with a large number of IoT devices. In FL, different IoT devices may have different resource capabilities (e.g., storage, computing, and battery), which can raise critical issues in the design of FL solutions, such as straggling effects and asynchronous model aggregation \cite{saputra2022federated}. In this regard, resource management problems in FL networks can be improved by taking into account the resource constraints of IoT devices, communication channel conditions, and local AI models. In addition to these constraints in FL, the cut layer selection needs to be considered in SL to balance on-device and on-server AI models and to optimize the performance metrics, such as completion latency and energy consumption. 

Several approaches have been investigated to facilitate resource management in \textcolor{black}{IoT} with distributed learning. The work in \cite{naderializadeh2021resource} proposes a MARL agent for a scalable and distributed framework of user association and power control in the downlink of wireless networks. Each agent has both local and global observations to decide the set of serving users and the transmit power for downlink transmissions. Moreover, the innovative design of observations and action spaces with a fixed-size deep neural network makes the proposed multi-agent DRL framework more scalable to network configurations, such as the number of radio towers and IoT devices. What is more, the proposed DRL approach can obtain better performance than the centralized alternative as the ``treating interference as noise" design philosophy in the centralized methods does not always guarantee the solution optimality. The work in \cite{vu2022joint} shows that cell-free massive multiple-input multiple-output (MIMO) would be a suitable technique to mitigate the straggler effect in FL when it is compared with conventional multiple access techniques, such as TDMA and FDMA. Specifically, this work develops an efficient solution for a resource allocation problem that aims to minimize the execution time of a global communication round and limit the number of global communication rounds. Based on practical simulation settings, the results show that the proposed resource allocation method in \cite{vu2022joint} would significantly reduce the execution time by $50\%$ compared with the baseline schemes, where the set of participating users is preset. A promising SL resource management solution is \cite{wu2022split}, where a cluster-based parallel learning method is proposed to minimize the training latency. In particular, IoT devices are grouped into different clusters, and the resource management problem of cut layer selection, user clustering, and radio spectrum allocation is studied. Simulations with both IID and non-IID data confirm a low training latency of the proposed parallel SL approach. In particular, the total training latency of parallel SL is $2.33$ times smaller than that of vanilla SL, and the per-round training latency is $3.68$ and $8.84$ times smaller than those of vanilla SL and FL, respectively. The proposed solution also greatly reduces the training latency over the benchmarks, i.e., $80.1\%$ and $56.9\%$ compared to the heuristic and random schemes. Further, it is noted that there have been studies focusing on the communication efficiency of distributed learning, such as quantized FL \cite{wang2021quantized}, FL with quantized compressive sensing \cite{oh2022communication}, binarized SL (an extreme quantization case) \cite{pham2022binarizing}, and ensemble distillation \cite{liu2022ensemble}. It is promising to consider these communication-efficient studies over wireless networks with optimized resource allocation \cite{chen2022energy}. 

\subsection{\textcolor{black}{Metaverse}}

{\color{black}
The Metaverse, developed through the integration of emerging technologies like AI, deep learning, blockchain, edge computing, and extended reality, is expected to become the next revolution of IoT that can bring an immersive experience to users in the virtual world via virtual technologies \cite{huynh2022artificial, wang2023survey}. Together with emerging technologies, such as blockchain, SemCom, and cloud edge computing, the Metaverse is seen as a key enabler of Web 3.0 and a new digital economy \cite{liu2022slicing4meta}. AI, particularly distributed learning, plays a vital role in the Metaverse, enabling various technical aspects such as neural interface, natural language processing, digital twin, blockchain, communication networking, and machine vision \cite{huynh2022artificial}. AI also has potential applications in the Metaverse, such as healthcare, manufacturing, smart cities, gaming, e-commerce, and decentralized finance. In this context, distributed learning algorithms can facilitate Metaverse services and projects, paving the way for the IoT revolution. 

Distributed learning algorithms can be employed on massive physical devices for content creation, promoting creativity, and enhancing user experiences. As such, they can enhance the Metaverse by enhancing virtual experiences and enabling immersive 3D environments. Research efforts and investments have surged in the nexus of distributed learning and IoT, particularly since Facebook's renaming in 2021. Nevertheless, the integration of distributed learning into the Metaverse still requires further research to overcome challenges and realize the Metaverse's potential.
\begin{itemize}
   \item \textit{Scalability}: The Metaverse should be developed using a decentralized architecture, through which distributed learning exploits computing resources and massive data on physical devices. Game-based incentive mechanisms are necessary to encourage high-quality and reliable contributions from distributed AI models, while users can enjoy the benefits of trained AI models and Metaverse services. For example, a coded distributed computing framework for vehicular Metaverse services \cite{jiang2022reliable}, with a two-layer incentive scheme, can increase the utility of service providers and average profit of selected vehicles by $17\%$ and $14\%$, respectively. However, deploying distributed learning for IoT Metaverse services requires a learning server, raising design issues like single points of failure and free-riding attacks. Blockchain is a promising solution due to its decentralization, immutability, and traceability features, allowing distributed learning algorithms to be deployed decentralized without a learning server, enabling a scalable and secure Metaverse \cite{nguyen2021federated_BC}.

   \item \textit{Learning algorithm and networks}: Learning algorithms and networks play a crucial role in processing data from the Metaverse to create and adapt virtual worlds and agents. However, finding meaningful and efficient ways to represent data, extract features, attributes, and semantics, encode them into vectors or tensors, and compress or decompose them is challenging. Representation learning is essential for reducing data dimensionality and complexity, enhancing performance and interpretability, and enabling transfer learning and generalization across different domains and tasks \cite{huynh2022artificial}. Besides, defining optimal policies or actions for agents in the Metaverse is also another challenge, including state space models, action spaces, reward functions, value functions, balancing exploration and exploitation, partial observability, and coordinating multiple agents. In this regard, RL is particularly useful for enabling autonomous and adaptive behavior, enhancing intelligence and interactivity, and creating engaging experiences for users  \cite{Daniel2022introduction}. Moreover, collaborative and distributed ways to train learning algorithms using data from multiple sources or devices in the Metaverse without compromising privacy or security are also critical issues. FL becomes crucial for leveraging massive and diverse data, enhancing scalability and efficiency, and protecting data and users' rights.
\end{itemize}
}

\subsection{Specifications and Standardization}
AI and ML in general and distributed learning, in particular, are considered the key to the design of future IoT and 6G systems \cite{ali2020white}, but standardization of distributed learning capabilities is still in its preliminary stage due to challenges such as different devices, operating systems (e.g., Linux, iOS, and Windows), hardware (e.g., Raspberry Pi, iPhone, and autonomous vehicles), and computing and storage capabilities. Moreover, existing distributed learning solutions lack a unified framework, reference guidelines, and a standard open-source library \cite{baresi2022open}. In this context, to standardize distributed learning for the future IoT, collaboration between academic and industry communities is essential, such as standardization bodies (e.g., 3GPP, ITU, and IEEE) and big techs (e.g., Samsung and KT Corp). Recently, the 3GPP Release 18 was approved at the December 2021 technical specification group (\#94-e) meeting, including the study of AI/ML for 5G advanced air interface \cite{RP-213468}. Herein, AI/ML algorithms are mainly developed for evaluation purposes  \cite{R1-2205695}, and several bodies have been established to develop and promote standardization activities. The O-RAN Alliance, established in August 2018, enables centralized training at non-real-time RAN intelligent controllers (RICs) and distributed execution and online inference at real-time RICs at the network edge  (e.g., end users or base stations), thereby enabling the deployment of AI algorithms at multiple timescales \cite{polese2022understanding}. Despite standardization efforts, research and development of future IoT and distributed learning are still ongoing. The first 3GPP specification for future 6G systems is expected to be developed in 2026-2027, and network operators will deploy the first 6G trial in 2028-2029 \cite{de2021survey}. The emergence of new IoT services and applications in the coming years will likely require extensive efforts from industry and academia to standardize distributed learning for IoT services and applications in future networks.

{\color{black}On another front, a shift towards future Industry 5.0 also requires four key elements \cite{Lu2022Jan,Chi2022Oct}: a human-centric approach, resilience and sustainability, collaboration of humans and machines, and hyper-customization. To deal with such challenges, it is anticipated that distributed learning engagement plays a bridge in enabling self-learning and knowledge sharing, enhancing creativity, innovation, and well-being through adaptive learning paths, intelligent tutors, and collaborative robots. Besides, this approach can provide robust and flexible models for coping with disruptions, uncertainties, and changes. Moreover, leveraging advanced technologies like AI, robotics, cloud computing, and IoT enables the enhancement of human-machine collaboration and promotes more efficient production. Furthermore, by using diverse models such as deep learning, GANs, and RL, distributed learning solutions can enhance customer experience and hyper-customization.  However, the standardization of distributed learning capabilities, especially FL, for Industry 5.0 has many challenges. Firstly, distributed learning frameworks raise communication overhead due to frequent communication between devices or agents and the central server or coordinator. Secondly, heterogeneity in computing capabilities, data distributions, availability, or reliability can cause inconsistency or bias in the global model. Thirdly, unstable activities of devices or agents can limit coordination, synchronization, aggregation, or resource allocation for a large number of dynamic devices or agents on a large scale. Fourthly, distributed learning frameworks are vulnerable to malicious attacks from external or internal adversaries that compromise data privacy or integrity. Fifthly, using a distributed learning approach may result in unfairness among devices or agents in terms of resource consumption or model quality. In this cases, some possible solutions should be considered: i) Applying compression, quantization, sparsification, or pruning to the model updates or parameters before sending them to the central server or coordinator; ii) realizing clustering, personalization, adaptation, or meta-learning to the devices or agents or the global model to account for their different characteristics or preferences; iii) Adopting decentralization, parallelization, federation, or partitioning to devices or agents or central servers or coordinators to distribute the workload or responsibility; iv) using encryption, differential privacy, homomorphic encryption, secure multiparty computation, or blockchain to data or leaning models to protect them from unauthorized access or manipulation; and v) developing resource allocation, incentive mechanisms, reputation systems, or fairness metrics to devices or agents or global models to balance their resource consumption or model quality.}
%

		
		
		
		
		

\section{Conclusion}
\label{sec:Conclusion}
AI will be playing a key role in overcoming the limitations of 5G IoT and ensuring the development of future 6G IoT. Motivated by the massive data generated by a massive number of IoT devices and their associated privacy concerns, distributed learning is a powerful AI technique for optimizing 6G IoT and enabling its emerging applications and services.
In this survey, we have carried out a comprehensive survey on the use of distributed learning for IoT services and applications. \textcolor{black}{We have started with a preliminary to AI and important distributed learning approaches: FL, SL, MARL, and distributed inference.} Then, we have turned our attention to an extensive review and discussion on the use of distributed learning for emerging IoT applications and services in future 6G networks. Based on our extensive review, we have unveiled that several challenges remain in the reviewed literature, and a number of promising research directions are open in this interesting area, including security and privacy, communication efficiency, resource allocation optimization, the next Internet revolution, specifications, and standardization. This research area will require more effort, and this survey will help researchers drive more innovations and advancements in the future.  

\balance


\end{document}